\documentclass[reqno,prd,nofootinbib,longbibliography]{revtex4-1}
\usepackage{graphicx,color}
\usepackage[skip=-70pt]{caption}
\usepackage{amssymb}
\usepackage{amsmath}
\usepackage{epstopdf}
\usepackage{yfonts}
\usepackage{subcaption}
\usepackage{ragged2e}
\DeclareCaptionJustification{justified}{\justifying}
\newcommand\lsim{\mathrel{\rlap{\lower4pt\hbox{\hskip1pt$\sim$}}
        \raise1pt\hbox{$<$}}}
\newcommand\gsim{\mathrel{\rlap{\lower4pt\hbox{\hskip1pt$\sim$}}
        \raise1pt\hbox{$>$}}}

\def\Pl{{\mathrm{Pl}}}
\def\be{\begin{equation}}
\def\ee{\end{equation}}
\def\GeV{{\,\mathrm{GeV}}}
\def\Gyr{\,\mbox{Gyr}}
\def\eV{{\,\mathrm{eV}}}
\def\keV{{\,\mathrm{keV}}}

\def\mnras{Mon.\ Not.\ R.\ Astr.\ Soc.}
\def\apjl{Astrophys.\ J.\ Lett.}
\def\apj{Astrophys.\ J.}
\def\aap{Astron. \& Astrophys.}
\def\kpc{\,\mbox{kpc}}
\def\yr{\,\mbox{yr}}
\def\Mpc{\,\mbox{Mpc}}
\def\pc{\,\mbox{pc}}
\def\kms{\,\mbox{km s}^{-1}}
\def\msun{\,M_\odot}
\def\bnabla{\mbox{\boldmath $\nabla$}}
\def\bfsigma{\mbox{\boldmath $\sigma$}}
\def\p{\partial}
\def\half{{\textstyle{\frac{1}{2}}}}

\def\ffrac#1#2{{\textstyle\frac{#1}{#2}}}
\def\mm{\textfrak{M}\,}
\def\nn{\textfrak{N}\,}
\def\fdm{\textsc{fdm}}

\begin{document}

\title{Ultralight scalars as cosmological dark matter}
\author{Lam Hui} 
\email{lhui@astro.columbia.edu}
\affiliation{Department of Physics, Columbia University, New York, NY 10027}
\author{Jeremiah P. Ostriker} 
\email{ostriker@princeton.edu, jpo@astro.columbia.edu}
\affiliation{Department of Astronomy, Columbia University, New York,
  NY 10027}
\affiliation{Department of Astrophysical Sciences, Princeton University, Princeton, NJ 08544}
\author{Scott Tremaine}
\email{tremaine@ias.edu}
\affiliation{Institute for Advanced Study, Princeton,
  NJ 08540}
\author{Edward Witten}
\email{witten@ias.edu}
\affiliation{Institute for Advanced Study, Princeton,
  NJ 08540}

\begin{abstract}

Many aspects of the large-scale structure of the universe can be described successfully using cosmological models in which $27\pm1\%$ of the critical mass-energy density consists of cold dark matter (CDM). However, few---if any---of the predictions of CDM models have been successful on scales of $\sim10\kpc$ or less.  This lack of success is usually explained by the difficulty of modeling baryonic physics (star formation, supernova and black-hole feedback, etc.). An intriguing alternative to CDM is that the dark matter is an extremely light ($m\sim 10^{-22}\eV$) boson having a de Broglie wavelength $\lambda\sim 1\kpc$, often called fuzzy dark matter (FDM). We describe the arguments from particle physics that motivate FDM, review previous work on its astrophysical signatures, and analyze several unexplored aspects of its behavior.  In particular, (i) FDM halos or sub-halos smaller than about $10^7 (m/10^{-22}\eV)^{-3/2}\msun$ do not form, and the abundance of halos smaller than a few times $10^{10} (m/10^{-22}\eV)^{-4/3}\msun$ is substantially smaller in FDM than in CDM; (ii)  FDM halos are comprised of a central core that is a stationary, minimum-energy solution of the Schr\"odinger--Poisson equation, sometimes called a ``soliton'', surrounded by an envelope that resembles a CDM halo. The soliton can produce a distinct signature in the rotation curves of FDM-dominated systems. (iii) The transition between soliton and envelope is determined by a relaxation process analogous to two-body relaxation in gravitating N-body systems, which proceeds as if the halo were composed of particles with mass $\sim\rho\lambda^3$ where $\rho$ is the halo density. (iv) Relaxation may have substantial effects on the stellar disk and bulge in the inner parts of disk galaxies, but has negligible effect on disk thickening or globular cluster disruption near the solar radius. (v) Relaxation can produce FDM disks but an FDM disk in the solar neighborhood must have a half-thickness of at least $\sim 300 (m/10^{-22}\eV)^{-2/3}\pc$ and a mid-plane density less than $0.2(m/10^{-22}\eV)^{2/3}$ times the baryonic disk density. (vi) Solitonic FDM sub-halos evaporate by tunneling through the tidal radius and this limits the minimum sub-halo mass inside $\sim30\kpc$ of the Milky Way to a few times $10^8 (m/10^{-22}\eV)^{-3/2}\msun$. (vii) If the dark matter in the Fornax dwarf galaxy is composed of CDM, most of the globular clusters observed in that galaxy should have long ago spiraled to its center, and this problem is resolved if the dark matter is FDM. (viii) FDM delays galaxy formation relative to CDM but its galaxy-formation history is consistent with current observations of high-redshift galaxies and the late reionization observed by {\it Planck}. If the dark matter is composed of FDM, most observations favor a particle mass $\gtrsim 10^{-22}\eV$ and the most significant observational consequences occur if the mass is in the range 1--$10\times 10^{-22}\eV$. There is tension with observations of the Lyman-$\alpha$ forest, which favor $m\gtrsim 10$--$20\times 10^{-22}\eV$ and we discuss whether more sophisticated models of reionization may resolve this tension.

\end{abstract}

\maketitle

\vskip2cm

\section{Introduction}

\noindent
The standard Lambda Cold Dark Matter (``$\Lambda$CDM'') model for the
mass-energy content of the universe and the development of cosmic
structure has been remarkably successful. In this model the universe
is geometrically flat and the largest contributors to the mass-energy
are dark energy, $68\pm1\%$ of the total, and dark matter, $27\pm1\%$
\cite{planck15}, both of unknown nature. Most likely the dark matter
consists of some undiscovered elementary particle(s), produced early in
the history of the universe, that is ``cold'' in the sense that the
effect of its velocity dispersion on structure formation is
negligible. Ordinary or baryonic matter is a minor constituent
($5\%$), and neutrinos and other light or zero-mass particles make an
even smaller contribution to the total mass-energy density. This
mixture of components, with small density fluctuations normalized to
the observed fluctuations in the cosmic microwave background radiation
field and allowed to grow via gravitational instabilities, can account
for the properties of the structures in the universe at most
well-observed scales and epochs. The initial spectral index of the
perturbations is slightly less than unity, $n_s = 0.965 \pm 0.006$, consistent with simple theories in which the universe passes through an
early inflationary stage. N-body simulations with initial linear
fluctuations having this spectral index show that the non-linear
dark-matter structures that develop---called ``halos'' if isolated and
``sub-halos'' if embedded in a larger halo---are hierarchical,
with every halo and sub-halo having embedded within it sub-halos, in a
roughly self-similar fashion. These simulations also show that halos
and sub-halos have singular density cusps at their centers, with the
density varying with radius roughly as $\rho(r)\propto r^{-1}$
\cite{dc91,nfw97}.

On scales larger than those of the stellar distribution in normal
galaxies---say $10\kpc$ or more---the predictions of $\Lambda$CDM have
been amply tested, and, although neither semi-analytic theory nor
numerical simulations can yet match all that we observe, the correspondence
between calculation and observation is now good enough to assure us
that {\it on large scales} the $\Lambda$CDM model is essentially
correct.  Particularly impressive is that the power spectrum of mass
fluctuations, determined at redshift $z \sim 10^3$ by observations of the
cosmic microwave background, correctly produces the power spectrum of
mass fluctuations at the present time, redshift $z \simeq 0$, to within a few percent even
though the amplitude at $z=0$ is some five orders of magnitude larger
than it was at $z\sim 10^3$.

On the other hand, on scales similar to those of galaxies, $\lesssim
10\kpc$, the CDM model has essentially not been tested. In fact, the
na\"ive predictions of the distribution of dark matter on these scales
are in most cases {\it inconsistent} with observations \cite{wein13}.  As one
example, the number density of galaxies varies with their total
stellar mass roughly as $dn(M_\star)\propto M_*^{-1.2}dM_*$, but the
predicted number density of halos increases with decreasing halo mass
much more steeply, $dn(M_h)\propto M_h^{-2}dM_h$. This apparent
discrepancy is usually attributed to ``baryonic physics'' that causes
the efficiency of transforming baryons into stars to be lower in
systems of lower mass.  Consequently thousands of optically invisible
low-mass halos ($M_h\lesssim 10^{8.5} \msun$) are predicted to exist
in our galaxy and others. A second example is the ``too big to fail'' problem: the high-luminosity satellite galaxies associated with the most massive sub-halos appear to be much less common than CDM would predict. A third issue is that the expected dark-matter density cusps in
the centers of galaxies have not been detected; this discrepancy is
usually attributed to gravitational stirring of the central regions of
galaxies consequent to supernova explosions, but in systems having a low
baryonic fraction that explanation is problematic.
 Thus the hierarchical nature of density fluctuations predicted by CDM is amply established on large scales such as clusters of galaxies but---to date---the expected distribution of sub-halos within the Milky Way and other galaxies has evaded detection. Given the
complexity of the physics, these discrepancies are not crippling
blows to the CDM model, but it can also be said that none of the
characteristic features associated with CDM on galaxy scales has ever
been detected.

A variant to CDM is warm dark matter (``WDM''), in
which the mass of the hypothetical dark-matter particle is
sufficiently small that its thermal velocity dispersion has a
significant influence on structure formation \cite{Viel2005}.
The linear power spectrum in WDM is greatly reduced below
the free-streaming scale, suppressing the formation of
low-mass halos or sub-halos \cite{Colin2000,Bode2001}.
The finite phase-space density
prevents the development of density cusps \cite{tg79}, 
though the implied central core size is generally small
\cite{Colin2000,Bode2001}.

An alternative hypothesis is that the dark matter is comprised of
very light bosons or axions, $m \sim 10^{-22}$--$10^{-21} \eV$ 
\cite{turner83,press1990,sin94,FDM,good00,peebles00,Amendola:2005ad,scb14}. All large-scale predictions are the same as in $\Lambda$CDM, but the
particle's large de Broglie wavelength (eq.\ \ref{eq:brog}) suppresses
small-scale structure. This material is sometimes termed wave dark
matter or fuzzy dark matter (``FDM''), a term introduced in a seminal paper by Hu, Barkana, and Gruzinov \cite{FDM}. A review of axion
cosmology is given by Marsh \cite{mar15}. 
Constraints on FDM from the cosmic microwave background are described in
\cite{hlo15}.
An attractive feature of FDM with a mass in this range is that a cosmic abundance $\Omega_m \sim 1$ can
arise in naturally occurring models, as we describe below. 
FDM, by virtue of its macroscopic de Broglie wavelength, 
possesses novel features in its nonlinear dynamics.
In this paper we assess the FDM hypothesis, we provide
some new calculations of the properties of FDM, and we propose a set of
tests to ascertain its consistency with observations. 
Both WDM and FDM, because they share a suppression of the
power spectrum on small scales, are subject to strong constraints from
the Lyman-$\alpha$ forest \cite{viel13}, which we also discuss below.

We note that all of the published (to date negative) experimental
searches for dark matter and its decay products would not have
detected the very light bosons that we are proposing for the principal
component of the dark matter, although experiments have been
suggested which have the potential to detect FDM (see discussion at
the end of \S\ref{sec:spin}). 

In \S\ref{sec:physics} we review the physical motivation for the FDM
hypothesis, in \S\ref{sec:FDMastrophysics}, \ref{sec:form}, \ref{sec:forest} we outline 
the astrophysical consequences,
tests and predictions of the model and in \S \ref{sec:summary} we summarize our
conclusions. Technical calculations are relegated to several Appendices.
Natural units, where $\hbar = c = 1$, are used in
\S\ref{sec:spin}, while factors of $\hbar$ and so on are kept explicit
in the rest of the paper---the only exception is that we use $\eV$ instead of
$\eV/c^2$ to denote mass throughout.

\section{The physics of FDM}

\label{sec:physics}

\subsection{Light fields of spin zero}

\label{sec:spin}

\noindent
The basic reason why it is natural in particle physics to have a very
light field of spin zero is that when the mass and self-couplings of a
spinless field $\phi$ are precisely zero, there is an extra
symmetry.  Thus the action
\begin{equation}
\label{thaction}
I=\ffrac{1}{2}\int d^4x\sqrt {-g} \,g^{\mu\nu}\partial_\mu \phi\,\partial_\nu
\phi 
\end{equation}
has the shift symmetry $\phi\to\phi+C$, with constant $C$.  This
symmetry is lost if a mass term $\frac{1}{2}m^2\phi^2$ or a more
general self-coupling $V(\phi)$ is added to the action, and likewise it could
be violated by the couplings of $\phi$ to other fields.

The candidate particle for FDM is not a precisely massless boson, but
one that is very nearly massless.  Thus $\phi$ must have an
approximate shift symmetry, not an exact one.  In fact, it is
suspected that in quantum gravity all continuous global symmetries,
such as the shift symmetry of a scalar field, are only 
approximate\footnote{As a prototype of this phenomenon, for many years it appeared that
the three separate lepton numbers $L_e$, $L_\mu$, and $L_\tau$ were 
conserved quantities in particle physics. We now know from observations of
neutrino oscillations that at least the separate lepton numbers $L_e-L_\mu$ and
$L_\mu-L_\tau$ are violated at a very low level.}.
So we expect the symmetry to be broken at some level.

There is at least one situation in which a very light spin-zero field
arises naturally: if the field $\phi$ is an angular variable so the
potential function $V(\phi)$ must be a periodic function of $\phi$.
Such periodic spin-zero fields often arise in naturally occurring
models, by which we mean models that were not invented for the purpose
of having approximate shift symmetries.  Moreover, it then often turns
out that the shift symmetries are violated only by an exponentially
small amount. These fields are sometimes called axion-like fields and
the corresponding particles are likewise called axion-like particles
because one of them could be a candidate for the QCD axion.

An axion-like candidate for FDM can be largely described by a simple
 model with two parameters\footnote{The assumptions that lead to a model
 of this form also greatly suppress the coupling of the axion to ordinary matter.
 Both for this reason and because it is so light, the FDM particle would not be detected
 in the usual searches for WIMP dark matter or even in QCD axion searches
 \cite{Sikivie:1983ip,Hoskins:2011iv,Ruz:2015uka}.} $\mu$ and $F$:
\begin{equation}
\label{twop}
I =\int d^4x\sqrt {-g}\,\Big[\half F^2g^{\mu\nu}\partial_\mu
  a\,\partial_\nu a - \mu^4(1-\cos a)\Big]. 
\end{equation}
Here $a$ is a dimensionless scalar field with a shift symmetry
$a\to
a+2\pi$.  Since $a$ is dimensionless, its kinetic energy involves the
constant $F$ (sometimes called the axion decay constant, but this
terminology is misleading in the context of FDM).  The periodicity
$a\cong a+2\pi$ would allow us to add to the action terms involving
higher harmonics of $a$ ($\cos ka$ or $\sin ka$ with integer $k$ and
$|k|>1$), but in models in which $\mu$ is small enough to be relevant
for our applications, those higher harmonics would have completely
negligible coefficients.  The mass of $a$ is
\begin{equation}
\label{genmass}
m=\frac{\mu^2}{F},
\end{equation}
and for FDM we want $m\sim 10^{-22}$--$10^{-21} \,\eV$. 

For example, all models of particle physics derived from string theory have at
least several periodic scalar fields such as $a$, and typical models
have many of them (dozens or even hundreds).  Various possible
applications of these fields have been considered (see for example \cite{Axiverse}).  With different assumptions about their masses,
they have been
proposed as candidates for the inflaton field of inflationary
cosmology; as potential QCD axions, whose existence may explain CP
conservation by nuclear forces; and as contributors to dark energy or
the cosmological constant and/or ingredients in a mechanism to explain
its smallness.  For our present purposes, we are interested in these
fields as candidates for FDM.

In most classes of model, $F$ lies within a relatively narrow range
bounded above by the reduced Planck mass $M_\Pl=1/\sqrt{8\pi G}=
2.435\times 10^{18}\GeV$, and below by the traditional ``grand unified''
scale of particle physics, $M_G\sim 1.1\times 10^{16}\GeV$:
\begin{equation}
\label{orange}
10^{18}\GeV \gtrsim F\gtrsim 10^{16}\GeV. 
\end{equation}
The lower bound is the actual value for the ``model-independent''
axion of the weakly coupled heterotic string \cite{Choi}.  For
applications to inflation, some authors have attempted to increase $F$
beyond the upper bound of $\sim 10^{18}\GeV$, but this has proved difficult \cite{BD}, except in models in which $\mu$ is so large that
the axion is not relevant.   The lower bound, which causes difficulty for some approaches to the
QCD axion, is less firm but is valid in many classes of models \cite{SW}.
As we shall see, for applications to FDM it is
satisfactory to take the range (\ref{orange}) at face value.
 
What about the second parameter $\mu$ in the action?  This is
generated by nonperturbative instanton effects of one kind or another,
depending on the model.  A rough formula is
 \begin{equation}
\label{zorf}
\mu^4\sim M_\Pl^2 \Lambda^2 e^{-S},
\end{equation}
where $S$ is the instanton action and $\Lambda$, which measures a
possible suppression of instanton effects due to supersymmetry, can
vary over a very wide range:
 \begin{equation}
\label{noroz}
10^{18}\GeV\gtrsim \Lambda \gtrsim 10^4\GeV. 
\end{equation}
Typical values are $\Lambda\sim 10^{18}\GeV$ (no suppression due to
supersymmetry), $\Lambda\sim 10^{11}\GeV$ (gravity-mediated
supersymmetry breaking), $\Lambda\sim 10^4\GeV$ (gauge-mediated
supersymmetry breaking).  The formula (\ref{zorf}) is only a very
rough one, but it is good enough to give an idea of how large $S$ must
be so that $m=\mu^2/F$ will be close to $10^{-22}\,\eV$.  Setting
$F=10^{17}\GeV$ (a different value in the range (\ref{orange}) would
make little difference, given the uncertainties), we find that to get
$m=10^{-22}\,\eV$, we need
 \begin{equation}
\label{actions}
S=\begin{cases} 165 & \mathrm{if}~\Lambda=10^4\GeV \cr
                                                     198 & \mathrm{if}~\Lambda=10^{11}\GeV\cr
                                                      230 &
                                                      \mathrm{if}~
                                                      \Lambda=10^{18}\GeV.\end{cases}
\end{equation}

The value of $S$ is rather model-dependent.  In some simple cases, one
finds $S\sim S_0=2\pi/\alpha_G$ where $\alpha_G$ is the Standard Model
gauge coupling extrapolated to energy $M_G\sim 1.1\times
10^{16}\GeV$.  In many classes of model, $S$ can be close to $S_0$
but not significantly bigger \cite{SW}. The value of $\alpha_G$ depends on the assumed
spectrum of elementary particles up to the unification scale.  The usual estimate based on
known particles only gives $\alpha_G\sim 1/25$, but for
example the model in \cite{SS} has $\alpha_G\sim 1/30$.  Some values
of $S_0$ are
\begin{equation}
\label{actionbase}
S_0=\begin{cases} 126 &\mathrm{for} ~\alpha_G=1/20\cr
                                                       157&\mathrm{for}~\alpha_G=1/25\cr
                                                       188&\mathrm{for}~\alpha_G=1/30.\end{cases}
\end{equation}
Thus the range of $S_0$ overlaps the desired range for $S$.  

These numbers should not be taken very seriously, even if the basic 
ideas are all correct, because $2\pi/\alpha_G$ is only a very rough
value for $S$. The reason for writing them was only to show that a mass of
$10^{-22}\, {\mathrm{eV}}$ is reasonable.  
In models with dozens or hundreds of axion-like
fields, it may readily happen that the largest values of $S$ are of
order $2\pi/\alpha_G$, but that most of the axions have much smaller
values of $S$. The axions with $S\ll 2\pi/\alpha_G$ are not candidates
for FDM.  In \cite{S} (a paper that was motivated by
applications to quintessence and dark energy), it is estimated that it
is difficult to make $S$ much larger than 200--300, but there is ample
uncertainty in this upper bound.

To estimate what value of $F$ will work well for FDM, we follow
classic reasoning \cite{PMW,AS,DF} that was originally developed to
estimate the contribution to dark matter of a QCD axion.  Analogous
estimates for FDM have been made by several authors
\cite{Axiverse,KM}. (The case of FDM is simpler as, in a minimal model of this type,\footnote{In a model
in which, beyond well-established physics, only one field $a$ is added,  the $\mu^4 \cos a$ interaction can be understood
as some sort of instanton effect (as described above), and in discussing the cosmic evolution, both $F$ and $\mu$ can be treated as constant
parameters. It is also possible to make a more complicated model with, for example, a new gauge force that becomes strong at an energy of order $\mu$,
generating the $\mu^4\cos a$ term.  Then,  as in the case of the QCD axion, at temperature $T>\mu$,
the $\mu^4\cos a$ must be replaced by $(\mu/T)^n \mu^4\cos a$, where
$n$ is a model-dependent constant.} one does not have
to estimate the turning on of strong coupling effects such as those of QCD.)  One
assumes that in the very early universe, the $a$ field was a constant
with some random initial value.  The logic in assuming a random
initial value is this.  Instead of taking the axion potential in Eq.\
(\ref{twop}) to be $\mu^4(1-\cos a)$, we could just have well have
assumed $\mu^4[1-\cos(a-a_0)]$ with some constant $a_0$.  The correct
value of $a_0$ depends on details of the mechanism that breaks the
shift symmetry at low energies.  In writing Eq.\ (\ref{twop}), we
shifted the field $a$ to set $a_0=0$, but whatever mechanism
determines the value of $a$ in the early universe has no way to
``know'' what value of $a$ will minimize the potential at low
energies.

Starting with a random initial value of $a$, one determines its
behavior in an expanding FRW universe with metric $d s^2=dt^2-R(t)^2
d{\bf x}^2$ by simply solving the classical equation of motion for a
field that depends on time only. This equation is
\begin{equation}
\label{eqmotion}
\ddot a+3 H \dot a+m^2\sin a =0, 
\end{equation}
with $H=\dot R/R$ the Hubble constant.  An approximation to the
 behavior of this equation is that $a$ is constant as long as $H\gtrsim m$,
and then oscillates with angular frequency $m$ (or $mc^2/\hbar$, if
one restores $\hbar$ and $c$).  The oscillations are damped as
$R^{-3/2}$.  In the period in which $a$ is oscillating, it can be
interpreted as describing a Bose condensate of ultralight particles of
zero spatial momentum.  The energy density of these particles scales
as $1/R^3$, like any other form of cold dark matter.  Indeed, the
axion ``condensate,'' which is just a fancy way to speak of the
classical axion field, behaves for many purposes as an exceptionally cold
form of CDM.

The temperature $T_0$ at which $H\sim m$ satisfies roughly
\begin{equation}
\label{zodd}
\frac{T_0^2}{M_\Pl}=m. 
\end{equation}
At that temperature, the total energy density of radiation is roughly
$T_0^4$ and the dark-matter density (with $a\sim 1$ in Eq.\ 
(\ref{twop})), is of order $\mu^4$.  As the universe expands, the
ratio of dark matter to radiation grows as $1/T$, and in the real
world, they are supposed to become equal at the temperature $T_1\sim
1\,\eV$ at which the universe becomes matter dominated.  So we want
\begin{equation}
\label{balform}
\frac{\mu^4}{T_0^4}\frac{T_0}{T_1}\sim 1. 
\end{equation}
Combining these formulas and using $\mu^4=F^2m^2$, we get
\begin{equation}
\label{goodform}
F\sim \frac{M_\Pl^{3/4}T_1^{1/2}}{m^{1/4}}\sim 0.5\times
10^{17}\GeV \, ,
\end{equation}
for $m \sim 10^{-22} \eV$. The fact that this is in the range described earlier in Eq.\ (\ref{orange}) is
an interesting coincidence, somewhat reminiscent of the WIMP miracle. In other words, the axion energy density today (normalized by the critical density) is:
\begin{equation}\label{eq:miracle}
\Omega_{\rm axion} \sim 0.1 \left( \frac{F}{\vphantom{\big|} 10^{17} {\,\rm GeV} }\right)^2 \left( \frac{m}{\vphantom{\big|} 10^{-22} {\,\rm eV}} \right)^{1/2} \, .
\end{equation}
The temperature $T_0$ at which the FDM field begins to oscillate is
\begin{equation}
\label{osctemp}
T_0\sim (m \, M_\Pl )^{1/2} \sim 500 \,\eV.
\end{equation}
This corresponds to a redshift $\sim 2 \times 10^6$.  This is after nucleosynthesis,
so FDM behaves during nucleosynthesis as a (negligible) contribution to dark energy.

In the axion model, is the dynamics of dark matter purely
gravitational or do we have to consider the axion self-interaction?  In the
two-parameter model of Eq.\ (\ref{twop}), once $F\sim 10^{17}\GeV$
and $m\sim 10^{-22}\,\eV$ are determined from observed properties of
dark matter, the coefficients of the nonlinear axion interactions are
determined.  Thus the axion equation of motion, including terms of
cubic order in $a$, is
\begin{equation}
\label{cubicapprox}
0= D_\mu D^\mu a+m^2 a -\frac{m^2}{6}a^3+\mathcal{O}(a^5). 
\end{equation}
To decide whether the $a^3$ term, which is attractive, is significant, we can proceed as follows.
In a body with a dimensionless gravitational potential
$\varepsilon$, the gravitational contribution to the equation is of
order $\varepsilon m^2a$.  The condition for the $a^3$ term to be
significant in comparison to gravity is therefore $m^2a^3\gtrsim
\varepsilon m^2 a$ or $a^2\gtrsim \varepsilon$.  We will evaluate this
condition in the very early universe and in today's universe.

At the temperature $T_0$ at which dark matter begins to oscillate
dynamically, we have $a^2\sim 1$ (since we have assumed a random
initial value of $a$) and $\varepsilon\sim 10^{-5}$ (the value of the
primordial cosmic fluctuations).  Thus $a^2\gg\varepsilon$ and the
self-interaction of $a$ dominates.  This continues until $a$
diminishes by a factor of about $10^{-5/2}$.  Since $a\sim R^{-3/2}$,
gravity dominates once $R$ increases by a factor of about $10^{5/3}$.
Thus gravity dominates below a temperature of roughly $10^{-5/3}T_0$,
and in particular gravity dominates by the time the temperature $T_1$
of radiation-matter equality is reached.

On the other hand, for a weakly bound object of
density $\rho$ and size $L$ in today's universe, we have $\varepsilon\sim G \rho L^2$.
As the dark-matter density is $\rho\sim F^2 m^2 a^2$, the condition
$a^2\gtrsim \varepsilon$ for the non-gravitational force to be
significant is $1\gtrsim GF^2m^2 L^2$ or
\begin{equation}
\label{criterion} 
L\lesssim\frac{\sqrt{8\pi} M_P}{F m}. 
\end{equation}
Quantitatively, for $F\sim 10^{17}\GeV$, $m\sim 10^{-22}\,\eV$, this
says that gravity dominates for a dark-matter dominated object of size
greater than roughly one parsec.

What are the difficulties with the axion approach to FDM?  We will just mention a couple of the more obvious issues.
One question is what to make of axion-like particles other than the
one that hypothetically makes up FDM.  Axions with a
larger value of $S$ (and therefore a smaller mass) than the FDM particle are really not a problem.  They simply make small
contributions to the dark matter or the dark energy in today's
universe.\footnote{ Axion-like fields that are oscillating in today's
  universe represent contributions to dark matter.  Those with masses
  less than today's Hubble constant are still not oscillating in the
  present universe and represent contributions to the dark energy.  It
  is argued in \cite{S} that axion contributions are too small to
  account for all of the dark energy except possibly in a situation
  with a very large number of axions contributing.  (In any event, we stress that
  an axion in the mass range appropriate to FDM definitely behaves as a contribution to
  dark matter, not dark energy.)}
Axions with a mass larger than about $10^4$ GeV (which corresponds to
$S\sim 50$ if, for example, $\Lambda\sim 10^{11}\GeV$) may decay
quickly enough to cause no cosmological difficulties.  However,
axion-like particles with masses in the large range
$10^4\GeV \gtrsim m \gtrsim 10^{-22}\,\eV$ could potentially create too much dark matter.
An optimist might hope that the world is described by a model in which
there are no axions with (roughly) $50<S<200$.  Another issue concerns
the tensor-to-scalar ratio $r$ of cosmological perturbations.  In a
model in which $r$ is large enough to be observed, under the simplest
assumptions a QCD axion in the range $F\gtrsim 10^{16}\GeV$
that is assumed in the above discussion  leads to isocurvature
fluctuations that are excluded by cosmological observations. (For an assessment,
see for example \cite{Marsh:2014qoa}. The
problem arises because the same mechanism that leads to quantum
fluctuations of order $H/2\pi$ for the inflaton field leads to
fluctuations of the same order for any sufficiently light axion field.
Independent fluctuations in two different scalar fields lead to
isocurvature perturbations.)  There is an analogous although less
severe problem for an FDM particle.  The problem is less
severe because the parameters of this particle have been adjusted so
that the maximum dark matter it can produce is comparable to what is
observed (by contrast a QCD axion with $F\gtrsim 10^{16}\GeV$ is at
risk of producing much too much dark matter). But there still is a
potential difficulty, if $r$ is observed.  We will have to wait to see
if this is a problem that needs to be solved.

If FDM exists, can it be detected in any way other than
by observing its gravitational effects? Conventional dark matter searches would
not find the FDM particle both because it is much too light and because
it couples much too weakly to ordinary matter (if it had strong nongravitational
couplings to ordinary matter, then allowing for quantum effects, it would not be as light
as described above).\footnote{
A derivative coupling of the axion to fermions is allowed, as is  a coupling $a\epsilon^{\mu\nu\alpha\beta}f_{\mu\nu}f_{\alpha\beta}$
where $f_{\mu\nu}$ is the electromagnetic field strength.  Such couplings will be proportional to $1/F$ in the sort of model
assumed above, which drastically suppresses their effects.
}
However, a number of proposals for
direct, or at least more direct, observation of ultralight axion-like
particles have been made in \cite{Axiverse}.  The value $m\sim
10^{-22}$--$10^{-21}\,\eV$ is not optimal for most of these proposals, but is at
the edge of what might be detected by observing certain effects
involving supermassive black holes.  It has been suggested (in
\cite{KM} and by P. Graham, private communication) that the
CASPEr-Wind version of the CASPEr axion experiment might ultimately
have the sensitivity to observe FDM.  This experiment is
described, though not by that name, in section 5A of
\cite{CASPEr}.  FDM might also be eventually detectable less directly by pulsar timing
observations \cite{kr14}.

\subsection{FDM as a superfluid}
\label{sec:FDMsuperfluid}

\noindent
From the above discussion, we see that for the purpose of studying
structures on galactic scales and above, we can ignore the
self-interaction of the axion-like particle. In other words, let us
consider the following action for a scalar\footnote{We use
a canonically normalized scalar field $\phi$, related to the dimensionless
field $a$ used in the last subsection by $\phi=Fa$.} $\phi$, minimally coupled
to the metric $g_{\mu\nu}$:
\begin{equation}\label{scalaraction}
S = \int \frac{d^4 x}{\hbar c^2} \sqrt{-g} \left[ \half
  g^{\mu\nu} \partial_\mu \phi\, \partial_\nu \phi - \half \frac{m^2 c^2}{\hbar^2} \phi^2 \right] \, ,
\end{equation}
where we have restored factors of $c$ and $\hbar$, with
$\phi$ having energy units.
We are interested in a mass $m$ that corresponds to an astronomically
relevant de Broglie wavelength:
\begin{equation}
\frac{\lambda}{2\pi}=\frac{\hbar}{mv}=1.92\kpc \left( \frac{10^{-22} \eV}{m} \right) \left( \frac{10\kms}{v} \right) 
\label{eq:brog}
\end{equation}
where $v$ is the velocity.  A collection of a large number of such
particles in the same state can be described by a classical scalar
field.
In the non-relativistic limit, it is helpful to express $\phi$ in
terms of a complex scalar $\psi$: 
\begin{equation}
\phi = \sqrt{\frac{\hbar^3 c}{2m}} \left(\psi \, e^{-i{mc^2 t /\hbar}} +
\psi^* \, e^{i{mc^2 t /\hbar}}\right) \, .
\end{equation}
As is well-known, the equation of motion for $\psi$ takes the form of the Schr\"odinger equation,
assuming $|\ddot\psi| \ll m c^2 |\dot\psi | / \hbar$:
\begin{equation}
\label{Schrodinger}
i \hbar\left( \dot\psi + \ffrac{3}{2} H\psi \right) =  \left( -
  \frac{\hbar^2}{2m R^2}\nabla^2
   + m \Phi \right) \psi \, ,
\end{equation}
where $\Phi({\bf r},t)$ is the gravitational potential and we have adopted the perturbed FRW metric:
\begin{equation}
ds^2 = \left( 1+ \frac{2\Phi}{c^2} \right) c^2 dt^2 - R^2(t) \left(1
  - \frac{2\Phi}{c^2} \right) d{\bf r}^2 \, .
\end{equation}
For many galactic dynamics applications it is sufficient to set the
scale factor $R(t)$ to unity, and the Hubble parameter $H\equiv \dot R/
R$ to zero.  The scalar $\psi$ should be interpreted as a classical
field, quantum fluctuations around which are small. The situation is
analogous to using the Maxwell equations to describe configurations
involving a large number of photons (see \cite{Guth2014} for a
discussion; see also \cite{WidrowKaiser,CU2014} for the use of the Schr\"odinger equation
in modeling large scale structure).

It is sometimes useful to think of the dark matter as a fluid, a superfluid in fact. Define
the fluid density $\rho$ and velocity ${\bf v}$ by:
\begin{equation}
\label{eq:madelung}
\psi \equiv \sqrt{\frac{\rho}{m}} e^{i\theta} \quad , \quad {\bf
  v}\equiv \frac{\hbar}{R\,m} \bnabla \theta
=\frac{\hbar}{2mi R}\left(\frac{1}{\psi}\bnabla\psi -\frac{1}{\psi^*}\bnabla\psi^*\right)\, .
\end{equation}
The vorticity of the flow $\bnabla\times{\bf v}$ vanishes, though the more physically relevant quantity
is the momentum density which has non-zero curl in general.
The following equations can be derived from the $\psi$ equation of
motion in comoving coordinates:
\begin{align}
\label{massconserv}
\dot\rho + 3 H \rho + \frac{1}{R}\bnabla \cdot (\rho {\bf v}) &= 0 \, , \\
\label{Euler} 
\dot {{\bf v}} + H {\bf v} + \frac{1}{R} ({\bf v} \cdot \bnabla) {\bf v} &= -\frac{1}{R} \bnabla \Phi + \frac{\hbar^2}{2 R^3 m^2} \bnabla
\left( \frac{\nabla^2 \sqrt{\rho}}{\sqrt{\rho}} \right) \, .
\end{align}
These are known as the Madelung equations, slightly generalized to an
expanding universe (see the Feynman lectures \cite{Feynman} for a discussion, and also \cite{spiegel80,Chavanis2011,suarez2011,uhlemann2014,mar15a}).
They strongly resemble the continuity and Euler equations of classical
fluid mechanics with the addition of the second term on the right of
Eq.\ (\ref{Euler}), commonly referred to as the quantum pressure
term. The quantum pressure gives rise to a certain ``stiffness'' of the FDM fluid that resists compression.

More precisely, the quantum pressure arises from a stress tensor $\bfsigma$,
\begin{equation}
\dot {{\bf v}} + H {\bf v} + \frac{1}{R} ({\bf v} \cdot \bnabla) {\bf
  v} = -\frac{1}{R} \bnabla \Phi + \frac{1}{R}\bnabla\cdot\bfsigma
\end{equation}
where
\begin{equation}
\sigma_{ij}=-\frac{\hbar^2}{4m^2R^2}\left(\frac{1}{\rho}\frac{\p\rho}{\p
    x_i}\frac{\p\rho}{\p x_j}-\frac{\p^2\rho}{\p x_i\,\p
    x_j}\right)=\frac{\hbar^2\rho}{4R^2m^2}\frac{\p^2\log\rho}{\p x_i\,\p x_j}.
\end{equation}
An equivalent form is
\begin{equation}
\p_t(\rho v_i)+4H\rho v_i+\frac{1}{R}\p_j
\Pi_{ij}+\frac{1}{R}\rho\p_i\Phi=0
\label{eq:fffggg}
\end{equation}
where the momentum flux density tensor is
\begin{equation}
\Pi_{ij}=\rho
v_iv_j-\sigma_{ij}=\frac{\hbar^2}{4mR^2}\left(\p_i\psi^*\p_j\psi
+\p_i\psi\p_j\psi^*-\psi^*\p_i\p_j\psi-\psi\p_i\p_j\psi^*\right)
\label{eq:momentum}
\end{equation}
up to the addition of a divergence-free tensor. 

The Madelung equations are well-suited to numerical simulations,
because standard hydrodynamics codes can be modified to incorporate
the quantum pressure \cite{scb14,schwabe16,Mocz}. 
As an example of the relation between the the fluid and scalar field viewpoints, a discussion of the collision  of streams is given in Appendix \ref{app:collisions}.

To minimize confusion, we would like to point out that despite
the appearance of $\hbar$ in many of the above formulae, all of the 
considerations above and in this paper can be 
understood purely in terms of classical field theory.  Indeed, $\hbar$ and the
 mass $m$ appear only in the form of the ratio $\hbar/m$.  All of
our formulas can be expressed in terms of this 
ratio without ever mentioning $\hbar$.  

\section{Astrophysics of FDM in the Milky Way and nearby galaxies}
\label{sec:FDMastrophysics}

\subsection{ Introductory remarks}

\noindent
The differences between the properties of a standard CDM universe and
a universe in which the dark matter is dominated by FDM will be most
prominent in dark-matter dominated systems where the de Broglie
wavelength $\lambda=h/(mv)$ is comparable to the system size
$r$. Since the virial velocity $v$ of halos or sub-halos generally
decreases as $r$ decreases this condition favors small systems, both
because $r$ is small and because $\lambda$ is large. Given the small
scales, the effects of FDM are most easily studied in nearby systems,
so most of the tests described below are best performed in the Milky
Way or other Local Group galaxies. On the other hand, in the standard
$\Lambda$CDM model of structure formation, small-scale systems form
earliest, so the differences between galaxy formation in CDM- and
FDM-dominated universes will be most dramatic at high redshift. Thus
the very near, in this section, and the very far, in \S\ref{sec:form} and \S\ref{sec:forest}, will be the foci of our attention. 

\subsection{Minimum size and maximum density}

The de Broglie wavelength for FDM is given by Eq.\ (\ref{eq:brog}).  Roughly speaking, $\lambda/(2\pi)$ cannot exceed the
virial radius $r \simeq GM/v^2$ of an equilibrium self-gravitating
system of mass $M$. Thus $r \gsim \hbar^2/(GMm^2)$. A more precise statement, derived in
Appendix \ref{app:speq}, is that the radius containing half the mass
of a spherically symmetric, time-independent, self-gravitating system
of FDM must satisfy the inequality
\begin{equation}
	r_{1/2} \ge  3.925\frac{\hbar^2}{GMm^2} = 0.335\kpc \,\frac{10^9M_\odot}{M}\left(\frac{10^{-22}\eV}{m}\right)^2.
\label{eq:one}
\end{equation}
The inequality is an equality if the system is in a stationary state
that minimizes the energy\footnote{In the approximation of the Schr\"odinger-Poisson equation there is a conserved particle number, and the soliton
solution minimizes the energy for a given particle number.  This particle number is not conserved in the full equations
governing a real scalar field interacting with gravity, so in that context the soliton is not absolutely stable, although its lifetime
is much greater than the Hubble time \cite{Eby2016}. }; this state is sometimes called a ``soliton''.  Similarly, the central density satisfies
\begin{equation}
\rho_c\le 0.0044 \left(\frac{Gm^2}{\hbar^2}\right)^3 M^4=7.05
\,M_\odot\pc^{-3}\left(\frac{m}{10^{-22}\eV}\right)^6\left(\frac{M}{10^9M_\odot}\right)^4.
\label{eq:rho}
\end{equation}

This upper limit to the density can be compared with the observed
central densities of dwarf spheroidal galaxies in the Local Group,
which are strongly dominated by dark matter even at their centers:
many have mass-to-light ratios $\gtrsim 100M_\odot/L_\odot$ inside
their half-light radii. Among 36 Local Group dwarf spheroidals the
maximum, mean and median density within the half-light radius are $5$,
0.5, and $0.1\msun\pc^{-3}$ \cite{mc12}, consistent with equation
(\ref{eq:rho}) if $m\simeq 10^{-22}\eV$ and the ground-state mass exceeds
$3\times10^8\msun$--$10^9\msun$. With these nominal values the halo
half-mass radius (\ref{eq:one}) is similar to the observed half-light
radii of dwarf spheroidal galaxies: the median and quartiles for the
Local Group sample are $0.25\genfrac{}{}{0pt}{}{+0.3}{-0.1}\kpc$.

If we assume that the central part of the halo is a soliton it is
possible to fit the kinematics of the stars in dwarf spheroidal
galaxies to determine the FDM particle mass. Using 8 dwarf
spheroidals, Chen et al.\ \cite{chen16} find masses between
$m=8\genfrac{}{}{0pt}{}{+5}{-3}\times 10^{-23}\eV$ (for Draco) and
$m=6\genfrac{}{}{0pt}{}{+7}{-2}\times 10^{-22}\eV$ (for Sextans). The
galaxy-to-galaxy scatter in $m$ can be ascribed to contamination by
foreground stars, the strong covariance between the observational
determination of the central density $\rho_c$ and the half-mass
radius $r_{1/2}$, and perhaps the assumption that the central part of
the halo is a soliton. Other estimates, based on fewer galaxies or more
approximate models, are roughly consistent \cite{scb14,mp15,cs16}\footnote{An exception is the recent paper by Gonz\'alez--Morales et al.\ \cite{gm16}, which finds $m<0.4\times10^{-22}\eV$ from analyzing the kinematics of stars in the Fornax and Sculptor dwarf galaxies.}.

\subsection{Relaxation in systems composed of FDM}

\label{sec:relax}

\noindent
Over distances large compared to the de Broglie wavelength, FDM
behaves similarly to CDM. Thus we expect the soliton to be
surrounded by an virialized halo with a Navarro--Frenk--White (NFW)
profile \cite{nfw97}. This expectation is demonstrated by simulations
\cite{scb14,sch14,schwabe16,velt2016}. Simulations yield a relation between the mass of the central soliton $M$ and the virial mass of the surrounding halo $M_{\rm vir}$, \cite{sch14},
\begin{equation} \label{eq:chrel}
M\simeq 2.7\times 10^{8}\msun \frac{10^{-22}\eV}{m}\left(\frac{M_{\rm vir}}{10^{10}\msun}\right)^{1/3};
\end{equation}
however, this relation is only well-tested over the limited range $10^9\msun \lesssim M_{\rm vir}\lesssim 10^{11}\msun$. These simulations show up to two orders of magnitude difference in
density between the solitonic core and the surrounding halo. At larger halo masses the density contrast between the solitonic core and the halo is expected to be even larger, but the soliton mass should fall below the estimate from Eq.\ (\ref{eq:chrel}), for reasons given below. 

An isolated CDM system that is in equilibrium (i.e., that satisfies
the time-independent collisionless Boltzmann equation) evolves only
through two-body relaxation \cite{BT}, on a timescale of order $t_{\rm
  relax}\sim 0.1 t_{\rm cr}(M/m)$ where $t_{\rm cr}\sim r/v$ is the
crossing time and we have neglected a logarithmic factor. For typical
galaxy masses $M$ and CDM particle masses $m_{\rm CDM}$, $t_{\rm relax}$ is many
orders of magnitude longer than the Hubble time. In contrast, an FDM system that is
not a soliton evolves by expelling probability density to infinity
since all eigenstates other than the solitonic ground state are
unstable. This process is sometimes called ``gravitational cooling''
\cite{ss94,hmt03,gul06,schwabe16}, and the wavy granularity that is the source of this relaxation is seen clearly in simulations \cite{scb14}. The granularity, which can also be thought of as an interference pattern, arises from two distinct sources: the finite number of modes or eigenstates in the halo---of order $(kr)^3$ where $2\pi/k=\lambda\sim h/(mv)$ is the de Broglie wavelength---and the spatial correlation in density fluctuations that arises because the FDM particles are bosons (similar to the Hanbury-Brown and Twiss effect for photons). The strength of the fluctuations and their dynamical effect can be estimated by the following crude arguments. If the
local density is $\rho$ then the FDM acts as quasiparticles with
effective mass $m_{\rm eff}\sim \rho(\half\lambda)^3$. Then the
relaxation time should be $t_{\rm relax}\sim 0.1 t_{\rm cr}M/m_{\rm eff}$ where $M\sim \frac{4}{3}\pi\rho r^3$ is the halo mass interior to radius $r$. We find\footnote{Thus FDM has less bound sub-structure than CDM but more unbound substructure.}
\begin{equation}
t_{\rm relax}(r)\sim \frac{0.4}{f_{\rm relax}}\frac{m^3v^2r^4}{\pi^3\hbar^3}\sim \frac{1\times
10^{10}\yr}{f_{\rm relax}}
\left(\frac{v}{\vphantom{\big|}100\kms}\right)^2\left(\frac{r}{5\kpc}\right)^4\left(\frac{m}{\vphantom{\big|}10^{-22}\eV}\right)^3,
\label{eq:trelax}
\end{equation}
where $f_{\rm relax}\lesssim 1$ is a dimensionless constant, the value of which remains to be estimated by simulations and/or more careful analytic arguments.

An FDM halo will develop a compact solitonic core from the mass originally in the halo inside radius $r_s$, where $r_s$ is given
approximately by the condition $t_{\rm relax}(r_s)=t_0$ and $t_0$ is the age of the halo. Low-mass halos have relatively small radii and virial velocities so the relaxation time is short and most of the halo is incorporated in the soliton. As the halo mass grows the virial radius and virial velocity increase, $r_{\rm vir}\sim M_{\rm vir}^{1/3}$ and $v_{\rm vir}\sim M_{\rm vir}^{2/3}$, so the relaxation time grows and only a small fraction of the halo mass resides in the central soliton. The soliton mass depends on the density and velocity distribution in the halo, which is approximately described by an NFW profile \cite{nfw97}. Thus the simple relation (\ref{eq:chrel}) between the virial mass and the soliton mass is probably an approximation that is only valid over a limited range of virial mass. The NFW density profile flattens at small radii so we expect the curve relating soliton mass to virial mass to flatten, and possibly even have negative slope, at large virial masses.

The fluctuating gravitational potential whose effects are described by Eq.\ (\ref{eq:trelax}) also leads to relaxation in the stellar components of the galaxy, although possibly with a different dimensionless constant. Some of the relevant dynamics has already been described in the context of dark matter composed of massive black holes \cite{lacey1985} or dark clusters \cite{carr1987}. To make a rough estimate of the importance of this process, we focus on the Milky Way at the distance of the Sun, taking $r\sim 10\kpc$, $v\sim 200\kms$, $\rho\sim 0.01\msun\pc^{-3}$. Then the de Broglie wavelength $\lambda\simeq 600\pc(10^{-22}\eV/m)$ and the mass of a typical quasiparticle is $m_{\rm eff}\sim \rho(\half\lambda)^3\sim 3\times 10^5\msun (10^{-22}\eV/m)^3$. The possible effects include:

\paragraph{Disruption of star clusters:} The fluctuating potential from FDM wavepackets exerts tidal forces that can pump energy into open and globular clusters. The disruption time can be estimated from Eq.\ (8.54) of \cite{BT},
\begin{equation}
t_{\rm dis}\simeq \frac{0.05}{f_{\rm relax}}\frac{\sigma_{\rm rel} m_{\rm cl} r_h^2}{\vphantom{\big|}Gm_{\rm eff}\rho a^3}.
\end{equation}
Here $\sigma_{\rm rel}\sim v$ is the one-dimensional dispersion in relative velocity, $m_{\rm cl}$ is the cluster mass, $r_h$ is the half-mass radius of the perturber, which we take to be $\half \lambda$, and $a$ is the radius or semimajor axis of the cluster star relative to the cluster center. This formula assumes that $r_h\gg a$, that the maximum velocity kick from a passing FDM quasiparticle is much less than the escape speed from the cluster, and that the passage time of the quasiparticle $r_h/v$ is shorter than the dynamical time in the cluster; all of these assumptions are reasonable for the examples discussed here. Inserting nominal parameters for globular clusters,
\begin{equation} \label{eq:tdis}
t_{\rm dis}=\frac{8.4\times10^{11}\yr}{f_{\rm relax}}\left(\frac{v}{\vphantom{\big|}200\kms}\right)^2\left(\frac{m_{\rm cl}}{\vphantom{\big|}3\times10^5\msun}\right)\left(\frac{m}{10^{-22}\eV}\right)\left(\frac{0.01\msun\pc^{-3}}{\rho}\right)^2\left(\frac{30\pc}{a}\right)^3.
\end{equation}
For the nominal parameters the disruption time is much too large to be of interest. However, the disruption time is shorter at smaller distances from the Galactic center because the FDM density $\rho$ is much larger and $t_{\rm dis}\sim \rho^{-2}$. Thus globular clusters close to the Galactic center could be shorn of their outer parts or even disrupted by FDM fluctuations if $f_{\rm relax}\sim 1$. The outer parts of one or two of the globular clusters in the Fornax dwarf galaxy could also be susceptible to disruption by this process.\footnote{T.\ Brandt, private communication.} Open clusters have a wide range of masses, radii, and ages but for the nominal values $m_{\rm cl}=300\msun$ and $a=2\pc$ we find $t_{\rm dis}\simeq 2\times 10^{11}\yr/f_{\rm relax}$ compared to a typical age of $3\times10^8\yr$; we conclude that for open clusters in the solar neighborhood the effects of FDM fluctuations are negligible. 

\paragraph{Disruption of wide binary stars:} The effect of FDM fluctuations on binary stars can be described by Eq.\ (\ref{eq:tdis}) if we replace the cluster mass $m_{\rm cl}$ by the total mass of the binary, which we take to be $2\msun$, and let $a$ represent the binary semimajor axis. The widest known binary stars have $a\simeq 0.1\pc$ so we obtain $t_{\rm dis}\simeq 1.5\times10^{14}\yr/f_{\rm relax}$, too long to be of interest. 

\paragraph{Thickening of the Galactic disk:} FDM fluctuations can pump energy into the orbits of disk stars. An approximate estimate of the characteristic time for the disk to double in thickness is
\begin{equation}
t_{\rm thick}=\frac{0.2}{f_{\rm relax}}\frac{\sigma_{\rm disk}^2v b_{\rm min}^2}{G^2m_{\rm eff}\rho z_{1/2}^2}.
\end{equation}
Here $\sigma_{\rm disk}$ and $z_{1/2}$ are the vertical velocity dispersion and half-thickness of the disk and $b_{\rm min}$ is the minimum impact parameter of encounters that contribute strongly to the thickening. We set $b_{\rm min}=\max(z_{1/2},\half\lambda)$. We have $\lambda > 2z_{1/2}$ for $m < 1.0\times 10^{-22}\eV (200\kms/v)(300\pc/z_{1/2})$; in this case
\begin{equation}
t_{\rm thick}=\frac{7.0\times10^{11}\yr}{f_{\rm relax}}\left(\frac{\sigma_{\rm disk}}{30\kms}\right)^2\left(\frac{v}{200\kms}\right)^2\left(\frac{m}{10^{-22}\eV}\right)\left(\frac{0.01\msun\pc^{-3}}{\rho}\right)^2\left(\frac{300\pc}{z_{1/2}}\right)^2.
\end{equation}
For larger masses, when $\lambda < 2z_{1/2}$,
\begin{equation}
t_{\rm thick}=\frac{7.0\times10^{11}\yr}{f_{\rm relax}}\left(\frac{\sigma_{\rm disk}}{30\kms}\right)^2\left(\frac{v}{200\kms}\right)^4\left(\frac{m}{10^{-22}\eV}\right)^3\left(\frac{0.01\msun\pc^{-3}}{\rho}\right)^2.
\end{equation}
These timescales are too long to be of interest in the solar neighborhood, but thickening by FDM fluctuations could convert disks into thicker structures such as pseudobulges in the inner parts of galaxies. 

\paragraph{Galactic bulges:} The bulges of disk galaxies have typical sizes of $\sim 1\kpc$ and thus would interact strongly with FDM fluctuations according to Eq.\ (\ref{eq:trelax}). However, this equation probably overestimates the relaxation rate inside the region where $t_{\rm relax}$ is less than the age of the galaxy because most of the FDM will be incorporated in the central soliton, which is stationary and does not contribute to relaxation. 

\paragraph{Orbital decay of supermassive black holes:} Most galaxies contain black holes of $10^6\msun$--$10^{10}\msun$ at their centers. When two galaxies with CDM halos merge, dynamical friction from the halo drains orbital energy from the black holes and they spiral in to the central few parsecs of the merger remnant \cite{BBR1980,Yu2002}. If the inspiral continues to even smaller orbital radii, until the decay timescale due to gravitational radiation becomes shorter than a Hubble time (typically at semimajor axes of 0.001--0.1 pc), then the two black holes will merge. The gravitational radiation from the late stages of the inspiral and the merger could be detectable by ground-based pulsar timing arrays and space-based interferometers, respectively. However, FDM fluctuations may inhibit the inspiral at kiloparsec scales. The effective mass of the FDM quasiparticles $m_{\rm eff}\sim \rho_{\rm FDM}(\half\lambda)^3$ can be rewritten using the relation $\rho\sim 3v^2/(4\pi Gr^2)$ if $\rho$ is interpreted as the mean density inside radius $r$ and FDM dominates this density so $\rho\sim\rho_{\rm FDM}$. Then 
\begin{equation}
m_{\rm eff}\sim \frac{3\pi^2\hbar^3}{4 Gr^2m^3v}=6\times10^7\msun\left(\frac{1\kpc}{r}\right)^2\left(\frac{200\kms}{v}\right)\left(\frac{10^{-22}\eV}{m}\right)^3.
\end{equation}
At small radii, but outside the solitonic core, the effective mass of the FDM fluctuations can be larger than the black-hole mass and in this case the fluctuations will pump energy into the black-hole orbits, opposing the drain of energy by dynamical friction from the hot stellar component (an additional effect in disk galaxies is dynamical friction from rotating stars and gas in the disk, which adds to the component of angular momentum normal to the disk). If the density of FDM is comparable to the density of stars and other baryons, the orbital decay could be slowed or even reversed; if the stellar system has density $\rho_\star$ and is not rotating, and the black hole has mass $m_{\rm BH}$, this reversal occurs if there is a radial region where $\rho_{\rm FDM}m_{\rm eff}\gtrsim \rho_\star m_{\rm BH}$. In this case the formation of sub-parsec black-hole binaries would be suppressed, and the rate of black-hole mergers could be much smaller than otherwise predicted. The predicted population of supermassive black holes orbiting at kiloparsec distances from the centers of galaxies would be extremely difficult to detect. 

These speculations and estimates require confirmation by more sophisticated analytic
arguments and simulations. They suggest that relaxation due to FDM fluctuations may substantially alter the structure of both the inner parts of galaxy-size FDM halos and their stellar components. 

\subsection{The cusp-core problem}

\noindent
A generic prediction of structure formation in CDM is that halos and
sub-halos should have singular density cusps at their centers, with
the density varying with radius roughly as $\rho(r)\sim r^{-1}$
\cite{dc91,nfw97}.  Baryonic processes such as adiabatic contraction,
infalling substructure, or density fluctuations due to supernova
feedback could modify this profile
\cite{nef96,esh01,gz02,rg05,tls06,mas08,rd08,goe10,gov10,cole11,pg12},
so the prediction of a central density cusp is most secure in dwarf
spheroidal galaxies, where the total density is strongly dominated by
dark matter at all radii \cite{mc12}.

In principle, observations of stellar kinematics can determine whether
the dark-matter density distribution at the centers of dwarf
spheroidals is a cusp as in CDM models or a core as in FDM. Although
the majority of studies favor cores over cusps, the
question has not yet been settled, largely because of degeneracies
between the observational signatures of the mass profile and the
velocity anisotropy profile
\cite{kle02,goe06,wu07,bat08,wal09,str10,wal11,ae12,aae13,bred13,jg13,str14,fat16}. 
The density profiles of low surface-brightness disk galaxies also
appear to have cores \cite{deblok01,deblok05,oh08}; see
\cite{deblok09} for a review.

We conclude that the kinematics of low surface-brightness galaxies,
both dwarf spheroidals and disks, is consistent with the cores
required by FDM and disfavors, but does not rule out, CDM. Definitive observations of dark-matter cusps down to a distance $r$ from the centers of galaxies would rule out FDM with a mass $m\lesssim \hbar/(vr)$ where $v$ is the velocity dispersion at $r$.  

\subsection{Lower bound on FDM halo masses}

\label{sec:maxmin}

We have seen (Eq.\ \ref{eq:one}) that self-gravitating time-independent FDM systems supported by quantum pressure have the unusual property that low-mass objects are larger than those of higher mass. This has important cosmological consequences that are independent of the details of the halo formation process and the spectrum of initial density perturbations, in particular a minimum mass for FDM halos. 

There are two arguments based on similar physical principles that produce the same lower bound on the halo mass. The first is based on the observation that if halos form by gravitational collapse they cannot be lower in density than the average universe in which they reside.  Let $\rho_{1/2}=\frac{1}{2}M/(\frac{4}{3}\pi r_{1/2}^3)$ be the mean density inside the half-mass radius (Eq.\ \ref{eq:one}). For comparison, the virial radius is commonly defined such that the mean density inside it is $\rho_{\rm vir}\equiv 200\rho_{\rm crit}$ where $\rho_{\rm crit}=3H^2/(8\pi G)$ is the critical density. If we require that $\rho_{1/2}>q\rho_{\rm vir}$ where $q$ is a factor $\gtrsim 1$ then
\begin{align}
M &> 2.8\,q^{1/4}\frac{H^{1/2}\hbar^{3/2}}{Gm^{3/2}}\nonumber \\
&=1.4\times10^7\msun\, q^{1/4}\left(\frac{H}{\vphantom{\big|}70\kms\Mpc^{-1}}\right)^{1/2}\left(\frac{10^{-22}\eV}{m}\right)^{3/2}.
\label{eq:jpo}
\end{align}

The second argument is based on the Jeans length, the minimum scale for gravitationally unstable density perturbations in a homogeneous background. This has been derived for FDM and gives for the critical (maximum) Jeans wavenumber \cite{kmz85,bianchi_1990,FDM,jk2008,Lee2010,Chavanis2011,suarez2011} 
\begin{equation}
k_J=\frac{2(\pi G \rho)^{1/4}m^{1/2}}{\hbar^{1/2}}
\end{equation}
where $\rho$ is the unperturbed matter density and $k_J$ is in physical, not comoving, coordinates. The corresponding Jeans length is $\lambda_J\equiv
2\pi/k_J$ and the Jeans mass is
\begin{align}
M_J&=\ffrac{4}{3}\pi\rho(\half\lambda_J)^3\\
&=1.5\times10^7\msun(1+z)^{3/4}
\left(\frac{\Omega_\fdm}{0.27}\right)^{1/4}\left(\frac{H_0}{\vphantom{\big|}70\kms\Mpc^{-1}}\right)^{1/2}\left(\frac{10^{-22}\eV}{m}\right)^{3/2}.
\label{eq:jeans}
\end{align}
Here $H_0$ is the Hubble constant and $\Omega_\fdm$ is the
fraction of the critical density in FDM.  Within the considerable uncertainties Eqs.\ (\ref{eq:jpo}) and (\ref{eq:jeans}) give the same result, a minimum halo mass of $(1\mbox{--}2)\times 10^7\msun(10^{-22}\eV/m)^{3/2}$, in fairly good agreement with an earlier estimate by \cite{sch14}, $4\times 10^7\msun(10^{-22}\eV/m)^{3/2}$, obtained by setting $M=M_{\rm vir}$ in Eq.\ (\ref{eq:chrel}).  For CDM the number of halos (and subhalos) is rising as $dn(M_h) \propto M_{h}^{-2}dM_h $ below $10^8 \msun$, so the contrast in the number of low-mass halos is dramatic. 

Dark-matter halos or sub-halos cannot easily exist with masses lower
than this limit, so dark-matter halos around globular clusters or similar ultra-compact dwarf galaxies are not expected in FDM models \cite{Lee2010}.

Whether or not halos form at or near this minimum mass depends on the initial density perturbation spectrum for FDM and its linear and nonlinear growth, a topic we defer to \S\ref{sec:form}. 

\subsection{A maximum soliton mass}

At sufficiently large mass $M$ the depth of the central potential of a solitonic core composed of FDM approaches $c^2$. From Eq.\ (\ref{eq:ground}) this occurs when $M\sim \hbar c/(Gm)$ and this represents a maximum mass for FDM solitons analogous to the Chandrasekhar mass for self-gravitating fermion systems. Calculations that employ general relativity \citep{kaup68,rb69,helfer2016} provide a more accurate value,
\begin{equation}\label{eq:mmax}
M_{\rm max}=0.633 \,\frac{\hbar c}{Gm}=8.46\times 10^{11}\msun\, \frac{10^{-22}\eV}{m}.
\end{equation}
FDM halos can and do exist with much larger masses. In such cases the mass of the central soliton is only a small fraction of the total halo mass; most of the halo behaves classically; and the mass spectrum and most other properties of halos would be the same as in CDM. An unresolved question is how the relaxation processes described in \S\ref{sec:relax} 
would affect the central structure in massive FDM halos. 
The maximum soliton mass is lowered when the self-interaction expected for an axion is taken into account \cite{Eby1,Eby2}.

\subsection{The missing satellite problem}

\noindent
Structure formation in standard $\Lambda$CDM cosmology is
approximately self-similar, with every gravitationally bound halo
containing bound sub-halos ``all the way down''. However, the expected distribution of CDM sub-halos in massive galaxies greatly
exceeds the number of small satellite galaxies observed to orbit luminous
galaxies \cite{kly99,moore99}. This discrepancy is often called the
``missing satellite'' problem. The most commonly proposed solutions to
this problem invoke baryonic physics, in particular (i) heating of the
halo gas by ultraviolet background radiation, which could suppress gas
accretion onto sub-halos; (ii) supernovae and stellar winds, which
could drive most of the gas out of the sub-halos. The conclusions to
be drawn from recent examinations of these processes
\cite{wetzel2016,nier16} are unclear. Heating and feedback certainly
reduce the stellar luminosity of the satellite galaxies associated
with sub-halos, but whether plausible parametrizations for these
processes can match the observations over the full range of halo and
sub-halo masses remains an open question. 
 
The ``missing satellite'' problem is reduced or resolved in FDM without appeal to baryonic physics, because the number of low-mass sub-halos is
expected to be much smaller in FDM than in CDM. There are two main reasons for this. The first is that sub-halos are more vulnerable to tidal disruption, both because of the upper limit
to their mass density (Eq.\ \ref{eq:rho}), and because FDM can tunnel through the potential
barrier centered on the tidal radius; thus an FDM sub-halo is {\em always} disrupted by a tidal field, no matter how weak, after a sufficiently long time.
A simplified model of this process for FDM solitons is described in
Appendix \ref{app:tide}. We find that an FDM soliton cannot survive on a circular orbit in a host system for $\gtrsim 10$
orbits\footnote{A typical massive galaxy like the Milky Way has a circular speed of about $200\kms$ so a satellite in a circular orbit of radius $r$ completes $15(30\kpc/r)$ orbits in a Hubble time.} unless its central density $\rho_c > 60\,\overline \rho_{\rm
  host}$; survival for a few hundred orbits requires $\rho_c > 100\,\overline \rho_{\rm
  host}$. Together with Eq.\ (\ref{eq:rho}) this implies that the
minimum mass of an FDM solitonic system that survives 10 orbits at radius $a$ inside a host of
mass $\mm$ interior to $a$ is given by
\begin{equation}
\label{eq:tides}
M>6.7\times10^8\msun\left(\frac{\mm}{\vphantom{\big|}10^{11}\msun}\right)^{1/4}
\left(\frac{10\kpc}{a}\right)^{3/4}\left(\frac{10^{-22}\,\mbox{eV}}{m}\right)^{3/2}. 
\end{equation}
This lower cutoff is consistent with numerical simulations
\cite{scb14}, which find that halo substructure is suppressed below a few
times $10^8\msun$. 

The second reason why there are fewer FDM sub-halos than CDM sub-halos is that the power spectrum of FDM density perturbations is suppressed at small masses relative to CDM. We defer a discussion of this topic to \S\ref{sec:form}. 

The sub-halo mass function can be
probed in several ways. One approach is to look for gravitational
lensing of background galaxies by the sub-halos \cite{ms98,li16}. The strongest current limits come from ALMA observations of high-redshift star-forming galaxies; at present these are only sensitive to sub-halo masses $\gtrsim 10^9\msun$ \cite{hez16}, but technical advances should reduce this limit to the interesting range. A second is to study the evolution of tidal streams, which we discuss in the next subsection. 

The ``too big to fail'' problem arises when standard abundance matching methods to relate galaxies and halos are used to predict the magnitude difference between primary and secondary galaxies in groups. These studies \cite{read2006,boylin2011} indicate that the expected number of luminous secondary galaxies is absent, and much work is being done to see if feedback from massive stars and black-hole outflows, or other baryonic physics, can account for the deficit. A fair, if brief, summary would be that while the latest high-resolution cosmological simulations \cite{wetzel2016} argue that the deficit will naturally be produced by radiatively driven winds and other feedback effects, considerably more work will need to be done by a variety of methods before this conclusion can be considered secure. In any event, FDM reduces the magnitude of the ``too big to fail'' problem, both because the abundance of sub-halos is smaller in FDM than CDM at all masses below $\sim 10^{10}\msun$ (\S\ref{sec:form}) and because sub-halo masses are generally estimated observationally by the maximum circular speed of the baryons, which is lower in FDM than CDM for halos of a given mass \cite{ms2014}.

A more speculative concern is the origin of globular clusters (GCs), which remains mysterious. With typical stellar masses of $10^5\msun$  and sizes of roughly $10\pc$, these old stellar systems surround all normal galaxies out to distances comparable to the virial radius. The total mass of GCs per galaxy seems to parallel the mass in dark matter more closely than the mass in baryons: the ratio of the mass in GCs to the mass in baryons is high at both ends of the galaxy mass spectrum, like the ratio of dark to baryonic mass. The presence of multiple stellar populations in GCs \cite{renzini2008} also suggests that these systems had or have escape velocities large enough to retain gas. In an early paper before the existence of dark matter was widely recognized,  Peebles and Dicke \cite{pd1968} speculated that GCs might have formed at the era of decoupling between matter and radiation, since they have the Jeans mass appropriate to that epoch ($z\simeq 10^3$). CDM has gravitationally unstable initial fluctuations on the scale of GCs but it is very difficult for the objects we see to have formed from these, for several reasons. First, the more massive sub-halos (with halo mass of $10^8\msun$, say) are absent from our inventory of galaxies and their absence is explained in CDM by arguing that their relatively small escape velocities allow feedback to expel gas so efficiently that star formation is suppressed. If this explanation is correct, then the formation of stars in systems like GCs that are 100 times less massive would certainly be prohibited. Furthermore the expected sizes of the halos associated with GCs would be several kpc, so large that contraction of even very slowly rotating baryons to the present size of GCs would produce disk systems, whereas observed GCs are nearly spherical. Peebles has suggested  \cite{peebles1984} that in the CDM model GCs should be embedded in dark halos, but despite extensive searches \cite{conroy2011,ibata2013,dia2014} there have been no reports of diffuse dark matter in or around GCs. It is possible that the dynamics of FDM or other exotic alternatives to CDM could play a central role in the formation of these important but puzzling systems, but at the moment this is only a speculation.

\subsection{Tidal streams}

\noindent
Sub-halos affect the evolution of the tidal streams shed by globular clusters as they
lose stars through external tidal forces and internal dynamical
evolutionary processes \cite{moore99}. The small-scale gravitational forces from
sub-halos can thicken, distort, and open gaps in streams; in extreme cases they may disperse the streams so rapidly that they disappear in much less than a Hubble time (see \cite{jc16} for a review).

First consider stream thickening. Suppose for simplicity that all sub-halos have the same mass $m$ and size $r_m$, and that the number density of sub-halos is $N$. If these move at speed $v$ relative to the stream, the probability of an encounter with impact parameter smaller than $b$ between a sub-halo and a point on the stream in an interval $t$ is
\begin{equation} \label{eq:enc}
p= \pi N b^2 vt.
\end{equation}
The differential tidal force across a stream of width $w$ due to a mass $m$ at distance $r$ is $F= kGm w/r^3$ where $k$ is of order unity, and integrating over the encounter yields an impulse $\Delta v=\int F\,dt=
2kGmw/(b^2v)$. According to Eq.\ (\ref{eq:enc}) the single closest encounter is given by $p=1$ or $b_{\rm min}^2=(\pi Nvt)^{-1}$, and this gives an impulse\footnote{Eq.\ (\ref{eq:chandra}) is closely related to formulae for the disruption of binary stars due to tidal forces from massive objects such as molecular clouds \cite{chandra1944,bht1985}.}
\begin{equation} \label{eq:chandra}
\Delta v=2\pi kw GNm t, \qquad t\lesssim t_0.
\end{equation}
An impulse $\Delta v$ will cause the stream width to increase as $\dot w=\Delta v$. 
The derivation of (\ref{eq:chandra}) treats the sub-halos as point masses, an approximation that is only valid when the minimum impact parameter is larger than the sub-halo radius, $b_{\rm min}\gtrsim r_m$. Thus it is only valid up to a time $t_0\equiv (\pi N v r_m^2)^{-1}$. At times smaller than this, the stream width will be dominated by the single closest encounter and if we assume this encounter occurs roughly midway through the time interval $t$ then 
\begin{equation}
w(t)\simeq w(0)(1+\half  \pi k GNmt^2), \qquad t \lesssim t_0.
\end{equation}
Over times larger than $t_0$, stream thickening becomes a diffusive process due to encounters with many sub-halos so $\Delta v$ and $w$ should be replaced by ensemble averages $\langle (\Delta v)^2\rangle^{1/2}$ and $\langle w^2\rangle^{1/2}$. A similar calculation yields
\begin{equation} \label{eq:diff}
\frac{d\langle (\Delta v)^2\rangle }{dt}=\frac{4\pi(kGm)^2N}{vr_m^2}\langle w^2\rangle, \qquad t\gtrsim t_0.
\end{equation}
Replacing $\langle(\Delta v)^2\rangle^{1/2}$ by $(d/dt)\langle w^2\rangle^{1/2}$ and integrating, we find
\begin{align}
\langle w^2(t)\rangle^{1/2}=\langle w^2(t_0)\rangle^{1/2}\exp\bigg[\bigg(\frac{2\pi N k^2G^2m^2}{vr_m^2}\bigg)^{1/3}(t-t_0)\bigg], \qquad t \gtrsim t_0.
\end{align}
Similar arguments can be used to derive the growth in the velocity dispersion of the stream. 

Even a zero-thickness stream will be distorted by the gravitational impulse from passing sub-halos. A zero-thickness stream is a one-dimensional manifold in phase space and must remain so; however it can be folded or crinkled so it appears to have non-zero thickness when projected onto a smaller number of phase-space dimensions (e.g., the two angular coordinates on the sky if it is detected as an overdensity in star counts). The statistical properties of this crinkling can be determined by recognizing that the gravitational force from a large number of sub-halos can be approximated as a Gaussian random field that is homogeneous in time and space on small scales, although most investigations so far rely heavily, and appropriately, on simulations as well \cite{jsh2002,yoon11,sbe16,erkal16}.

The dynamical effect of a sub-halo on a stream that is easiest to detect is probably the opening of a ``gap'' by a close encounter with a single massive sub-halo---strictly, this is a fold rather than a gap since as described above a zero-thickness stream is only distorted, not broken, by conservative forces \cite{carl09,erkal15}.

Even once the effects of substructure are detected, it may be difficult to distinguish CDM from FDM, for several reasons: (i) most of the disruptive influence on streams comes from the most
massive sub-halos, which are present in both CDM and FDM \cite{carl09}; (ii) sub-structure in the baryonic disk can also disrupt the streams \cite{agvw16};
(iii) in CDM the mass fraction in sub-halos is a strong function of radius, typically ranging from a few per cent at the virial radius, to 0.1\% at $30\kpc$, to $0.01\%$ at $r<10\kpc$ \cite{diemand07}; (iv) numerical simulations suggest that many long, thin streams can survive for a Hubble time even in a CDM halo \cite{ngan16}, so the difference between CDM and FDM will lie in the abundance of old streams rather than their presence or absence. 

Despite these concerns, we note that the power of tidal streams to discriminate among halo models will improve dramatically over the next several years with astrometry from the Gaia spacecraft, which will discover many more
streams and provide much more accurate kinematic observations of them. 

\subsection{Galactic disks}

\noindent
The relaxation time (\ref{eq:trelax}) is sufficiently short that
solitonic ``dark disks'' could coalesce around the inner parts of
baryonic galactic disks. 

A useful first approximation to the structure of solitonic disks is
obtained by neglecting the baryonic contribution to the gravitational
potential and computing the ground state of the Schr\"odinger--Poisson
equation in one dimension (Eq.\ \ref{eq:sp} with $\nabla^2\to
d^2/dz^2$). The density $\rho_\fdm(z)$ in this state is symmetric
around $z=0$ and can be characterized by the central density
$\rho_\fdm(0)$, the surface density $\Sigma_\fdm=\int_{-\infty}^\infty
dz\,\rho_\fdm(z)$ and the half thickness $z_{1/2}$, defined by
$\int_0^{z_{1/2}}dz\,\rho_\fdm(z)=\half\int_0^\infty
dz\,\rho_\fdm(z)$. We find
\begin{align}
\rho_\fdm(0)&=0.984\left(\frac{G\Sigma_\fdm^4m^2}{\hbar^2}\right)^{1/3}=0.104\msun\pc^{-3}\left(\frac{\Sigma_\fdm}{100\msun\pc^{-2}}\right)^{4/3}\left(\frac{m}{10^{-22}\eV}\right)^{2/3}\nonumber
\\
z_{1/2}&=0.2744\left(\frac{\hbar^2}{G\Sigma_\fdm m^2}\right)^{1/3}=260\pc
\left(\frac{100\msun\pc^{-2}}{\Sigma_\fdm}\right)^{1/3}\left(\frac{10^{-22}\eV}{m}\right)^{2/3}.
\label{eq:fdmdisk}
\end{align}

In the solar neighborhood, the total surface density within
$\pm1.1\kpc$ of the Galactic midplane is $68\pm4\msun\pc^{-2}$, of
which $51\pm4\msun\pc^{-2}$ is composed of baryons (gas and stars)
\cite{br13}.  The total density in the Galactic midplane is
$\rho(0)=0.10\pm0.01\msun\pc^{-3}$, most or all of which
is baryonic \cite{hf00}. The first of these results implies that at
most $20\msun\pc^{-2}$ is present in a dark-matter disk; then Equation
(\ref{eq:fdmdisk}) implies that for $m=10^{-22}\eV$, $z_{1/2}\gtrsim
450\pc$ and $\rho(0)\lesssim 0.012\msun\pc^3$, large enough and small
enough respectively that the assumption that the solitonic disk
dominates the gravitational potential is inconsistent\footnote{An
  unrelated problem with a solitonic disk of this kind is that it is
  likely unstable at horizontal wavelengths large compared to its thickness.}.

In the opposite limit, the gravitational potential is dominated by
baryons rather than FDM. Since the scale height of the solitonic disk
is large, we can approximate the baryonic contribution as arising from
a zero-thickness sheet of surface density $\Sigma_b$. Then
\begin{align}
\rho_\fdm(0)&=1.141\,\frac{\Sigma_\fdm}{\Sigma_b}\left(\frac{G\Sigma_b^4m^2}{\hbar^2}\right)^{1/3}\nonumber
\\
&=0.120\msun\pc^{-3}\frac{\Sigma_\fdm}{\Sigma_b}\left(\frac{\Sigma_b}{100\msun\pc^{-2}}\right)^{4/3}\left(\frac{m}{10^{-22}\eV}\right)^{2/3}\nonumber
\\
z_{1/2}&=0.2399\left(\frac{\hbar^2}{G\Sigma_b m^2}\right)^{1/3}=228\pc
\left(\frac{100\msun\pc^{-2}}{\Sigma_b}\right)^{1/3}\left(\frac{10^{-22}\eV}{m}\right)^{2/3}.
\label{eq:fdmdiska}
\end{align}
Taking $\Sigma_b=51\msun\pc^{-2}$ and $\Sigma_\fdm=17\msun\pc^{-2}$
we obtain $\rho_\fdm(0)=
0.016\msun\pc^{-3}(m/10^{-22}\eV)^{2/3}$ and $z_{1/2}=285\pc(m/10^{-22}\eV)^{-2/3}$. We conclude that a possible
solitonic disk of FDM with $m=10^{-22}\eV$ has thickness comparable to the baryonic disk
and makes only a small ($\lesssim 20\%$) relative contribution to the density
in the solar neighborhood; for $m=10^{-21}\eV$ the central density would be comparable to the density of stars and gas and the disk would be thinner than the stellar disk. For comparison, the local density expected
from a CDM halo is $\sim 0.01\msun\pc^{-3}$ \cite{bt12}. The thin dark
disk proposed by \cite{rr14}, with surface density $\sim 10\msun\pc^{-2}$ and thickness
$\sim10\pc$, could not be composed of FDM unless the particle mass
$m\gtrsim 5\times 10^{-20}\eV$, much too large to explain the other
small-scale structure issues addressed in this paper.  

\subsection{Dynamical friction}

\noindent
The only low-luminosity satellite galaxy of the Milky Way that
contains globular clusters is the Fornax dwarf spheroidal, the most luminous satellite in this class, which has
five. The timescale for orbital decay of a point mass $m$ in a
background of density $\rho$ and velocity dispersion $\sigma$ is
roughly $\sigma^3/(G^2\rho m)$ \cite{chandra43,BT}. The mass density
in Fornax is high enough, and the velocity dispersion low enough, that
dynamical friction should have caused most of the clusters to spiral to the
center of the galaxy and merge to form a prominent nucleus, which is
not seen \cite{tre76}. A similar problem is present in faint dwarf
elliptical galaxies in several nearby galaxy clusters
\cite{lotz01}. Various explanations have been put forward to explain
the discrepancy (e.g., \cite{oh2000}); the most widely discussed
possibility is that the drag from dynamical friction is reduced,
eliminated, or even reversed if the stellar system hosting the
clusters has a homogeneous (constant-density) core.  This effect,
sometimes called ``core stalling'', is seen in a variety of N-body
simulations \cite{goe06,in11,cole12} but the physics behind it remains
murky \cite{tw84,rgm06,in11}.

Here we ask how dynamical friction is modified if the dark matter in
dwarf galaxies is comprised of FDM rather than CDM
\cite{good00}. There are three distinct effects: (i) many exotic
dark-matter models, including FDM, produce cores at the centers of
halos rather than the central cusp found in CDM
\cite{tg79,kw00,dh01,sal06,vnd11,mac12,bk15} and thus modify the rate of
orbital decay according to the classic Chandrasekhar formula
\cite{chandra43} through changes in the dark-matter density and
velocity dispersion, the orbital speed, etc.; (ii) as reviewed above,
core stalling can reduce or eliminate the drag from dynamical friction
in a homogeneous core compared to the value predicted by Chandrasekhar; (iii)
standard estimates of the drag from dynamical friction must be
modified to account for the large de Broglie wavelengths of FDM
particles, as shown in Appendix \ref{app:fric}. Here we focus on the
last of these effects, also investigated by \cite{lora12}. 

The orbital decay timescale for an object of mass $m_{\rm cl}$ in a circular
orbit of radius $r$ in a host system of density $\rho(r)$
and enclosed mass $\mm(r)$ is
given by Eq.\ (\ref{eq:tdf}),
\begin{equation}
\tau=  \frac{37.5\Gyr}{\vphantom{\big|}C}\left(\frac{\mm(r)}{\vphantom{\big|}10^8\msun}\frac{1\kpc}{r}\right)^{3/2}\frac{10^5\msun}{m_{\rm cl}}\frac{0.01\msun\pc^{-3}}{\rho(r)},
\label{eq:forfric}
\end{equation}
where the dimensionless constant $C$ is plotted in Figure
\ref{Ffriction}, and the fiducial values have been chosen to
approximately match Fornax and its globular clusters. For a more
careful comparison between frictional decay times in FDM and CDM we adopt masses for the five Fornax clusters from
\cite{cole12} and set their orbital radii equal to the projected radii
from the center of Fornax multiplied by $2/\surd{3}$ (the ratio of the radius to the median projected radius in a spherical distribution). For CDM (i) we take the Fornax
density from the ``steep cusp'' model of \cite{cole12}, which has a
logarithmic density slope near the center $d\log\rho/d\log r=-1$ as in
the NFW profile \cite{nfw97} that characterizes CDM halos; (ii) we set the constant $C$ in Eq.\
(\ref{eq:forfric})  to $0.5\log [2v^2r/(Gm_{\rm cl})]$ following Eq.\
(\ref{eq:cch}); the factor 0.5 is a crude empirical correction that
arises because the classical dark-matter particles have velocities
comparable to the globular clusters and only particles traveling
slower than the test object contribute to the frictional force
\cite{BT}. For FDM we use the ``large core'' model of \cite{cole12}, and determine $C$ from Eq.\ (\ref{eq:casymp}) using $k=mv/\hbar$ where the velocity $v$ is determined by assuming the cluster is on a circular orbit.

\begin{table}[ht]
\centering
\caption{Orbital decay times for the globular clusters in Fornax for
  cold dark matter (CDM) and fuzzy dark matter (FDM)}
\vspace{1.15in}
\begin{tabular}{ c|c|c|c|c|c|c|c} 
       & projected radius& cluster mass & \multicolumn{2}{c|}{CDM} & \multicolumn{3}{c}{FDM}
       \\[2pt]
\hline
$n$ & $r_\perp$ (kpc) & $m_{\rm cl}$ ($\msun$) &\ \ $C$\ \ & $\ \tau$ (Gyr) & $\quad kr \quad$ & $C$ &
\ $\tau$ (Gyr)\\[2pt]
 \hline
1 & 1.6   & $3.7\times 10^4$ & 4.29 & 112   & 8.90 & 2.46   &  215   \\
2 & 1.05 & $1.82\times10^5$  & 3.32 & 9.7    & 5.04 & 1.88   & 12 \\
3 & 0.43 & $3.63\times10^5$  & 2.45 & 0.62  & 0.97 & 0.29   & 2.2  \\
4 & 0.24 & $1.32\times10^5$  & 2.50 & 0.37  &  0.31 & 0.033 & 10 \\
5 & 1.43 & $1.78\times10^5$  & 3.46 & 21.3  &  7.79 & 2.32   & 31 \\[5pt]
\multicolumn{8}{p{0.55\textwidth}}{\small {\sc notes:} Projected
separations and globular cluster masses are
taken from \cite{cole12}. Orbital decay times $\tau$ are determined
from Eq.\ (\ref{eq:tdf}) using the ``steep cusp'' and ``large
core'' density distributions from \cite{cole12} for CDM and FDM
respectively. The dimensionless wavenumber $kr=mvr/\hbar$ is evaluated assuming $m=3\times10^{-22}\eV$ and $v^2=G\mm(r)/r$, appropriate for a circular orbit. The dimensionless constant
$C$ is determined using Eq.\ (\ref{eq:casymp}) for FDM, and as
described in the text for CDM.}
\end{tabular}
\label{tab:for}
\end{table}

The resulting decay times are shown in Table \ref{tab:for} for a particle mass of $3\times 10^{-22}\eV$. In all
cases shown in the Table the orbital decay times are longer in a FDM halo than in a CDM
halo. The shortest decay time in the FDM halo exceeds 2 Gyr, compared
to 0.4 Gyr in the CDM halo. Four of the five clusters have decay
times of 10 Gyr or more in the FDM halo, thus solving the puzzle of
why the globular clusters in Fornax have survived. For other particle masses the decay time scales roughly as $m^{-2}$ for clusters 3 and 4, which have the smallest orbits and the shortest decay times. Observations of a larger sample of dwarf galaxies containing globular clusters could further test the possibility that dynamical friction is suppressed compared to the expectations from CDM. Note that the dynamical friction issue 
could in principle be decoupled from the issue of predicting the density profile. With sufficiently high quality data, the density profile of the host galaxy can be observationally determined, sidestepping debates about the impact of baryonic effects. Given the measured profile, CDM or FDM makes predictions for the dynamical friction timescale that can be checked against observations.

Dynamical friction from a CDM halo is also expected to drain
the angular momentum from the central bar found in the majority of
disk galaxies \cite{sell80,wein85,ds00,wk02,dbs09,ath14,sell14}. This
expectation is in tension with the observation that most bars are
rapidly rotating, in that the corotation radius is not far
outside the outer edge of the bar \cite{agu15}. This tension could be
resolved if the dark matter is efficiently trapped into resonances
with the bar \cite{pwk16}, or if the halo has a large core radius so
the density of the halo within several kpc of the galaxy center is
lower than expected from CDM. The FDM hypothesis could contribute
to the resolution of this problem since friction would also be reduced if
the half-mass radius of the central FDM core were larger than the bar
radius ($kr\lesssim 1$ in Figure \ref{Ffriction}).

\subsection{The most massive halos}

The virial masses of the large halos associated with the richest clusters of galaxies approach $M_{\rm vir}=1$--$2\times10^{15}\msun$, If the relation (\ref{eq:chrel}) can be extrapolated to these large halo masses, it implies a central soliton mass
\begin{equation} \label{eq:chrel2}
M\simeq 1.3\times 10^{10}\msun \frac{10^{-22}\eV}{m}\left(\frac{M_{\rm vir}}{10^{15}\msun}\right)^{1/3}.
\end{equation}
This is still well below the maximum soliton mass for FDM particles with no self-interactions (Eq.\ \ref{eq:mmax}). 
The corresponding half-mass radius is given by Eq.\ (\ref{eq:one}),
\begin{equation}
r_{1/2}\simeq 25\pc \frac{10^{-22}\eV}{m}\left(\frac{10^{15}\msun}{M_{\rm vir}}\right)^{1/3}.
\end{equation}
Structures with this mass and size would be a unique signature of FDM. We have argued in \S\ref{sec:relax} that the extrapolation yielding Eq.\ (\ref{eq:chrel2}) probably overestimates the soliton mass at large halo masses, but it is worthwhile to investigate the observable consequences that would result if this extrapolation were correct. 

We first ask whether dense solitons at cluster centers might already have been detected, but interpreted as supermassive black holes. The nearest moderately rich cluster is the Virgo cluster. The luminous galaxy M87 is located near the center of the cluster and contains a central dark object of mass $(6.6\pm0.4)\times
10^9\msun$ as determined by observations of stellar kinematics \cite{geb11}. The resolution of these observations is about 0.2--0.3 arcsec or 17--25
pc and the upper limit to the size of the central object is probably twice as large. For comparison, the soliton in a halo of mass $M_{\rm vir}\simeq 2\times10^{14}\msun$, roughly appropriate for Virgo, would have $M\simeq 7\times 10^9 (10^{-22}\eV/m)\msun$ and $r_{1/2}\simeq 40 (10^{-22}\eV/m)\pc$, roughly consistent with these observations. A second prominent cluster is the Coma cluster, about five times more distant than Virgo. The luminous galaxy NGC 4889 near the center of Coma has a central dark object of mass 0.6--$3.7\times10^{10}\msun$ \cite{mcconnell11}. The spatial resolution of these observations is $\sim 100\pc$ so the soliton expected to form at the center of Coma could easily masquerade as a black hole. 
Despite these appealing numerical coincidences, we do not favor the hypothesis that the central dark objects in M87 and NGC 4889 are solitons rather than black holes. There is an active galactic nucleus and a relativistic jet at the center of M87 and a wide range of observational evidence and theoretical arguments suggest that these phenomena are always associated with black holes. Moreover, the masses are similar to those of central dark objects in other galaxies with similar properties that are at the centers  of rich clusters \cite{kormendy2013}. 

Where, then, are the massive, dense solitons predicted by FDM? There are two main possibilities: (i) The galaxies we have examined may not be at the centers of their respective cluster halos. 
The Virgo cluster has a double structure, and although M87 is in the core of the denser sub-cluster it is displaced from its center in both position and velocity \cite{bing87}. The Coma cluster also appears to consist of two sub-clusters that have not yet relaxed into equilibrium \cite{fitchett87}. (ii) As we have discussed, the soliton mass--virial mass relation in Eqs. (\ref{eq:chrel}) and (\ref{eq:chrel2}) may overestimate the soliton mass at the halo masses $\sim 10^{15}\msun$ considered here. For example, Eq.\ (\ref{eq:trelax}) implies that in a halo with typical velocity $v=10^3\kms$, typical for rich clusters, the relaxation time is less than $10^{10}\yr$ only at radii $r\lesssim 1.5\kpc\, f_{\rm relax}^{1/4}(10^{-22}\eV/m)^{3/4}$, and the halo mass inside that radius is insufficient to produce a soliton of $10^{10}\msun$. 

A related question is whether the soliton can survive if there is a supermassive black hole at its center. A non-rotating black hole of mass $M_\bullet$ traveling at speed $v$ through a uniform scalar field of mass $m$ and density $\rho$ accretes mass at a rate \cite{unruh1976}
\begin{align}
\frac{dM_\bullet}{dt}=&
\frac{32\pi^2 (GM_\bullet)^3 m\rho}{ \vphantom{\big|}\hbar c^3 v[1-\exp(-\xi)]},
\quad\mbox{where}\quad \xi=\frac{2\pi GM_\bullet m}{\vphantom{\big|}\hbar v}
\\
=&\left\{\begin{array}{ll}
\frac{\displaystyle 16\pi (GM_\bullet)^2\rho}{\displaystyle \vphantom{\Big|}c^3},& \quad\xi\ll1, \\
\frac{\displaystyle 32\pi^2 (GM_\bullet)^3 m\rho}{\displaystyle \vphantom{\Big|}\hbar c^3 v},& \quad \xi\gg 1.\end{array}\right.
\end{align}

For an approximate analysis we take $\rho$ and $v$ to be the central density and virial velocity of the soliton (Eqs.~\ref{eq:ground} and \ref{eq:spfact}). Then
\begin{equation}
\xi= 19.07\,\frac{M_\bullet}{M}
\end{equation}
where $M$ is the soliton mass, independent of the mass of the FDM particle. The mass accretion rate is 
\begin{equation}\label{eq:bhgrow}
\frac{dM_\bullet}{dt}=\left\{\begin{array}{ll}
\frac{\displaystyle 2.5\times10^7\msun}{\displaystyle 10^{10}\yr}\left(\frac{\displaystyle M_\bullet}{\displaystyle  10^9\msun}\right)^2\left(\frac{\displaystyle m}{\displaystyle  10^{-22}\eV}\right)^6\left(\frac{\displaystyle M}{\displaystyle 10^{10}\msun}\right)^4, & \quad \xi\ll1 \\
\frac{\displaystyle 4.75\times10^7\msun}{\displaystyle  10^{10}\yr}\left(\frac{\displaystyle M_\bullet}{\displaystyle  10^9\msun}\right)^3\left(\frac{\displaystyle m}{\displaystyle 10^{-22}\eV}\right)^6\left(\frac{\displaystyle M}{\displaystyle 10^{10}\msun}\right)^3, & \quad \xi\gg1. \end{array}
\right. 
\end{equation}

From the first of these equations we find that a seed black hole of mass
$M_{\bullet i}\ll M$ grows to $\xi\sim 1$ in a time 
\begin{equation}
t=4\times
  10^{14}\yr\left(\frac{10^{-22}\eV}{m}\right)^6\frac{10^6\msun}{M_{\bullet i}}\left(\frac{10^{10}\msun}{M}\right)^4.
\end{equation}
For the nominal parameters this rate is negligible, i.e., small seed black holes embedded in the soliton do not grow significantly. However, the rate depends strongly on the particle mass and could be significant  for $m\gtrsim 5\times10^{-22}\eV$.
Once the soliton contains a black hole more massive than a few percent of the soliton mass, so $\xi\gg1$, the last of Eqs.\ (\ref{eq:bhgrow}) implies that a black hole of initial mass $M_{\bullet i}$ swallows the soliton on a timescale
\begin{equation}
t=1\times
  10^{11}\yr\left(\frac{10^{-22}\eV}{m}\right)^6\left(\frac{10^9\msun}{M_{\bullet i}}\right)^2\left(\frac{10^{10}\msun}{M}\right)^3.
\end{equation}
Once again, this rate is not important for the nominal parameters but could be significant for particle masses a few times larger than the nominal value of $10^{-22}\eV$. 

\section{FDM and galaxy formation}
\label{sec:form}

\noindent
In contrast to CDM, in which small density fluctuations are unstable
on all spatial scales inside the horizon, FDM is unstable only for
masses larger than the Jeans mass, Eq.\ (\ref{eq:jeans}). As discussed in \S\ref{sec:maxmin}, the Jeans mass defines a lower limit to the mass of FDM halos and sub-halos. 

A stronger, but cosmology-dependent, constraint on the abundance of halos in FDM arises because the linear power spectrum of density fluctuations in FDM is 
suppressed relative to CDM at small scales. The degree of suppression is expressed by the ratio of the FDM power spectrum to the CDM power
spectrum. This ratio at $z=0$ is less than $0.5$ for
wavenumbers greater than and halo masses less than \cite{FDM,sch16}
\begin{align} \label{eq:khalf}
k_{1/2}&=4.5\Mpc^{-1}\left(\frac{m}{\vphantom{\big|}10^{-22}\eV}\right)^{4/9},
\\
M_{1/2}&=\ffrac{4}{3}\pi\rho \left(\frac{\pi}{k_{1/2}}\right)^3=
5\times 10^{10}\msun\,
\frac{\Omega_\fdm}{\vphantom{\big|}0.27}\left(\frac{10^{-22}\eV}{m}\right)^{4/3}.
\end{align}
Linear theory therefore predicts a sharp cutoff in the initial masses of FDM structures considerably above the limit given by Eqs.\ (\ref{eq:jpo}) or (\ref{eq:jeans}). However, non-linear numerical computations show that these overdensities fragment, leaving a spectrum of lower mass self-gravitating objects down to the scale given by those two equations, in which the number density of sub-halos is reduced in FDM relative to CDM by a factor $\sim (3M/M_{1/2})^{2.4}$ \cite{sch16}. It is important to verify these conclusions with different simulations and different numerical techniques, but for now we assume that they are correct. 

These arguments show that FDM tends to suppress formation of small
galaxies at high redshift compared to CDM. Thus it becomes a serious
question as to whether the FDM model would allow galaxies to be formed
that were capable of reproducing the high-redshift galaxy luminosity
function, re-ionizing the universe at $z = 8$--9 \cite{planck16}, and
producing the early, small-scale structure probed by the Lyman--$\alpha$
forest .

Because the power spectrum of density fluctuations in FDM is
suppressed for halo masses much larger than the Jeans mass ($M_{1/2}\gg
M_J$) one can approximate the formation of structure in FDM using CDM
simulations in which the initial power spectrum is that of FDM
(although of course this approximation cannot capture the formation of
solitons at the centers of these halos). Using
this approach, Schive et al.\ \cite{sch16} and \cite{cora16} have shown that FDM can
reproduce the UV luminosity function of galaxies in the redshift range
4--10, assuming a plausible relation between galaxy luminosity and
halo mass, if $m>10^{-22}\eV$. They also show that FDM is consistent
with current observations of the reionization history for plausible
estimates of the production rate of ionizing photons, about three
times larger than required for CDM\footnote{These results are
  consistent with a semi-analytic study by Bozek et
  al. \cite{bozek15}, once this is updated to use the 2016 Planck value
  for the reionization optical depth \cite{planck16}.}. 

\section{Lyman-alpha forest constraints on WDM and FDM}
\label{sec:forest}

The Lyman-$\alpha$ forest offers an additional probe of the power
spectrum. Viel et al.\ \cite{viel13} examined constraints on warm dark
matter from the forest and concluded that the mass of a hypothetical
WDM particle must exceed $m_{\textsc{wdm}}=3.3\mbox{\,keV}$ at the 2-$\sigma$ level. The 3-$\sigma$, 4-$\sigma$ and 9-$\sigma$ lower limits are
$2.5, 2$ and $1 {\,\rm keV}$ respectively. WDM and FDM do not have exactly the same matter power spectrum.
However, they share the same power spectrum on large scales,
and both exhibit a precipitous drop in power below a respective characteristic scale. Thus we can roughly translate an observational bound on WDM to a
corresponding one on FDM by matching $k_{1/2}$ for FDM from
Eq.~(\ref{eq:khalf})\footnote{We thank
Rennan Barkana for a private communication on this point, and for
discussions on the impact of reionization on WDM/FDM constraints.
The procedure of matching $k_{1/2}$ was also discussed in \cite{ms2014}.} to the
analogous quantity for WDM \cite{viel13,ms2014}\footnote{We focus on $m_{\rm WDM} = 2.5\keV$ as this case is treated in detail by Viel et al.},
\begin{equation}
\label{khalf}
k_{1/2}=12.5\Mpc^{-1}\left(\frac{m_{\textsc{wdm}}}{\vphantom{\big|}2.5\mbox{\,keV}}\right)^{1.11}
\left(\frac{\Omega_{\textsc{wdm}}}{\vphantom{\big|}0.27}\right)^{-0.11}\left(\frac{H_0}{\vphantom{\big|}70\kms\Mpc^{-1}}\right)^{2.22}.
\end{equation}
A WDM mass of $1, 2, 2.5, 3.3$ keV translates into
a FDM mass $m \sim 1, 6, 10, 20 \times 10^{-22}$ eV respectively.
Thus the range of FDM mass we are interested in is
disfavored by the Lyman-$\alpha$ forest data, and strongly so at the low-mass end.
It would be useful, however, to verify this conclusion with actual FDM numerical
simulations of the forest. It is possible that FDM has density fluctuations on scales relevant
to the forest (see \S\ref{sec:relax} and \cite{scb14}) that are not present in a WDM model.
It is also prudent to examine the assumptions underlying
the WDM constraints, which we now do. 

It has been understood starting from the 1990s that 
the statistical properties of the Lyman-$\alpha$ forest can be
explained by the $\Lambda$CDM model with a minimal
set of additional assumptions
\cite{Cen:1994da,Hernquist:1995uma,Zhang:1995zh,HGZ97,Croft:1997jf,
GH97,Croft:1998pe,H98}. The most important of these is that the neutral hydrogen
density $n_{\rm HI}$ responsible for the Lyman-$\alpha$ absorption obeys 
\begin{eqnarray}
\label{nHI}
n_{\rm HI} \propto (1 + \delta_b)^2 T^{-0.7} J^{-1} \,
\end{eqnarray}
where $\delta_b$ is the fractional overdensity of baryons, $T$ is the temperature, 
and $J$ is the amplitude of the ionizing background
(suitably averaged over frequencies, weighted by the ionization cross-section).
Photoionization equilibrium is assumed to be a good approximation, which leads to the factors
of $(1 + \delta_b)^2$ (ionized fraction is close to unity) and
$T^{-0.7}$ (approximate scaling of the recombination rate with the temperature). 
There is a tight correlation between temperature and density
in a low-density photoionized medium, $T = T_0 (1 + \delta_b)^{\gamma-1}$,
where $\gamma \sim 1$--$1.6$ \cite{HG97}. It is generally assumed that $T_0$, $\gamma$ and $J$
have negligible spatial fluctuations; fluctuations in $n_{\rm HI}$
thus reflect fluctuations in baryon density $\delta_b$.
The baryon fluctuations in the intergalactic medium are assumed to evolve
under gravity, counteracted by pressure. 
Feedback processes, such as galactic winds or outflows,
are assumed to have negligible impact on the forest.

Lyman-$\alpha$ constraints on WDM fall into two categories:
those that are primarily from high-redshift, high-resolution data, and
those that come from data at lower redshifts and larger scales.
We start with a discussion of the first category, exemplified by
the thorough analysis of Viel et al.\ \cite{viel13}, who derived constraints
from data at $z \sim 4.2$--$5.4$ covering scales $k \sim 0.4$--$9 {\,\rm
  Mpc}^{-1}$. 
Their strongest constraint comes from the highest redshifts, $z \sim 5.4$,
and the smallest scales, $k \sim 4$--$9 \Mpc^{-1}$
(see Figure 12 of \cite{viel13}). 
WDM with a mass of $2.5 \keV$
causes a 10\% suppression of power on these scales, in tension with the data\footnote{\label{WDMeffect}
Keeping all other parameters fixed, at this mass, scale and redshift WDM suppresses the power by $\sim 50\%$. However, allowing other parameters (such as the temperature)
to vary to best fit the data, the net suppression is closer to
$\sim 10\%$. Note  that $k_{1/2}$ in
Eq. (\ref{khalf}) refers to the suppression scale for the
linear matter power spectrum; the nonlinear mapping from mass
to flux moves the suppression scale to a lower $k$ for
the flux power spectrum.
}.
At high redshifts, two effects are potentially
significant: (i) the larger average neutral hydrogen density
and the smaller number of ionizing sources leads
to enhanced spatial fluctuations of the ionizing background
$J$; (ii) there can be significant fluctuations in the temperature, remnants of
an inhomogeneous reionizing process as one approaches
the epoch of reionization (recent Planck constraints
suggest reionization on average occurred around
$z \sim 8$ \cite{planck16}).

An attempt was made by Viel et al.\ to account for the first effect using a simulation
in which the ionizing background is sourced by rare quasars,
with the accompanying clustering and Poisson fluctuations. 
However, the simulation did not include spatial fluctuations due
to the inhomogeneous distribution of absorbers.
There appears to be a variety of possible modifications
to the flux power spectrum from a fluctuating ionizing background,
depending on assumptions about the density and clustering
of the ionizing sources and absorbers.
For instance, \cite{McDonald04} showed that while quasars as sources
tend to add power on large scales without significantly affecting
the small scale power, Lyman-break galaxies tend to suppress
power on large scales $k \sim 0.1$--$0.5 \Mpc^{-1}$ and add
power on small scales $k \gsim 1 \Mpc^{-1}$, especially at high redshifts
(see Figures 5 and 7 of \cite{McDonald04}). 
Even in the case of quasars as sources, different investigators
arrive at different conclusions about the scale dependence of the flux
power spectrum modification
(see for instance \cite{Croft2003} who included inhomogeneous radiative transfer,
and \cite{McDonald04,MeiksinWhite} who applied an average attenuation).
In fitting the forest data, the common practice is to assume a
template for the effect of a fluctuating ionizing background---meaning the {\it shape} (scale dependence) of the flux
power spectrum modification---and allow the {\it amplitude} of the
modification to be a free parameter. 
In this way, \cite{viel13} obtained a bound on the amplitude
of a particular quasar template, and showed that it does not
have a large impact on the WDM constraint.
It would be useful to repeat the analysis using other templates,
including those that might be more degenerate with the effect of
WDM in their scale dependence
(such as those from galaxies as ionizing sources, and
those that account for inhomogeneous radiative transfer).

The second astrophysical effect to consider is spatial fluctuations
in temperature due to patchy reionization.
Reionization typically raises the temperature
of the intergalactic medium to a few times $10^4$ K.
The gas generally cools afterwards (except for late time HeII reionization
that could temporarily reverse this trend).
Because different patches of the medium 
reionize at different times due to fluctuations in the
distribution of ionizing sources and absorbing materials, 
the temperature could vary by a factor of a few at
redshift $\sim 5$ 
if the reionization of different patches spans the redshift
range $\sim 7$--10 \cite{HH03,Aloisio2015}.
Let us stress that by temperature variations, we are not referring
to the fact that temperature depends on density---such a dependence
is expected, even in the absence of inhomogeneous reionization,
and is captured by the temperature-density relation $T = T_0 (1+
\delta_b)^{\gamma-1}$ described earlier. 
Here we are interested in an extra source of
fluctuations in temperature. It can be described in a number of ways.
One way is to phrase this in terms of fluctuations in temperature
that exist even at the same local density, say the mean density $T_0$
\footnote{At redshift $\gsim 5$, the relevant
density is actually $\delta_b < 0$ for pixels that are not completely
saturated. We use the temperature at the mean density
$T_0$ as a convenient proxy, since spatial fluctuations in $T_0$
imply temperature fluctuations at another density too.
It is worth emphasizing that inhomogeneous reionization
leads to spatial fluctuations in the exponent $\gamma$ and in fact the
whole temperature-density relation. 
}. A more sophisticated description would correlate the local
temperature to the density smoothed on a larger scale (scale of
the reionization patch). A patch with a higher  large-scale density might reionize
earlier, due to an abundance of sources, leading to a lower
temperature at the observed redshift.
The opposite could happen: the same patch might reionize later, due to
shielding by an abundance of absorbers, leading to a higher
temperature.

Less attention has been paid to this second effect, that of patchy reionization on
the flux power spectrum.
An exception is \cite{Cen2009}\footnote
{We thank Uros Seljak for pointing out this 
and other references, and for thoughtful discussions
on the Lyman-$\alpha$ forest.}
who compared how two
simulations, with early and late reionization respectively, predict
different flux power spectra at $z \gsim 4$.
They found that the late reionization scenario
increases the power on large scales $k \lsim 5 \Mpc^{-1}$
and suppresses it on small scales $k \gsim 5 \Mpc^{-1}$.
Could further investigations, exploring different assumptions about
the sources and absorbers, reveal qualitatively different behavior, 
just as in the case of the first effect?
Let us describe one possibility, motivated by the observations of Aloisio
et al.\ \cite{Aloisio2015}.

By examining the
average absorption at $z \sim 5.5$--5.9 along different lines of
sight, Aloisio et al.\ found evidence that $T_0$ fluctuates on a scale of $\sim 36$ Mpc,
suggesting reionization patches of this size.
Such fluctuations leave two imprints on the flux power spectrum.
Fluctuations in temperature lead to fluctuations in the recombination
rate, and therefore in the neutral hydrogen density 
(Eq. \ref{nHI}). 
The affected scale is roughly $k \lsim \pi / 36 \Mpc \sim
0.09 \Mpc^{-1}$ -- this is beyond the largest scales probed
in existing 1D power spectrum analyses
\cite{Palanque2013}.
Cross power spectrum might
be useful in this respect.
A second imprint, perhaps surprisingly, is on small scales, and arises because of smoothing. 
There are two known sources of smoothing, thermal
broadening (an effect in 1D) and gas pressure, i.e., Jeans
smoothing (an effect in 3D).
Thermal broadening dominates the smoothing of the observed
(1D) flux power spectrum \cite{ZHT2001,ZSH2003}, although for
our purpose there is no real need to distinguish between the two.
A finite temperature leads to a smoothing of the flux power spectrum:
$P_f (k) \rightarrow P_f (k) \, W(k, T_0)$ ,
where $W(k, T_0)$ is the smoothing kernel\footnote{The smoothing kernel strictly speaking should depend on
$\gamma$ and even reionization history \cite{GH97}, in addition to $T_0$.
We use $T_0$ as a proxy for the dependence of the smoothing kernel
on the thermal state of the gas.}.
At a temperature around $10^4 {\,\rm K}$,
the smoothing kernel suppresses power at $k \gsim 5 \Mpc^{-1}$;
the higher the temperature, the larger the affected scales.
A measurement of the flux power spectrum from many reionization
patches effectively probes the {\it averaged} smoothing kernel
$\langle W(k, T_0) \rangle$.
This need not be the same as the smoothing kernel at the averaged
temperature $W(k, \langle T_0 \rangle)$.
In general $\langle W(k, T_0) \rangle$ might even
differ in shape from the smoothing kernel at {\it any} temperature.
We expect $\langle W(k, T_0) \rangle$ to retain more small-scale
power: at high $k$'s, the power is dominated by patches where the
temperature fluctuates below the mean. The magnitude of this
effect depends on the range of temperature in these
reionization patches. We estimate that the
effect could be
at the $\sim 10 \%$ level on scales relevant for the WDM constraint,
if the difference between the highest and lowest temperatures
exceeds $10^4$ K. It would be useful to
investigate this more carefully using numerical simulations.

It is worth noting that, the effect of patchy reionization aside,
simply modifying the thermal history could impact the WDM constraint
also. It was shown by \cite{Garzilli2015} that
allowing for a non-monotonic temperature evolution
(expected if HeII reionization occurs at intermediate redshifts)
weakens the $2\sigma$ lower bound on WDM mass from $3.3 \keV$ to 
$2.1 \keV$,
using the same data as \cite{viel13}.

We close by discussing WDM constraints from Lyman-$\alpha$ forest data
on large scales and low redshifts ($k \sim 0.06$--$1.6 \Mpc^{-1}$, $z
\sim 2.2$--$4.4$), for instance Seljak et al.\ \cite{Seljak2006}, and more recently, 
Baur et al.\ \cite{Baur2015}---the latter gave a $2\sigma$ lower limit of
$m_{\rm WDM} > 4.1 \keV$. 
At these lower redshifts, the effects discussed above, patchy
reionization and fluctuations in the ionizing background, are reduced.
On the other hand, the effect of WDM is
also smaller on these relatively large scales: the difference in power
between CDM and WDM ($m_{\rm WDM} = 2.5 \keV$ for instance) 
is at the few percent level if the other parameters are fixed
\cite{Baur2015}; allowing them to float to
their respective best-fit values would further diminish the difference
between the two models. At this level of precision, the effect of 
fluctuations in the ionizing background might not be negligible.
Moreover, \cite{Viel2012} showed that
galactic outflows can also impact the flux power spectrum at this level
(see also \cite{Adelberger2003}).
Typically, these astrophysical effects are accounted for by
template fitting: a template based on numerical simulations of the
effect is introduced which fixes the scale (and often redshift)
dependence of the flux power modification; the amplitude of the
modification is treated as a free parameter, and WDM constraints
are obtained by marginalizing over this parameter.
As emphasized earlier in the context of fluctuations in the ionizing
background, it is helpful to explore the diversity of possible
templates. In the case of galactic outflows, it would be
useful to know how robust the assumed redshift and scale dependence 
is to uncertainties in the physics of galaxy formation.

In summary, it is possible that the existing Lyman-$\alpha$ forest
constraints on WDM would be relaxed if the diversity of modifications
to the flux power spectrum, from a number of astrophysical effects,
are taken into account.
We believe effects similar to those we have discussed could 
weaken the corresponding FDM lower bound to 
a few times $10^{-22}\eV$. 
Further analyses of simulations and data are required to
determine if this is the case.
It would also be helpful to carry out the flux power spectrum analyses
directly on FDM simulations, to test the validity of (or to bypass)
the translation between WDM and FDM constraints.

\section{Summary}
\label{sec:summary}

The standard Lambda Cold Dark Matter or $\Lambda$CDM model for the nature of dark matter and the growth of cosmic structure in the universe describes a wide variety of observations with remarkable precision. To date, experimental searches for the particle or particles comprising CDM have been unsuccessful or controversial, but ``absence of evidence is not evidence of absence'' and many ongoing experiments offer the hope of future success. A larger concern is that few if any of the predictions of the CDM scenario on scales smaller than $\sim 10\kpc$ have been successful. While all large-scale predictions of $\Lambda$CDM appear to be valid, the simple-minded predictions of this model are inconsistent with the relatively small number of low-mass field and satellite galaxies, the absence of dark-matter cusps at the centers of galaxies dominated by dark matter, the absence of dark matter around normal globular clusters, the weakness of dynamical friction in dwarf galaxies, and several other phenomena. Complex dynamics or baryonic physics may ultimately prove able to account for all of these, but it is worthwhile to investigate whether some well-motivated model for the properties of the dark matter could replicate the successes of CDM on large scales and in addition predict properties of the dark-matter distribution and dynamics on small scales (and correspondingly at early times) that are in better agreement with the observations than is CDM. 

One possibility that has been considered by a number of investigators is an extremely light boson having a de Broglie wavelength $\lambda$ of kpc scale at the typical internal velocities in galaxies (Eq.\ \ref{eq:brog}). A fluid with these properties would resist compression to smaller scales and could not form equilibrium self-gravitating halos with mass smaller than about $10^7\msun$ (Eqs.\ \ref{eq:jpo} and \ref{eq:jeans}), so it provides a model that is eminently falsifiable by comparison to observations at early times and small scales. We have summarized the work of others (see also the recent review by Marsh \cite{mar15}) and added several new calculations, concluding that, if the particle mass lies in the range $m =1$--$10\times 10^{-22}\eV$, the FDM scenario is an attractive alternative to CDM with new and powerful tests soon to be available and new calculations and simulations needed to elucidate its behavior. For several of the astrophysical applications the lower end of this mass range is preferable but current observations of the Lyman-$\alpha$ forest, though their interpretation is somewhat model-dependent, favor the upper end of the range and may require masses as large as 10--$20\times 10^{-22}\eV$. We note in the text and in what follows the many possible observations that could cleanly demonstrate that FDM is not the dominant form of dark matter.

A possible concern is that it is unnatural to have a boson with a mass $\sim 10^{-22}$--$10^{-21}\eV$, many orders of magnitude less than the masses of known elementary particles. In fact, a spinless field of very small mass that also interacts very weakly with ordinary matter is fairly natural in particle physics, because there is extra symmetry in the limit of a massless field of spin zero that interacts only with gravity.  Moreover, there are natural mechanisms to generate a nonzero but exponentially small mass for such a field.   In models with multiple fields of this type, which are relatively well-motivated (see for example \cite{Axiverse}), it is plausible that one of these fields might have a mass in the required range (see Eq. \ref{eq:miracle}).

Simulations show that FDM halos are comprised of an envelope that resembles the NFW profile found in CDM \cite{nfw97}, surrounding a central core or peak that is a stationary, minimum-energy solution of the Schr\"odinger--Poisson equation, sometimes called a ``soliton''. As described in \S\ref{sec:relax}, the density of the soliton can be much larger than the density of the surrounding halo, and produces a distinct signature in the rotation curve of FDM-dominated systems that contain baryonic disks. The sharp peak in the rotation curve due to a massive central soliton may be observable in ultra-diffuse galaxies such as Dragonfly 44 \cite{vandokkum2016}. On the other hand the cuspy density profile $\rho\sim r^{-1}$ expected in CDM dominated systems should not exist in FDM because the soliton has a homogeneous core. 

In contrast to isolated CDM halos, FDM halos of density $\rho$ relax in a manner analogous to gravitating N-body systems, as if they were composed of quasiparticles with mass $\sim\rho(\lambda/2)^3$ (\S\ref{sec:relax}). We suggest that relaxation both adds mass to the central soliton and expels mass to infinity on a timescale given roughly by Eq.\ (\ref{eq:trelax}), but this suggestion needs to be tested by simulations. Relaxation due to FDM quasiparticles or wavepackets is probably too weak to affect the disk thickness in the solar neighborhood or to disrupt globular clusters or wide binary stars at or beyond the solar radius, but it may thicken the inner parts of galactic stellar disks and pump energy into stellar bulges. We also suggest that relaxation could produce FDM disks in the central regions of galaxies. The resistance of FDM to compression put a stringent lower bound on the scale height of an FDM disk; in the solar neighborhood any such disk could not easily have a half thickness (the distance from the midplane containing half the mass) less than several hundred parsecs or local density greater than $\sim 0.02\msun\pc^{-3}$ (for $m=10^{-22}\eV$; the thickness and density scale as $m^{-2/3}$ and $m^{2/3}$ respectively). 

As noted earlier, dark-matter halos significantly less massive than about $10^7(m/10^{-22}\eV)^{-3/2}\msun$ are not possible in FDM and the limit for sub-halos is even stronger because of tunneling of FDM through the tidal radius of the sub-halo (Eq.\ \ref{eq:tides}). Thus the elucidation of the properties of halo substructure within our Galaxy (by studying stellar streams) or along lines of sight through other galaxies (via gravitational lensing) and the study of the dark-matter distribution in low-mass galaxies offer promising tests that could distinguish FDM from CDM. 

It has long been a puzzle that the five globular clusters in the Fornax dwarf galaxy have not yet spiraled to the center of that system due to dynamical friction. If the dark matter in Fornax is FDM rather than CDM, our calculations show a substantial increase in the timescale for dynamical friction, and thus FDM essentially solves the problem of the survival of these clusters. Further observational work on the globular cluster systems of other dwarf spheroidal galaxies would be rewarding.

The results discussed so far do not depend directly on the power spectrum of cosmic perturbations, but the expected cutoff of the FDM spectrum at small scales (Eq.\ \ref{eq:khalf}) has many further implications. It reduces early galaxy formation below what would have been expected from the CDM model, but this modified formation history is consistent with current observations at $4\lesssim z\lesssim 10$ \cite{sch16} and the late reionization, $z\sim 8$, recently reported by {\it Planck} \cite{planck16}. There is a substantial reduction in the number of halos expected in the local universe as compared to CDM for halos below a few times $10^{9}(m/10^{-22}\eV)^{-4/3}\msun$ but a sufficient number remain to provide for the local dwarf galaxies, which have estimated halo masses of roughly $3\times10^9\msun$ \cite{fat16}.
Thus the ``too big to fail'' and ``missing satellite'' problems are addressed directly by FDM, but until our understanding of baryonic physics and feedback is more complete we will not know whether the change in the properties of the dark-matter sub-halos implied by FDM is either required or useful.

In summary, the hypothesis that the principal component of the ubiquitous dark matter is an ultra-light axion is an attractive and testable alternative to CDM, having no serious inconsistencies with current data if the particle mass  $m \gtrsim 10^{-22}\eV$. There are  significant and attractive observational consequences if the mass is in the range 1--$10\times 10^{-22}\eV$. There is tension with   observations of the Lyman-$\alpha$ forest, which favor masses 10--$20\times10^{-22}\eV$ or higher. More sophisticated calculations of reionization and of structure formation in FDM are required to determine whether this variety of constraints is consistent with observations. 

\section*{Acknowledgements}

We thank Rennan Barkana, Andrei Beloborodov, Tim Brandt, Tom Broadhurst, Justin Khoury, Yuri Levin, Adam Lidz, David Marsh, Matt McQuinn, Alberto Nicolis, Riccardo Penco, Massimo Pietroni, Hsi-Yu Schive, Uros Seljak, Sergei Sibiriyakov, Pierre Sikivie, David Spergel and Matteo Viel for their comments and insights. This research was supported in part by NASA grants NNX14AM24G and NXX16AB27G, NSF grants AST-1406166 and PHY-1606531, and DOE grant DE-SC0011941.

\appendix

\section{The quantum virial theorem}

\noindent
In this Appendix we derive the quantum-mechanical analog
to the classical virial theorem \cite{Chavanis2011}. 

We begin with the Schr\"odinger equation
\begin{equation}
i\hbar\frac{\p\psi}{\p t}=H\psi=-\frac{\hbar^2}{2m}\nabla^2\psi+m\Phi({\bf
  r},t)\psi
\end{equation}
where $H$ is the Hamiltonian operator, $\Phi({\bf r},t)$ is the
gravitational potential, and $m|\psi({\bf r},t)|^2=\rho({\bf r},t)$, the
mass density. The moment of inertia is $I=\half m\int d{\bf
  r}\,r^2|\psi|^2$. Then 
\begin{equation}
 \dot I=-\frac{im}{2\hbar}\int d{\bf r}\,r^2\,(\psi^*H\psi - \psi H \psi^*)= -\frac{im}{2\hbar}\int d{\bf r}\,\psi^*[r^2,H]\psi
\end{equation}
where $[\cdot,\cdot]$ is the commutator. Since $[r^2,\Phi]=0$ this
  simplifies to
\begin{equation}
 \dot I=\ffrac{1}{4}i\hbar\int d{\bf r}\,\psi^*[r^2,\nabla^2]\psi.
\end{equation}
Similarly,
\begin{equation}
\ddot I=\ffrac{1}{4}\int d{\bf r}\,\psi^*[[r^2,\nabla^2],H]\psi.
\end{equation}
Using the result $[[r^2,\nabla^2],H]=-4m({\bf
  r}\cdot\bnabla\Phi)-4(\hbar^2/m)\nabla^2$ we have 
\begin{align}
\ddot I&=-\int d{\bf r}\,\psi^*\Big(m\psi\,{\bf r}\cdot\bnabla\Phi +
\frac{\hbar^2}{m}\nabla^2\psi \Big)\nonumber \\
&=-m\int d{\bf r}\,|\psi|^2\,{\bf r}\cdot\bnabla\Phi
+\frac{\hbar^2}{m}\int d{\bf r}\,|\bnabla\psi|^2.
\end{align}
At this point it is useful to write the wavefunction in the form
$\psi=\sqrt{\rho/m}\exp(i\theta)$ with $\theta$ real, and define ${\bf
  v}\equiv \hbar\bnabla\theta/m$ (cf.\ Eq.\ \ref{eq:madelung}). Then
\begin{align}
\ddot I&=-\int d{\bf r}\,\rho\,{\bf r}\cdot\bnabla\Phi
+ \int d{\bf r}\,\rho v^2+\frac{\hbar^2}{m^2}\int d{\bf
  r}\,|\bnabla\sqrt{\rho}\,|^2\nonumber \\
&\equiv V+2K + 2Q.
\label{eq:qvt}
\end{align}
This is the quantum virial theorem. The term $K\equiv \half\int d{\bf r}\,\rho v^2$
is analogous to kinetic energy, although the analogy is exact only in
the classical limit. The term $Q$ can be called the quantum energy. If the system is self-gravitating then
\begin{equation}
\Phi({\bf r})=-G\int \frac{d{\bf r}'\,\rho({\bf r}')}{|{\bf r}-{\bf
    r}'|}
\end{equation}
so
\begin{equation}
V=-G\int d{\bf r}d{\bf r}'
\frac{\rho({\bf r})\rho({\bf r}')\,{\bf r}\cdot({\bf r}-{\bf
    r}')}{|{\bf r}-{\bf r}'|^3}=-\frac{G}{2}\int d{\bf r}d{\bf r}'
\frac{\rho({\bf r})\rho({\bf r}')}{|{\bf r}-{\bf r}'|}=W
\end{equation}
where $W$ is the total potential energy of the system. The total
energy of the system is $K+W+Q$, which is conserved. 

In a steady state $\ddot I=0$, and since $K\ge 0$ we have
\begin{equation}
\frac{Q}{|W|}\le \frac{1}{2}
\end{equation}
which sets a lower limit on the particle mass $m$ for an equilibrium system with a
given density distribution $\rho({\bf r})$. The limit is saturated
($Q/|W|=\half$) if the phase of the wavefunction is position-independent
so $\bnabla\theta=0$, a condition that is satisfied in the soliton.

\section{Eigenstates of the Schr\"odinger--Poisson equation}

\label{app:speq}

\noindent
A useful standard for comparison to theoretical models and simulations of FDM halos is provided by the eigenstates of the time-independent Schr\"odinger--Poisson equation, which are
solutions of the combined equations
\begin{align}
-\frac{\hbar^2}{2m}\nabla^2\psi+ m\Phi\psi=mE\psi, \qquad \nabla^2\Phi=4\pi Gm|\psi|^2.
\label{eq:sp}
\end{align}
Here $\psi({\bf r})$ and $\Phi({\bf r})$ are the wavefunction and gravitational potential, and $E$ is the energy per unit mass of the eigenstate. We consider isolated systems so we assume that $\psi,\Phi$
approach zero as $|{\bf r}|\to \infty$ and that they are regular near the origin. The total mass $M=m\int d{\bf r}\,|\psi|^2$ and it is straightforward to show that $ME=2W + K + Q$; thus the energy eigenvalue $E$ is {\em not} the total energy per unit mass of the system. Using the time-independent virial theorem we can show that the energy eigenvalue is related to the potential energy by $W=\frac{2}{3}ME$. 

We restrict our attention to spherical solutions\footnote{An interesting unsolved problem is whether there are non-spherical
solutions of the time-independent Schr\"odinger--Poisson equation.}. Then the wavefunction can be taken to be real and 
\begin{align}
-\frac{\hbar^2}{2mr^2}\frac{d}{dr}r^2\frac{d\psi}{dr} +m\Phi\psi=mE\psi, \qquad 
\frac{1}{r^2}\frac{d}{dr}r^2\frac{d\Phi}{dr}=4\pi Gm\psi^2.
\label{eq:spos}
\end{align}

Following other authors \cite{kaup68,rb69,bgj98,hmt03,guzman2004,Chavanis2011} we compute numerically the solutions of these equations, and label the
eigenstates $n=0,1,2,\cdots$ in order of increasing energy. The
eigenstate labeled by $n$ has $n$ nodal radii at which the density is
zero. Because the Schr\"odinger--Poisson equations admit a scaling
invariant, all systems corresponding to a given level $n$ form a
one-parameter family that can be specified by the total mass
$M$. Then quantities such as the central density, central potential,
half-mass radius, virial velocity, energy eigenvalue, and potential energy can be written
\begin{equation}
\rho_c=\left(\frac{Gm^2}{\hbar^2}\right)^3M^4 \rho_n, \quad
\Phi_c=-\left(\frac{GMm}{\hbar}\right)^2 \phi_n, \quad
r_{1/2}=\frac{\hbar^2}{GMm^2}f_n, 
\label{eq:ground}
\end{equation}
\begin{equation}
v_{\rm vir}=(-W/M)^{1/2}=\frac{GMm}{\hbar}w_n^{1/2}, \quad E=-\left(\frac{GMm}{\hbar}\right)^2\epsilon_n, \quad W=-\frac{G^2M^3m^2}{\hbar^2}w_n
\label{eq:spfact}
\end{equation}
The dimensionless constants $\rho_n$, $\phi_n$, $f_n$, $\epsilon_n$,
and $w_n$ are given in
Table \ref{tab:sp}. Note that $w_n=\frac{2}{3}\epsilon_n$. 

\begin{table}[ht]
\centering
\caption{Properties of the lowest eigenstates of the
  Schr\"odinger--Poisson equation (Eq.\ \ref{eq:spfact})}
\vspace{1.15in}
\begin{tabular}{ c|c|c|c|c|c} 
$n$ & $\rho_n$ & $\phi_n$ & $f_n$ & $\epsilon_n$ & $w_n$ \\
 \hline
0 & 0.00440   & 0.3155 & 3.9251   & 0.16277  & 0.10851  \\
1 & 0.000180 & 0.07146 & 23.562 & 0.03080  & 0.02053 \\
2 & 0.000031 & 0.03139 & 60.903 & 0.012526  & 0.008351 \\
3 & $9.400 \times10^{-6}$ & 0.01772 & 116.18 &  0.006747 & 0.004498 \\
4 & $3.733 \times 10^{-6}$ & 0.01240 & 178.60 &  0.004209 & 0.002806 \\ [1ex]
\end{tabular}
\label{tab:sp}
\end{table}

We see that the central density is a strongly decreasing function of
the level number $n$ so the densest system is the ground state,
$n=0$. The ground state is linearly stable but the $n^{\rm th}$
excited state has $n$ spherically symmetric unstable modes which decay to the ground
state through the dispersion of probability density to infinity; see
\S\ref{sec:relax} and references \cite{ss94,hmt03,gul06}. Thus the
ground state, sometimes called a ``soliton'', is the long-term
attractor for any FDM system.

\section{Quantum-mechanical treatment of the tidal radius}

\label{app:tide}

\noindent
To derive the classical formula for the tidal radius, consider a mass $M$ in a circular orbit of radius $a$ around a point-mass host and
work in a rotating Cartesian reference frame that is centered on $M$, with the
$z$-axis pointing in the same direction as the orbital angular
momentum and the $x$-axis pointing away from the host. Let $a$ and
$\Omega$ be the orbital radius and angular speed. Then for $r=(x^2+y^2+z^2)^{1/2}\ll a$ the equations of motion for a test
particle are (see for example Eq.\ 8.97 of \cite{BT})
\begin{equation}
\ddot x=2\Omega\dot y +3\Omega^2x-\frac{GMx}{r^3}, \quad \ddot
y=-2\Omega\dot x -\frac{GMy}{r^3}, \quad \ddot
z=-\Omega^2z-\frac{GMz}{r^3}.
\end{equation}
There is an equilibrium solution ($\dot x=\dot y=\dot z=0$)
at $x=\pm r_t$, $y=z=0$ where $r_t=(GM/3\Omega^2)^{1/3}$ is the tidal
radius. If the host has mass $\mm$ then 
\begin{equation}
r_t=a\left(\frac{M}{3\mm}\right)^{1/3}.
\end{equation}

We will consider a simplified system, in which
the mass $M$ is subjected to a spherically symmetric tidal potential
$\Phi_t=-\ffrac{3}{2}\Omega^2r^2=-\ffrac{3}{2}G\mm r^2/a^3$. Then we can work in an inertial
frame centered on $M$, in which the equations of motion are
\begin{equation}
\ddot {\bf r}=-3\Omega^2{\bf r}-\frac{GM}{r^3}{\bf
  r}=GM\left(\frac{1}{r_t^3}-\frac{1}{r^3}\right){\bf r};
\label{eq:simpletide}
\end{equation}
these have equilibrium solutions at $r=r_t$ and capture much of the
dynamics relevant to tidal disruption. 

The Schr\"odinger-Poisson equation analogous to (\ref{eq:simpletide}) is 
\begin{align}
-\frac{\hbar^2}{2m^2r^2}\frac{d}{dr}r^2\frac{d\psi}{dr} +(\Phi-\ffrac{3}{2}\Omega^2r^2)\psi=E\psi, \qquad \frac{1}{r^2}\frac{d}{dr}r^2\frac{d\Phi}{dr}=4\pi Gm\psi^2.
\label{eq:sposom}
\end{align}
In contrast to the case of particles such as CDM, which can orbit forever if they are inside the tidal radius, FDM is described by the wave equation (\ref{eq:sposom}) which allows matter with
$r\ll r_t$ to tunnel through the potential barrier centered on $r_t$
and escape to infinity. Thus all self-gravitating systems in an
external tidal field are eventually tidally disrupted. It is worth stressing that by tunneling, we do not mean tunneling in the sense of quantum field theory---classical field theory provides an adequate description of all phenomena discussed in this paper (see comments at the end of \S \ref{sec:FDMsuperfluid}).

\begin{figure}[htb]
\begin{center}
\includegraphics[trim=0mm 0mm 0mm 0mm, clip, scale=.5, angle=0]{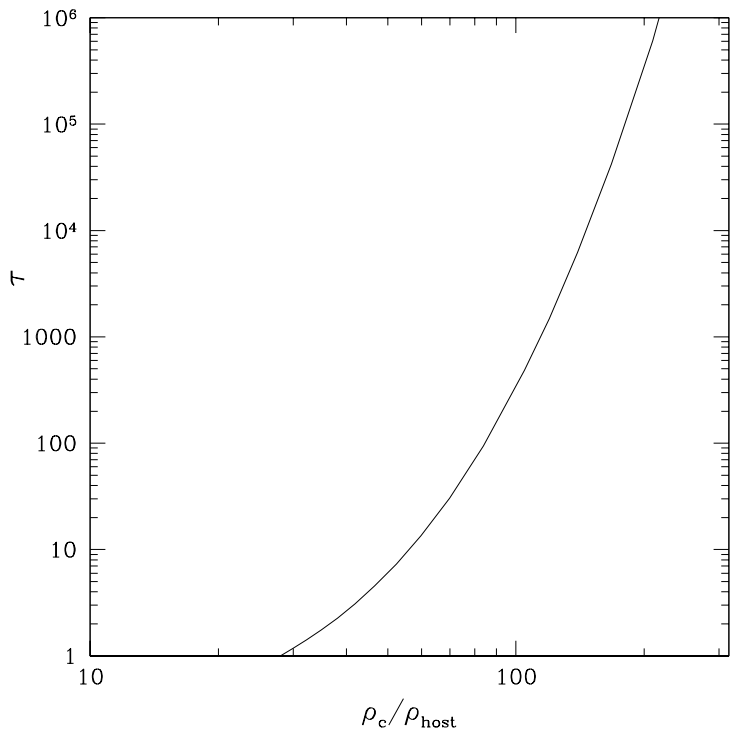}
\caption
{The lifetime of a stellar system in the ground state of the
  Schr\"odinger--Poisson equation when a spherically symmetric tidal
  field is present. The vertical axis gives the dimensionless lifetime
  $\tau$ in units of the orbital period (eq.\ \ref{eq:taudef}) and the
  horizontal axis gives the ratio of the central density to the mean
  density of the host averaged over the orbital radius. 
}
\label{Ftides}
\end{center}
\end{figure}

To determine the disruption timescale, we solve Eqs.\ 
(\ref{eq:sposom}) numerically. We impose an outgoing wave boundary
condition as $r\to\infty$; using the WKB approximation we can show
that this must have the form
\begin{equation}
\psi(r)\to \frac{A}{r^{3/2}}\exp\left(i\frac{\sqrt{3}m\Omega}{2\hbar}r^2\right)
\label{tide:asymp}
\end{equation}
where $A$ is a constant. We impose regular boundary conditions at the origin ($\psi'(r)\sim r$ and $\Phi'(r)\sim r$) and allow the energy eigenvalue $E$ to be complex; then the rate of mass loss is
\begin{equation}
\dot M=2M\mbox{\,Im}\,(E).
\end{equation}
We may then define a
dimensionless tidal disruption time by
\begin{equation}
\tau\equiv \frac{M\Omega}{2\pi\dot M},
\label{eq:taudef}
\end{equation}
so a system will lose a substantial fraction of its mass due to tides
in $\tau$ orbits. 

In Figure \ref{Ftides} we show $\tau$ as a function of
$\rho_c/\rho_{\rm host}$ where $\rho_c$ is the central density of the
satellite and $\rho_{\rm
  host}=3\mm/(4\pi a^3)=3\Omega^2/(4\pi G)$ is the mean density of the host galaxy
averaged over the distance $a$ to the satellite. 

\section{Quantum-mechanical treatment of dynamical friction}

\label{app:fric}

\noindent
The goal of this Appendix is to compute the dynamical friction on a
point-mass object (the ``test object'') moving through the FDM superfluid. As the object
travels through the fluid, a gravitational wake forms behind it, and
the associated overdensity exerts a drag. The wave nature of FDM is 
expected to suppress the overdensity, reducing the drag.

Let us work in a frame in which the test object, of mass $m_{\rm cl}$, is stationary
at the origin, and the dark-matter fluid flows past with a velocity
${\bf v}$ and a uniform density $\rho$ in the far past. The dark matter
interacts with the object gravitationally, but its self-gravity is
ignored (the validity of this approximation is discussed following Eq.\ \ref{eq:casymp}). This is the classic Coulomb scattering problem.  The time-independent Schr\"odinger
equation for the wavefunction $\psi$ has the solution
(e.g., \cite{MottMassey}):
\begin{equation}
\label{MM}
\psi = \nn e^{ikz} M[i\beta, 1, ik(r-z)]\, ,
\end{equation}
where $M$ is a confluent hypergeometric function. Here $z$ is the
coordinate parallel to ${\bf v}=v\,\hat{\bf z}$, $\hbar k = mv$ is the
associated momentum, and $r$ is the radial distance from the point
mass. The dimensionless parameter
\begin{equation}
\beta \equiv \frac{Gm_{\rm cl}m^2}{\hbar^2k}=\frac{Gm_{\rm cl}m}{\hbar
  v}=2\pi\frac{Gm_{\rm cl}}{v^2\lambda};
\label{eq:betadef}
\end{equation}
apart from the factor of $2\pi$ this is the ratio of the
characteristic length scale $Gm_{\rm cl}/v^2$ (the impact parameter at which
significant deflection of the orbit occurs in the classical limit) to the de
Broglie wavelength $\lambda$ (Eq.\ \ref{eq:brog}). We are interested
in the regime $\beta\ll1$, whereas the classical description of
dynamical friction \cite{chandra43,BT} is in the regime $\beta\gg1$. If 
we normalize the wavefunction so that $|\psi|^2$ is the
density, then the normalization $\nn$ is determined by the
condition that $|\psi|^2=\rho$ as $-z=r\to\infty$. Up to an unimportant
phase,
\begin{equation}
\label{normalization}
\nn = \sqrt{\rho} \, e^{\pi\beta/2} \, |\Gamma(1-i\beta)|
\end{equation}
where $\Gamma$ is the gamma function.

The dynamical friction force is given by a surface
integral of the fluid's momentum flux density tensor (Eq.\ \ref{eq:momentum}):
\begin{equation}
\label{Fric}
{\bf F}=F\hat{\bf z}\quad\mbox{where}\quad F = -\oint dS_j \, \Pi_{jz} \, .
\end{equation}
The system is assumed to be in a steady state with no Hubble expansion, so we may set the Hubble constant $H=0$ and the scale factor $R=1$. Then using Eq.\ (\ref{eq:fffggg}) and the divergence theorem we can show that
\begin{equation}
\label{Frica}
F = \int d{\bf r}\,\rho \frac{\p\Phi}{\p z}
\end{equation}
with $\Phi = - GM/r$. Not surprisingly, $-F\hat{\bf z}$ is the 
integral of the gravitational force of the object on the surrounding
fluid. 

We take the surface to be that of a sphere of radius $r$
centered on the test object. We write the frictional force as
\begin{equation}
F=\frac{4\pi G^2m_{\rm cl}^2\,\rho}{v^2}\,C(\beta,kr).
\label{eq:cdef}
\end{equation}
Then using the solution (\ref{MM}) and
Eqs. (\ref{Fric}) and (\ref{Frica}) respectively, we obtain the two equivalent
expressions 
\begin{align}
C&=\frac{e^{\pi\beta}|\Gamma(1-i\beta)|^2}{2\beta}\int_0^{2kr}dq\,\bigg[
\Big(\frac{kr}{\beta}+1\Big)\Big(\frac{q}{kr}-1\Big)|M(i\beta,1,iq)|^2
 \nonumber\\
&\qquad 
+q\,\mbox{Im\,}[M(i\beta,1,iq)^\ast M(i\beta+1,2,iq)] +\beta
q|M(i\beta+1,2,iq)|^2\bigg]
\label{eq:cdef1}\\
&=\frac{e^{\pi\beta}|\Gamma(1-i\beta)|^2}{2\beta}\int_0^{2kr}dq\,|M(i\beta,1,iq)|^2\Big(\frac{q}{kr}-2-\log\frac{q}{2kr}\Big).
\label{eq:cdef2}
\end{align}
The Coulomb scattering problem has infrared divergence ($C\to\infty$
as $r\to\infty$) so we keep $r$ finite; we may 
think of $r$ as representing either the size of the test object's
orbit or the size of its host system, whichever is smaller. 

The analogous classical expression\footnote{In this analog we assume
  that an object of mass $m_{\rm cl}$ travels at speed $v$ through a background of much
  less massive objects, all at rest, having total density $\rho$. Of
  course this analog is unrealistic since the background system would
  be gravitationally unstable on all scales.} can be obtained by determining the momentum
flow through a sphere of radius $r$ from hyperbolic Keplerian 
orbits that are parallel to the $z$-axis as $z\to-\infty$:
\begin{equation}
C_{\rm
  cl}=\frac{1+\Lambda}{\Lambda}\tanh^{-1}\frac{\sqrt{\Lambda^2+2\Lambda}}{1+\Lambda}
  -\sqrt{1+2/\Lambda}
\label{eq:couloga}
\end{equation}
where
\begin{equation} 
\Lambda\equiv\frac{kr}{\beta}=\frac{v^2r}{Gm_{\rm cl}};
\label{eq:coulogb}
\end{equation}
this factor is usually called the ``Coulomb logarithm''. It is straightforward to verify numerically that Eqs.\ (\ref{eq:cdef1}) and (\ref{eq:cdef2}) agree with Eq.\ (\ref{eq:couloga}) in the limit in which $kr\to\infty$ and $\beta\to\infty$ while $\Lambda=kr/\beta$ is constant. 

If the test object is much less massive than the host system in which
it resides ($m_{\rm cl} \ll \mm$) then the classical virial
theorem\footnote{The classical virial theorem applies because the
  test object behaves classically.} implies that $v^2\sim
G\mm/r$ so $\Lambda \sim \mm/m_{\rm cl} \gg 1$. Then 
\begin{equation}
C_{\rm  cl}=\log 2\Lambda -1 +\frac{1}{\Lambda}\log 2\Lambda
+\mbox{O}(\Lambda^{-2}).
\label{eq:cch}
\end{equation}
This formulation differs slightly from the usual treatment due to Chandrasekhar \cite{chandra43,BT}, who imposed an
infrared cutoff by placing an upper bound on the impact parameter
rather than the distance from the test object. Thus we should not
expect our result to agree exactly with Chandrasekhar's even in the
classical limit $kr, \beta \to\infty$ (although both give $C_{\rm
  Ch}=\log\Lambda + \mbox{O}(1)$ as $\Lambda\to\infty$). An exact
comparison would require solving a wavepacket scattering problem. 

If the host system is composed of FDM, the frictional force depends on
the dimensionless number $kr=mvr/\hbar$. It is useful to re-write $kr$
in terms of the half-mass radius $r_{1/2}$ and the virial
velocity $v_{\rm vir}$  of the halo. For the ground state, we have from
Eq.\ (\ref{eq:ground})
\begin{equation}
kr=f_0 w_0^{1/2} \frac{v}{v_{\rm vir}}\frac{r}{r_{1/2}}=1.29
\frac{v}{v_{\rm vir}}\frac{r}{r_{1/2}}.
\label{eq:krdef}
\end{equation}
Thus $kr$ is of order unity; then Eq.\ (\ref{eq:betadef}) implies that
$\beta\sim m_{\rm cl}/\mm$ so $\beta\ll1$ when the test object is
much less massive than the host system. A Taylor series expansion of
the confluent hypergeometric function gives
\begin{equation}
M(i\beta,1,iq)= 1 -\beta \,\mbox{Si\,}(q) -i\beta\mbox{\,Cin\,}(q)
+\mbox{O}(\beta^2),
\end{equation}
where $\mbox{Si\,}(z)=\int_0^z \sin t\, dt/t$ and  and
$\mbox{Cin\,}(z)=\int_0^z(1-\cos t)\,dt/t$ are sine and cosine integrals. Then the integral
(\ref{eq:cdef2}) is straightforward:
\begin{equation}
C=\mbox{Cin\,}(2kr)+\frac{\sin 2kr}{2kr} -1 +\mbox{O}(\beta).
\label{eq:casymp}
\end{equation}
For $kr\ll1$, $C\to\frac{1}{3}(kr)^2$. For $kr\gg1$, the behavior of $C$ can be derived from Eq. (\ref{eq:cdef1}), using
results from \cite{BFK2004,Karule1990}:
\begin{equation}
C = {\rm ln\,} (2kr) - 1 - {\,\rm Re\,}\Psi(1 + i\beta) + \mbox{O}(1/kr) \, ,
\end{equation}
where $\Psi$ is the digamma function, with the following asymptotics: 
${\,\rm Re\,}\Psi (1+i\beta) \rightarrow {\,\rm ln\,}\beta$ for large $\beta$,
and $-\gamma_E$ for small $\beta$
($\gamma_E$ is the Euler--Mascheroni constant). This expression has an accuracy of better than $10 \%$ down to $kr = 1$, if $\Lambda > 100$.

In deriving these results we have relied on the approximation that the self-gravity of the dark matter can be ignored. This approximation is also used in the classical calculations of dynamical friction, where it is justified because the strongest contribution to the friction comes from scales much smaller than the system size\footnote{To see this, note that for $\Lambda\gg1$ the frictional force varies as $\log 2\Lambda$ according to Eq.\ (\ref{eq:cch}). Since $\Lambda \sim r$ according to Eq.\ (\ref{eq:coulogb}) each octave in $r$ contributes equally to the friction. A classical calculation that includes self-gravity is given by \cite{wein89}.}. The approximation is less well-justified for FDM since the strongest contribution comes from scales comparable to the size of the solitonic core---according to Eq.\ (\ref{eq:casymp}) the force is proportional to $C$ which is $\sim (kr)^2$ for $kr\ll1$. 

\begin{figure}[htb]
\begin{center}
\includegraphics[trim=0mm 0mm 0mm 0mm, clip, scale=.5, angle=0]{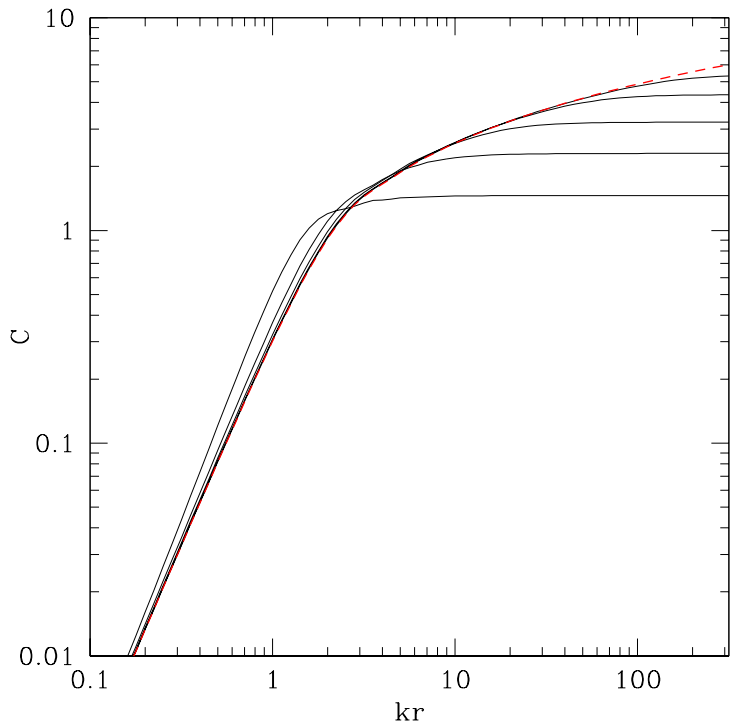}
\caption
{The dynamical friction force on a test object of mass $m_{\rm cl}$ traveling at
  speed $v$ through FDM with density $\rho$ is given
  by Eq.\ (\ref{eq:cdef}), with the dimensionless function $C$
  plotted in this figure. The horizontal axis is $kr$ where $k=mv/\hbar$ is the
  wavenumber of the FDM particles in the rest frame of the test
  object, and $r$ is an upper cutoff to the distance from the test
  object that approximately represents the smaller of the radius of
  the orbit and the size of the host
  system. Each curve is for a fixed value of $\Lambda=v^2r/Gm_{\rm cl}$, which by the virial
  theorem is approximately the ratio of the mass of the host to the
  mass of the test object; from bottom to top
  $\Lambda=3,10,30,100,300$. The dashed red line shows the
  asymptotic behavior in the limit $\Lambda\to\infty$ (Eq.\
  \ref{eq:casymp}). 
}
\label{Ffriction}
\end{center}
\end{figure}

Figure \ref{Ffriction} shows $C(kr/\Lambda,kr)$ (Eq.\
\ref{eq:cdef}) as a function of the dimensionless wavenumber $kr$, for
$\Lambda=3,10,30,100,300$. Recall that the Coulomb logarithm $\Lambda$
is given by Eq.\ (\ref{eq:coulogb}) and is of order the
ratio $\mm/m_{\rm cl}$ of the mass of the host galaxy to the mass of the test object. The solid lines are computed from Eq.\ (\ref{eq:cdef2})
and the dashed red line shows the asymptotic behavior as
$\Lambda\to\infty$ (Eq.\ \ref{eq:casymp}). 
Figure \ref{Ffriction} shows that when $kr\sim 1$, the factor $C$ is
also near unity for a wide range of $\Lambda$. If on the other hand
the test object orbits near the center of the host system then
$kr\ll1$ and $C\to \frac{1}{3}(kr)^2$. 

To estimate the timescale for orbital decay, we assume that the test object is on a circular orbit
of radius $r$. Its velocity is given by $v^2=G\mm(r)/r$ where $\mm(r)$
is the mass of the host system interior to $r$, and its angular momentum
is $L=m_{\rm cl} rv$. Dynamical friction exerts a torque $|F|r$ in the direction
opposite to the orbital motion, where $F$ is given
by Eq.\ (\ref{eq:cdef}), so the decay timescale is
\begin{equation}
\tau\equiv \frac{L}{r|F|}=\frac{\mm(r)^{3/2}}{\vphantom{\big|}4\pi G^{1/2}m_{\rm cl}\rho r^{3/2}C}.
\label{eq:tdf}
\end{equation}
At orbital radii $r\ll r_{1/2}$, for a solitonic host galaxy, we can further replace $C$ with $\frac{1}{3}(kr)^2=\frac{1}{3}(mvr/\hbar)^2=\frac{1}{3}(m/\hbar)^2G\mm(r)r$, the density $\rho$ with its central value $\rho_c$ from Eq.\ (\ref{eq:ground}), and $\mm(r)$ with $\frac{4}{3}\pi\rho_cr^3$. We obtain 
\begin{equation}
\tau=\frac{3^{1/2}}{2\pi^{1/2}\rho_0^{1/2}w_0^{1/2}f_0^3}\frac{v_{\rm
    vir}r_{1/2}^3}{\vphantom{\big|}Gm_{\rm cl}r}=0.370 \,\frac{v_{\rm vir}r_{1/2}^3}{\vphantom{\big|}Gm_{\rm cl}r}.
\label{eq:tdff}
\end{equation}

\section{The collision of streams}
\label{app:collisions}

When two streams of CDM collide, they pass through
each other. What should we expect for FDM? On the one hand, it can
be thought of as a fluid, thus multiple-streaming cannot occur, i.e.,
the density and velocity fields are single-valued.
On the other hand, FDM is described by the Schr\"odinger equation,
where the linear superposition of wavefunctions is permitted (for
simplicity, we ignore gravity in this Appendix). 
To illuminate this issue, it is instructive to study two simple analytic solutions.

The free Schr\"odinger equation in one spatial dimension
(ignoring gravity and cosmic expansion)
$i \hbar\, \partial_t \psi = - \hbar^2{\partial_x^2 \psi / (2m)}$
admits a Gaussian superposition of plane waves as a solution:
\begin{eqnarray}
\psi (t, x) \propto \int dk \, {\rm exp\,} \left[-\ffrac{1}{2} x_0^2
  k^2\right] \, {\rm exp\,} \left[{i \left( kx - \frac{\hbar^2k^2}{2m}t \right)}\right] \, , 
\end{eqnarray}
where the real number $x_0$ characterizes the width of the
Gaussian. The corresponding density and velocity fields are derived from Eqs.\ (\ref{eq:madelung}):
\begin{eqnarray}
\label{FDMevol}
&& \rho(t, x) = \frac{M_{\rm tot}}{\pi^{1/2} x_0\left( 1 + \tilde t \, {}^2
\right)^{1/2}} \exp
\left(- \frac{\tilde x^2}{1 + \tilde t \, {}^2}
\right)  \, ,
\nonumber \\
&& v(t, x) = \frac{\hbar\,\tilde t \, \tilde x}{ \vphantom{\Big|}m x_0 \left(1 + \tilde t \, {}^2\right)} \, ,
\end{eqnarray}
where $M_{\rm tot}$ is the integrated mass, and
$\tilde t$ and $\tilde x$ are the dimensionless time and distance:
\begin{eqnarray}
\label{txrescaled}
\tilde t \equiv \frac{\hbar t }{m x_0^2 } \quad , \quad
\tilde x \equiv \frac{x}{x_0} \, .
\end{eqnarray}
This solution describes a flow converging upon the origin at $t < 0$,
and diverging from it at $t > 0$.
The density profile is narrowest at $t=0$,
with a width of $x_0$. 
It is instructive to consider what CDM would predict
if one were to initiate $\rho$ and $v$ with values
that match the above at some early time $t_{\rm i}$ ($< 0$).
In other words, a CDM particle with initial position
$q$ has the trajectory:
\begin{eqnarray}
x(t) = q + v(t_{\rm i}, q) (t - t_{\rm i}) \, .
\end{eqnarray}
The velocity is unchanged as the particle free streams
and the density follows by mass conservation:
\begin{eqnarray}
\label{CDMevol}
&& \rho(t, x) =\frac{M_{\rm tot} }{\pi^{1/2} x_0}
\frac{ (1 + \tilde t_i \, {}^2)^{1/2}}{\vphantom{\Big|}| 1 + \tilde t  \, \tilde t_i | }
\exp\left[ -  \frac{ (1 + \tilde t_i \, {}^2)}{\vphantom{\Big|}(1 + \tilde t  \, \tilde
    t_i)^2 }  \tilde x \, {}^2 \right] \, ,
\nonumber \\
&& v(t, x) = \frac{\hbar\, \tilde t_i \, \tilde x}{\vphantom{\Big|} m x_0 \left( 1 + \tilde t \, \tilde t_i \right)} \, ,
\end{eqnarray}
where $\tilde t_i \equiv \hbar t_i / (m x_0^2)$. 
Assuming $|\tilde t_i| \gg 1$, one can see that
the FDM and CDM predictions approximately match for $|\tilde t | \gsim 1$.
For $|\tilde t |\lesssim 1$, however, they can be quite
different---especially at $\tilde t = - 1/ \tilde t_i$
when the CDM density (\ref{CDMevol}) blows up at the origin, due to the formation
of a caustic. In contrast, the FDM density (\ref{FDMevol}) remains regular at all times.

These differences in behavior between FDM and CDM can be interpreted in two ways.
As a fluid, FDM is affected by quantum stress or
pressure. The spatial gradient in the density produces pressure that opposes the
formation of a density peak that is too narrow;
the pressure causes a bounce at the point of collision which
eventually erases the peak.
Alternatively, one can interpret FDM as a collection of
particles, albeit with a long de Broglie wavelength. 
Wavepackets describing these particles do superimpose.
The particles stream past each other in both FDM and CDM, but 
the fuzziness associated with the 
macroscopic de Broglie wavelength smooths out
the caustic in FDM. 
Both interpretations are valid.

Suppose one observes a collision of two streams at a relative velocity
$v_0$. Based on the above simple example, we expect fuzziness effects
to be noticeable for
$|x|\lsim x_0 \, , \, |t|\lsim t_0$ 
where
\begin{eqnarray}
\label{bulletc}
x_0 = \frac{\hbar}{mv_0}=192\pc\frac{10^{-22}\eV}{m} \frac{100\kms}{v_0}
\quad , \quad
t_0 = \frac{\hbar}{mv_0^2}=1.87 \times 10^6\yr
\frac{10^{-22} \eV}{m}\left(\frac{100\kms}{v_0} \right)^2 \, .
\end{eqnarray}
Thus, for clusters of galaxies, where $v_0\sim 10^3\kms$, we do not expect
fuzziness to play an important role.
Ultimately, gravity should be included in the analysis. 
When bound objects could form under a collision, 
the difference between FDM and CDM would no longer be transitory. 

\begin{figure}[htb]
\begin{center}
\includegraphics[trim=0mm 0mm 0mm 0mm, clip, scale=.5, angle=0]{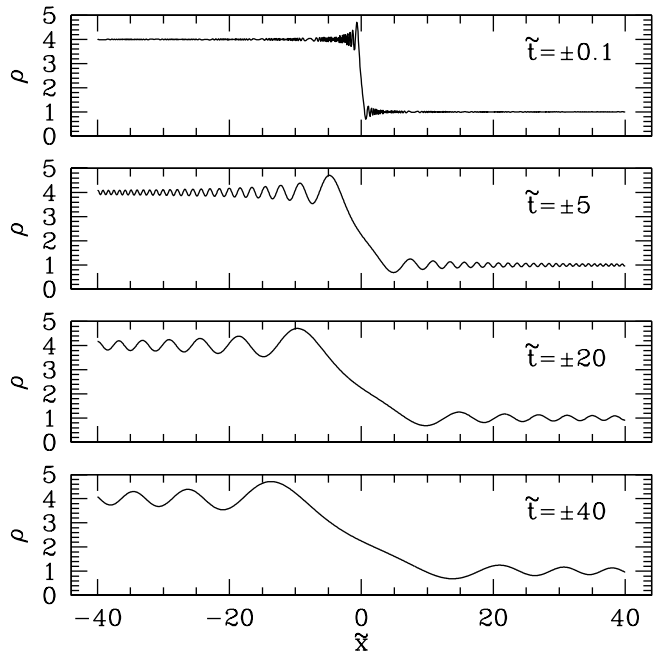}
\caption
{Density evolution in a self-similar FDM solution, according to Eq.\  (\ref{psistep}). 
The dimensionless time $\tilde t$ and distance $\tilde x$ are
defined in Eq. (\ref{txrescaled}), where $x_0$ is an arbitrary
length scale. The overall normalization of
the density $\rho$ is also arbitrary. With the choice
${\cal A} = {\cal B}$ made here, the ratio of the asymptotic density
at $x \rightarrow \pm \infty$ is $({\cal A} + {\cal B})^2/{\cal B}^2
= 4$. 
}
\label{stepshock}
\end{center}
\end{figure}

For additional insight into the nature of collisions, it is useful to
consider self-similar solutions. 
The free Schr\"odinger equation has the symmetry
$x \rightarrow \lambda x$, $t \rightarrow \lambda^2 t$. 
It is thus natural to look for a solution which is a function of
$x/\sqrt{t}$ alone:
\begin{eqnarray}
\psi(t, x) = \frac{{\cal A}}{\sqrt{2\pi}} e^{-i\pi/4}
\int_{x \sqrt{m/\hbar t}}^\infty e^{i \xi^2 /2} d\xi + {\cal B} \, ,
\end{eqnarray}
where ${\cal A}$ and ${\cal B}$ are constants.
(For $t < 0$, we choose $x \sqrt{m/\hbar t} = i x \sqrt{m/\hbar |t|}$.)
The solution has simple spatial asymptotics:
\begin{eqnarray}
\psi(t, x \rightarrow -\infty) = {\cal A} + {\cal B} \quad , \quad
\psi(t, x \rightarrow \infty) = {\cal B} \, .
\end{eqnarray}
We choose ${\cal A}$ and ${\cal B}$ to be real.
The asymptotic behavior of the density $\rho=|\psi|^2$ is 
$\rho \rightarrow ({\cal A} + {\cal B})^2$ as $x \rightarrow - \infty$
and $\rho \rightarrow {\cal B}^2$ as $x \rightarrow \infty$.
There is thus a jump in density. Moreover, the jump becomes increasingly
sharp as $t \rightarrow 0$. This resembles what one expects for
shock formation in a normal fluid.
Let us write out $\psi$ more explicitly as:
\begin{eqnarray}
\label{psistep}
\psi(t, x) = \frac{{\cal A}}{2} + {\cal B} - 
\frac{{\cal A}}{2} (1 \mp i) {\cal C}\left( x \sqrt{\frac{m}{\pi\hbar |t|}}
\right)
- \frac{{\cal A}}{2} (1 \pm i) {\cal S}\left( x \sqrt{\frac{m}{\pi \hbar|t|}}
\right) \, ,
\end{eqnarray}
where the upper/lower sign is for a positive/negative $t$.
The functions ${\cal C}$ and ${\cal S}$ are the Fresnel integrals, defined by
\begin{eqnarray}
{\cal C}(\theta) = \int_0^\theta {\,\rm cos\,} (\half \pi \xi^2) d\xi
\quad , \quad
{\cal S}(\theta) = \int_0^\theta {\,\rm sin\,} (\half \pi \xi^2) d\xi \, .
\end{eqnarray} 
The implied density $\rho = |\psi |^2$  is depicted in Figure  \ref{stepshock}. 

The resemblance to a shock (in its rest frame) is superficial.
The evolution in the case of FDM is time-reversal symmetric, whereas there is an irreversible entropy production in 
a normal shock.
From a fluid perspective, the quantum pressure of FDM
reacts against the steepening density profile around $x = 0$,
ultimately causing a bounce that reverses the evolution.
From a particle perspective, the two colliding streams
go past each other after $t=0$---much as CDM would---thereby smoothing out the jump in density around $x=0$. 
The difference from CDM is the absence of caustics where
density formally diverges.
The FDM density profile has characteristic oscillations
on the scale $x \sim \sqrt{\hbar|t|/m}$. 

\bibliography{references}

\begin{thebibliography}{195}%
\makeatletter
\providecommand \@ifxundefined [1]{%
 \@ifx{#1\undefined}
}%
\providecommand \@ifnum [1]{%
 \ifnum #1\expandafter \@firstoftwo
 \else \expandafter \@secondoftwo
 \fi
}%
\providecommand \@ifx [1]{%
 \ifx #1\expandafter \@firstoftwo
 \else \expandafter \@secondoftwo
 \fi
}%
\providecommand \natexlab [1]{#1}%
\providecommand \enquote  [1]{``#1''}%
\providecommand \bibnamefont  [1]{#1}%
\providecommand \bibfnamefont [1]{#1}%
\providecommand \citenamefont [1]{#1}%
\providecommand \href@noop [0]{\@secondoftwo}%
\providecommand \href [0]{\begingroup \@sanitize@url \@href}%
\providecommand \@href[1]{\@@startlink{#1}\@@href}%
\providecommand \@@href[1]{\endgroup#1\@@endlink}%
\providecommand \@sanitize@url [0]{\catcode `\\12\catcode `\$12\catcode
  `\&12\catcode `\#12\catcode `\^12\catcode `\_12\catcode `\%12\relax}%
\providecommand \@@startlink[1]{}%
\providecommand \@@endlink[0]{}%
\providecommand \url  [0]{\begingroup\@sanitize@url \@url }%
\providecommand \@url [1]{\endgroup\@href {#1}{\urlprefix }}%
\providecommand \urlprefix  [0]{URL }%
\providecommand \Eprint [0]{\href }%
\@ifxundefined \urlstyle {%
  \providecommand \doi  [0]{\begingroup \@sanitize@url \@doi}%
  \providecommand \@doi [1]{\endgroup \@@startlink {\doibase
  #1}doi:\discretionary {}{}{}#1\@@endlink }%
}{%
  \providecommand \doi  [0]{doi:\discretionary{}{}{}\begingroup
  \urlstyle{rm}\Url }%
}%
\providecommand \doibase [0]{http://dx.doi.org/}%
\providecommand \Doi [0]{\begingroup \@sanitize@url \@Doi }%
\providecommand \@Doi  [1]{\endgroup\@@startlink{\doibase#1}\@@Doi}%
\providecommand \@@Doi [1]{#1\@@endlink}%
\providecommand \selectlanguage [0]{\@gobble}%
\providecommand \bibinfo  [0]{\@secondoftwo}%
\providecommand \bibfield  [0]{\@secondoftwo}%
\providecommand \translation [1]{[#1]}%
\providecommand \BibitemOpen [0]{}%
\providecommand \bibitemStop [0]{}%
\providecommand \bibitemNoStop [0]{.\EOS\space}%
\providecommand \EOS [0]{\spacefactor3000\relax}%
\providecommand \BibitemShut  [1]{\csname bibitem#1\endcsname}%
\bibitem [{\citenamefont {Ade}\ \emph {et~al.}(2015)\citenamefont {Ade} \emph
  {et~al.}}]{planck15}%
  \BibitemOpen
  \bibfield  {author} {\bibinfo {author} {\bibfnamefont {P.A.R.}\ \bibnamefont
  {Ade}} \emph {et~al.} (\bibinfo {collaboration} {Planck}),\ }\bibfield
  {title} {\enquote {\bibinfo {title} {{Planck 2015 results. XIII. Cosmological
  parameters}},}\ }\href@noop {} { (\bibinfo {year} {2015})},\ \Eprint
  {http://arxiv.org/abs/1502.01589} {arXiv:1502.01589} \BibitemShut {NoStop}%
\bibitem [{\citenamefont {Dubinski}\ and\ \citenamefont
  {Carlberg}(1991)}]{dc91}%
  \BibitemOpen
  \bibfield  {author} {\bibinfo {author} {\bibfnamefont {J.}~\bibnamefont
  {Dubinski}}\ and\ \bibinfo {author} {\bibfnamefont {R.~G.}\ \bibnamefont
  {Carlberg}},\ }\bibfield  {title} {\enquote {\bibinfo {title} {{The structure
  of cold dark matter halos}},}\ }\Doi {10.1086/170451} {\bibfield  {journal}
  {\bibinfo  {journal} {Astrophys. J.},\ }\textbf {\bibinfo {volume} {378}},\
  \bibinfo {pages} {496--503} (\bibinfo {year} {1991})}\BibitemShut {NoStop}%
\bibitem [{\citenamefont {Navarro}\ \emph {et~al.}(1997)\citenamefont
  {Navarro}, \citenamefont {Frenk},\ and\ \citenamefont {White}}]{nfw97}%
  \BibitemOpen
  \bibfield  {author} {\bibinfo {author} {\bibfnamefont {J.~F.}\ \bibnamefont
  {Navarro}}, \bibinfo {author} {\bibfnamefont {C.~S.}\ \bibnamefont {Frenk}},
  \ and\ \bibinfo {author} {\bibfnamefont {S.D.M.}\ \bibnamefont {White}},\
  }\bibfield  {title} {\enquote {\bibinfo {title} {{A universal density profile
  from hierarchical clustering}},}\ }\Doi {10.1086/304888} {\bibfield
  {journal} {\bibinfo  {journal} {Astrophys. J.},\ }\textbf {\bibinfo {volume}
  {490}},\ \bibinfo {pages} {493--508} (\bibinfo {year} {1997})},\ \Eprint
  {http://arxiv.org/abs/astro-ph/9611107} {arXiv:astro-ph/9611107} \BibitemShut
  {NoStop}%
\bibitem [{\citenamefont {{Weinberg}}\ \emph {et~al.}(2015)\citenamefont
  {{Weinberg}}, \citenamefont {{Bullock}}, \citenamefont {{Governato}},
  \citenamefont {{Kuzio de Naray}},\ and\ \citenamefont {{Peter}}}]{wein13}%
  \BibitemOpen
  \bibfield  {author} {\bibinfo {author} {\bibfnamefont {D.~H.}\ \bibnamefont
  {{Weinberg}}}, \bibinfo {author} {\bibfnamefont {J.~S.}\ \bibnamefont
  {{Bullock}}}, \bibinfo {author} {\bibfnamefont {F.}~\bibnamefont
  {{Governato}}}, \bibinfo {author} {\bibfnamefont {R.}~\bibnamefont {{Kuzio de
  Naray}}}, \ and\ \bibinfo {author} {\bibfnamefont {A.H.G.}\ \bibnamefont
  {{Peter}}},\ }\bibfield  {title} {\enquote {\bibinfo {title} {{Cold dark
  matter: controversies on small scales}},}\ }\Doi {10.1073/pnas.1308716112}
  {\bibfield  {journal} {\bibinfo  {journal} {Proceedings of the National
  Academy of Science},\ }\textbf {\bibinfo {volume} {112}},\ \bibinfo {pages}
  {12249--12255} (\bibinfo {year} {2015})},\ \Eprint
  {http://arxiv.org/abs/1306.0913} {arXiv:1306.0913} \BibitemShut {NoStop}%
\bibitem [{\citenamefont {{Viel}}\ \emph {et~al.}(2005)\citenamefont {{Viel}},
  \citenamefont {{Lesgourgues}}, \citenamefont {{Haehnelt}}, \citenamefont
  {{Matarrese}},\ and\ \citenamefont {{Riotto}}}]{Viel2005}%
  \BibitemOpen
  \bibfield  {author} {\bibinfo {author} {\bibfnamefont {M.}~\bibnamefont
  {{Viel}}}, \bibinfo {author} {\bibfnamefont {J.}~\bibnamefont
  {{Lesgourgues}}}, \bibinfo {author} {\bibfnamefont {M.~G.}\ \bibnamefont
  {{Haehnelt}}}, \bibinfo {author} {\bibfnamefont {S.}~\bibnamefont
  {{Matarrese}}}, \ and\ \bibinfo {author} {\bibfnamefont {A.}~\bibnamefont
  {{Riotto}}},\ }\bibfield  {title} {\enquote {\bibinfo {title} {{Constraining
  warm dark matter candidates including sterile neutrinos and light gravitinos
  with WMAP and the Lyman-{$\alpha$} forest}},}\ }\Doi
  {10.1103/PhysRevD.71.063534} {\bibfield  {journal} {\bibinfo  {journal}
  {\prd},\ }\textbf {\bibinfo {volume} {71}},\ \bibinfo {eid} {063534}
  (\bibinfo {year} {2005})},\ \Eprint
  {http://arxiv.org/abs/arXiv:astro-ph/0501562} {arXiv:astro-ph/0501562}
  \BibitemShut {NoStop}%
\bibitem [{\citenamefont {{Col{\'{\i}}n}}\ \emph {et~al.}(2000)\citenamefont
  {{Col{\'{\i}}n}}, \citenamefont {{Avila-Reese}},\ and\ \citenamefont
  {{Valenzuela}}}]{Colin2000}%
  \BibitemOpen
  \bibfield  {author} {\bibinfo {author} {\bibfnamefont {P.}~\bibnamefont
  {{Col{\'{\i}}n}}}, \bibinfo {author} {\bibfnamefont {V.}~\bibnamefont
  {{Avila-Reese}}}, \ and\ \bibinfo {author} {\bibfnamefont {O.}~\bibnamefont
  {{Valenzuela}}},\ }\bibfield  {title} {\enquote {\bibinfo {title}
  {{Substructure and halo density profiles in a warm dark matter cosmology}},}\
  }\Doi {10.1086/317057} {\bibfield  {journal} {\bibinfo  {journal} {\apj},\
  }\textbf {\bibinfo {volume} {542}},\ \bibinfo {pages} {622--630} (\bibinfo
  {year} {2000})},\ \Eprint {http://arxiv.org/abs/arXiv:astro-ph/0004115}
  {arXiv:astro-ph/0004115} \BibitemShut {NoStop}%
\bibitem [{\citenamefont {{Bode}}\ \emph {et~al.}(2001)\citenamefont {{Bode}},
  \citenamefont {{Ostriker}},\ and\ \citenamefont {{Turok}}}]{Bode2001}%
  \BibitemOpen
  \bibfield  {author} {\bibinfo {author} {\bibfnamefont {P.}~\bibnamefont
  {{Bode}}}, \bibinfo {author} {\bibfnamefont {J.~P.}\ \bibnamefont
  {{Ostriker}}}, \ and\ \bibinfo {author} {\bibfnamefont {N.}~\bibnamefont
  {{Turok}}},\ }\bibfield  {title} {\enquote {\bibinfo {title} {{Halo formation
  in warm dark matter models}},}\ }\Doi {10.1086/321541} {\bibfield  {journal}
  {\bibinfo  {journal} {\apj},\ }\textbf {\bibinfo {volume} {556}},\ \bibinfo
  {pages} {93--107} (\bibinfo {year} {2001})},\ \Eprint
  {http://arxiv.org/abs/arXiv:astro-ph/0010389} {arXiv:astro-ph/0010389}
  \BibitemShut {NoStop}%
\bibitem [{\citenamefont {Tremaine}\ and\ \citenamefont {Gunn}(1979)}]{tg79}%
  \BibitemOpen
  \bibfield  {author} {\bibinfo {author} {\bibfnamefont {S.}~\bibnamefont
  {Tremaine}}\ and\ \bibinfo {author} {\bibfnamefont {J.~E.}\ \bibnamefont
  {Gunn}},\ }\bibfield  {title} {\enquote {\bibinfo {title} {{Dynamical role of
  light neutral leptons in cosmology}},}\ }\Doi {10.1103/PhysRevLett.42.407}
  {\bibfield  {journal} {\bibinfo  {journal} {Phys. Rev. Lett.},\ }\textbf
  {\bibinfo {volume} {42}},\ \bibinfo {pages} {407--410} (\bibinfo {year}
  {1979})}\BibitemShut {NoStop}%
\bibitem [{\citenamefont {Turner}(1983)}]{turner83}%
  \BibitemOpen
  \bibfield  {author} {\bibinfo {author} {\bibfnamefont {M.~S.}\ \bibnamefont
  {Turner}},\ }\bibfield  {title} {\enquote {\bibinfo {title} {{Coherent scalar
  field oscillations in an expanding universe}},}\ }\Doi
  {10.1103/PhysRevD.28.1243} {\bibfield  {journal} {\bibinfo  {journal} {Phys.
  Rev.},\ }\textbf {\bibinfo {volume} {D28}},\ \bibinfo {pages} {1243--1247}
  (\bibinfo {year} {1983})}\BibitemShut {NoStop}%
\bibitem [{\citenamefont {{Press}}\ \emph {et~al.}(1990)\citenamefont
  {{Press}}, \citenamefont {{Ryden}},\ and\ \citenamefont
  {{Spergel}}}]{press1990}%
  \BibitemOpen
  \bibfield  {author} {\bibinfo {author} {\bibfnamefont {W.~H.}\ \bibnamefont
  {{Press}}}, \bibinfo {author} {\bibfnamefont {B.~S.}\ \bibnamefont
  {{Ryden}}}, \ and\ \bibinfo {author} {\bibfnamefont {D.~N.}\ \bibnamefont
  {{Spergel}}},\ }\bibfield  {title} {\enquote {\bibinfo {title} {{Single
  mechanism for generating large-scale structure and providing dark missing
  matter}},}\ }\Doi {10.1103/PhysRevLett.64.1084} {\bibfield  {journal}
  {\bibinfo  {journal} {Physical Review Letters},\ }\textbf {\bibinfo {volume}
  {64}},\ \bibinfo {pages} {1084--1087} (\bibinfo {year} {1990})}\BibitemShut
  {NoStop}%
\bibitem [{\citenamefont {Sin}(1994)}]{sin94}%
  \BibitemOpen
  \bibfield  {author} {\bibinfo {author} {\bibfnamefont {S.-J.}\ \bibnamefont
  {Sin}},\ }\bibfield  {title} {\enquote {\bibinfo {title} {{Late time
  cosmological phase transition and galactic halo as Bose liquid}},}\ }\Doi
  {10.1103/PhysRevD.50.3650} {\bibfield  {journal} {\bibinfo  {journal} {Phys.
  Rev.},\ }\textbf {\bibinfo {volume} {D50}},\ \bibinfo {pages} {3650--3654}
  (\bibinfo {year} {1994})},\ \Eprint {http://arxiv.org/abs/hep-ph/9205208}
  {arXiv:hep-ph/9205208} \BibitemShut {NoStop}%
\bibitem [{\citenamefont {Hu}\ \emph {et~al.}(2000)\citenamefont {Hu},
  \citenamefont {Barkana},\ and\ \citenamefont {Gruzinov}}]{FDM}%
  \BibitemOpen
  \bibfield  {author} {\bibinfo {author} {\bibfnamefont {W.}~\bibnamefont
  {Hu}}, \bibinfo {author} {\bibfnamefont {R.}~\bibnamefont {Barkana}}, \ and\
  \bibinfo {author} {\bibfnamefont {A.}~\bibnamefont {Gruzinov}},\ }\bibfield
  {title} {\enquote {\bibinfo {title} {{Cold and fuzzy dark matter}},}\ }\Doi
  {10.1103/PhysRevLett.85.1158} {\bibfield  {journal} {\bibinfo  {journal}
  {Phys. Rev. Lett.},\ }\textbf {\bibinfo {volume} {85}},\ \bibinfo {pages}
  {1158--1161} (\bibinfo {year} {2000})},\ \Eprint
  {http://arxiv.org/abs/astro-ph/0003365} {arXiv:astro-ph/0003365} \BibitemShut
  {NoStop}%
\bibitem [{\citenamefont {Goodman}(2000)}]{good00}%
  \BibitemOpen
  \bibfield  {author} {\bibinfo {author} {\bibfnamefont {J.}~\bibnamefont
  {Goodman}},\ }\bibfield  {title} {\enquote {\bibinfo {title} {{Repulsive dark
  matter}},}\ }\Doi {10.1016/S1384-1076(00)00015-4} {\bibfield  {journal}
  {\bibinfo  {journal} {New Astron.},\ }\textbf {\bibinfo {volume} {5}},\
  \bibinfo {pages} {103--107} (\bibinfo {year} {2000})},\ \Eprint
  {http://arxiv.org/abs/astro-ph/0003018} {arXiv:astro-ph/0003018} \BibitemShut
  {NoStop}%
\bibitem [{\citenamefont {Peebles}(2000)}]{peebles00}%
  \BibitemOpen
  \bibfield  {author} {\bibinfo {author} {\bibfnamefont {P.J.E.}\ \bibnamefont
  {Peebles}},\ }\bibfield  {title} {\enquote {\bibinfo {title} {{Fluid dark
  matter}},}\ }\Doi {10.1086/312677} {\bibfield  {journal} {\bibinfo  {journal}
  {Astrophys. J.},\ }\textbf {\bibinfo {volume} {534}},\ \bibinfo {pages}
  {L127--L129} (\bibinfo {year} {2000})},\ \Eprint
  {http://arxiv.org/abs/astro-ph/0002495} {arXiv:astro-ph/0002495} \BibitemShut
  {NoStop}%
\bibitem [{\citenamefont {Amendola}\ and\ \citenamefont
  {Barbieri}(2006)}]{Amendola:2005ad}%
  \BibitemOpen
  \bibfield  {author} {\bibinfo {author} {\bibfnamefont {L.}~\bibnamefont
  {Amendola}}\ and\ \bibinfo {author} {\bibfnamefont {R.}~\bibnamefont
  {Barbieri}},\ }\bibfield  {title} {\enquote {\bibinfo {title} {{Dark matter
  from an ultra-light pseudo-Goldstone-boson}},}\ }\Doi
  {10.1016/j.physletb.2006.08.069} {\bibfield  {journal} {\bibinfo  {journal}
  {Phys. Lett.},\ }\textbf {\bibinfo {volume} {B642}},\ \bibinfo {pages}
  {192--196} (\bibinfo {year} {2006})},\ \Eprint
  {http://arxiv.org/abs/hep-ph/0509257} {arXiv:hep-ph/0509257} \BibitemShut
  {NoStop}%
\bibitem [{\citenamefont {Schive}\ \emph
  {et~al.}(2014){\natexlab{a}}\citenamefont {Schive}, \citenamefont {Chiueh},\
  and\ \citenamefont {Broadhurst}}]{scb14}%
  \BibitemOpen
  \bibfield  {author} {\bibinfo {author} {\bibfnamefont {H.-Y.}\ \bibnamefont
  {Schive}}, \bibinfo {author} {\bibfnamefont {T.}~\bibnamefont {Chiueh}}, \
  and\ \bibinfo {author} {\bibfnamefont {T.}~\bibnamefont {Broadhurst}},\
  }\bibfield  {title} {\enquote {\bibinfo {title} {{Cosmic structure as the
  quantum interference of a coherent dark wave}},}\ }\Doi {10.1038/nphys2996}
  {\bibfield  {journal} {\bibinfo  {journal} {Nature Phys.},\ }\textbf
  {\bibinfo {volume} {10}},\ \bibinfo {pages} {496--499} (\bibinfo {year}
  {2014}{\natexlab{a}})},\ \Eprint {http://arxiv.org/abs/1406.6586}
  {arXiv:1406.6586} \BibitemShut {NoStop}%
\bibitem [{\citenamefont {Marsh}(2016)}]{mar15}%
  \BibitemOpen
  \bibfield  {author} {\bibinfo {author} {\bibfnamefont {D.J.E.}\ \bibnamefont
  {Marsh}},\ }\bibfield  {title} {\enquote {\bibinfo {title} {{Axion
  cosmology}},}\ }\Doi {10.1016/j.physrep.2016.06.005} {\bibfield  {journal}
  {\bibinfo  {journal} {Phys. Rept.},\ }\textbf {\bibinfo {volume} {643}},\
  \bibinfo {pages} {1--79} (\bibinfo {year} {2016})},\ \Eprint
  {http://arxiv.org/abs/1510.07633} {arXiv:1510.07633} \BibitemShut {NoStop}%
\bibitem [{\citenamefont {Hlozek}\ \emph {et~al.}(2015)\citenamefont {Hlozek},
  \citenamefont {Grin}, \citenamefont {Marsh},\ and\ \citenamefont
  {Ferreira}}]{hlo15}%
  \BibitemOpen
  \bibfield  {author} {\bibinfo {author} {\bibfnamefont {R.}~\bibnamefont
  {Hlozek}}, \bibinfo {author} {\bibfnamefont {D.}~\bibnamefont {Grin}},
  \bibinfo {author} {\bibfnamefont {D.J.E.}\ \bibnamefont {Marsh}}, \ and\
  \bibinfo {author} {\bibfnamefont {P.~G.}\ \bibnamefont {Ferreira}},\
  }\bibfield  {title} {\enquote {\bibinfo {title} {{A search for ultralight
  axions using precision cosmological data}},}\ }\Doi
  {10.1103/PhysRevD.91.103512} {\bibfield  {journal} {\bibinfo  {journal}
  {Phys. Rev.},\ }\textbf {\bibinfo {volume} {D91}},\ \bibinfo {pages} {103512}
  (\bibinfo {year} {2015})},\ \Eprint {http://arxiv.org/abs/1410.2896}
  {arXiv:1410.2896} \BibitemShut {NoStop}%
\bibitem [{\citenamefont {Viel}\ \emph
  {et~al.}(2013){\natexlab{a}}\citenamefont {Viel}, \citenamefont {Becker},
  \citenamefont {Bolton},\ and\ \citenamefont {Haehnelt}}]{viel13}%
  \BibitemOpen
  \bibfield  {author} {\bibinfo {author} {\bibfnamefont {M.}~\bibnamefont
  {Viel}}, \bibinfo {author} {\bibfnamefont {G.~D.}\ \bibnamefont {Becker}},
  \bibinfo {author} {\bibfnamefont {J.~S.}\ \bibnamefont {Bolton}}, \ and\
  \bibinfo {author} {\bibfnamefont {M.~G.}\ \bibnamefont {Haehnelt}},\
  }\bibfield  {title} {\enquote {\bibinfo {title} {{Warm dark matter as a
  solution to the small scale crisis: new constraints from high redshift
  Lyman-$\alpha$ forest data}},}\ }\Doi {10.1103/PhysRevD.88.043502} {\bibfield
   {journal} {\bibinfo  {journal} {Phys. Rev.},\ }\textbf {\bibinfo {volume}
  {D88}},\ \bibinfo {pages} {043502} (\bibinfo {year} {2013}{\natexlab{a}})},\
  \Eprint {http://arxiv.org/abs/1306.2314} {arXiv:1306.2314} \BibitemShut
  {NoStop}%
\bibitem [{\citenamefont {Sikivie}(1983)}]{Sikivie:1983ip}%
  \BibitemOpen
  \bibfield  {author} {\bibinfo {author} {\bibfnamefont {P.}~\bibnamefont
  {Sikivie}},\ }\bibfield  {title} {\enquote {\bibinfo {title} {{Experimental
  tests of the invisible axion}},}\ }\bibfield  {booktitle} {\emph {\bibinfo
  {booktitle} {{11th International Symposium on Lepton and Photon Interactions
  at High Energies Ithaca, New York, August 4-9, 1983}}},\ }\Doi
  {10.1103/PhysRevLett.51.1415, 10.1103/PhysRevLett.52.695.2} {\bibfield
  {journal} {\bibinfo  {journal} {Phys. Rev. Lett.},\ }\textbf {\bibinfo
  {volume} {51}},\ \bibinfo {pages} {1415--1417} (\bibinfo {year} {1983})},\
  \bibinfo {note} {[Erratum: Phys. Rev. Lett.\ 52, 695 (1984)]}\BibitemShut
  {NoStop}%
\bibitem [{\citenamefont {Hoskins}\ \emph {et~al.}(2011)\citenamefont {Hoskins}
  \emph {et~al.}}]{Hoskins:2011iv}%
  \BibitemOpen
  \bibfield  {author} {\bibinfo {author} {\bibfnamefont {J.}~\bibnamefont
  {Hoskins}} \emph {et~al.},\ }\bibfield  {title} {\enquote {\bibinfo {title}
  {{A search for non-virialized axionic dark matter}},}\ }\Doi
  {10.1103/PhysRevD.84.121302} {\bibfield  {journal} {\bibinfo  {journal}
  {Phys. Rev.},\ }\textbf {\bibinfo {volume} {D84}},\ \bibinfo {pages} {121302}
  (\bibinfo {year} {2011})},\ \Eprint {http://arxiv.org/abs/1109.4128}
  {arXiv:1109.4128} \BibitemShut {NoStop}%
\bibitem [{\citenamefont {Ruz}\ \emph {et~al.}(2015)\citenamefont {Ruz},
  \citenamefont {Vogel},\ and\ \citenamefont {Pivovaroff}}]{Ruz:2015uka}%
  \BibitemOpen
  \bibfield  {author} {\bibinfo {author} {\bibfnamefont {J.}~\bibnamefont
  {Ruz}}, \bibinfo {author} {\bibfnamefont {J.~K.}\ \bibnamefont {Vogel}}, \
  and\ \bibinfo {author} {\bibfnamefont {M.~J.}\ \bibnamefont {Pivovaroff}}
  (\bibinfo {collaboration} {CAST}),\ }\bibfield  {title} {\enquote {\bibinfo
  {title} {{Recent constraints on axion-photon and axion-electron coupling with
  the CAST experiment}},}\ }\bibfield  {booktitle} {\emph {\bibinfo {booktitle}
  {{Proceedings, 13th International Conference on Topics in Astroparticle and
  Underground Physics (TAUP 2013): Asilomar, California, September 8-13,
  2013}}},\ }\Doi {10.1016/j.phpro.2014.12.025} {\bibfield  {journal} {\bibinfo
   {journal} {Phys. Procedia},\ }\textbf {\bibinfo {volume} {61}},\ \bibinfo
  {pages} {153--156} (\bibinfo {year} {2015})}\BibitemShut {NoStop}%
\bibitem [{\citenamefont {Arvanitaki}\ \emph {et~al.}(2010)\citenamefont
  {Arvanitaki}, \citenamefont {Dimopoulos}, \citenamefont {Dubovsky},
  \citenamefont {Kaloper},\ and\ \citenamefont {March-Russell}}]{Axiverse}%
  \BibitemOpen
  \bibfield  {author} {\bibinfo {author} {\bibfnamefont {A.}~\bibnamefont
  {Arvanitaki}}, \bibinfo {author} {\bibfnamefont {S.}~\bibnamefont
  {Dimopoulos}}, \bibinfo {author} {\bibfnamefont {S.}~\bibnamefont
  {Dubovsky}}, \bibinfo {author} {\bibfnamefont {N.}~\bibnamefont {Kaloper}}, \
  and\ \bibinfo {author} {\bibfnamefont {J.}~\bibnamefont {March-Russell}},\
  }\bibfield  {title} {\enquote {\bibinfo {title} {{String axiverse}},}\ }\Doi
  {10.1103/PhysRevD.81.123530} {\bibfield  {journal} {\bibinfo  {journal}
  {Phys. Rev.},\ }\textbf {\bibinfo {volume} {D81}},\ \bibinfo {pages} {123530}
  (\bibinfo {year} {2010})},\ \Eprint {http://arxiv.org/abs/0905.4720}
  {arXiv:0905.4720} \BibitemShut {NoStop}%
\bibitem [{\citenamefont {Choi}\ and\ \citenamefont {Kim}(1985)}]{Choi}%
  \BibitemOpen
  \bibfield  {author} {\bibinfo {author} {\bibfnamefont {K.}~\bibnamefont
  {Choi}}\ and\ \bibinfo {author} {\bibfnamefont {J.~E.}\ \bibnamefont {Kim}},\
  }\bibfield  {title} {\enquote {\bibinfo {title} {{Compactification and axions
  in E(8)$\times$E(8)-prime superstring models}},}\ }\Doi
  {10.1016/0370-2693(85)90693-8} {\bibfield  {journal} {\bibinfo  {journal}
  {Phys. Lett.},\ }\textbf {\bibinfo {volume} {B165}},\ \bibinfo {pages}
  {71--75} (\bibinfo {year} {1985})}\BibitemShut {NoStop}%
\bibitem [{\citenamefont {Banks}\ \emph {et~al.}(2003)\citenamefont {Banks},
  \citenamefont {Dine}, \citenamefont {Fox},\ and\ \citenamefont
  {Gorbatov}}]{BD}%
  \BibitemOpen
  \bibfield  {author} {\bibinfo {author} {\bibfnamefont {T.}~\bibnamefont
  {Banks}}, \bibinfo {author} {\bibfnamefont {M.}~\bibnamefont {Dine}},
  \bibinfo {author} {\bibfnamefont {P.~J.}\ \bibnamefont {Fox}}, \ and\
  \bibinfo {author} {\bibfnamefont {E.}~\bibnamefont {Gorbatov}},\ }\bibfield
  {title} {\enquote {\bibinfo {title} {{On the possibility of large axion decay
  constants}},}\ }\Doi {10.1088/1475-7516/2003/06/001} {\bibfield  {journal}
  {\bibinfo  {journal} {JCAP},\ }\textbf {\bibinfo {volume} {0306}},\ \bibinfo
  {pages} {001} (\bibinfo {year} {2003})},\ \Eprint
  {http://arxiv.org/abs/hep-th/0303252} {arXiv:hep-th/0303252} \BibitemShut
  {NoStop}%
\bibitem [{\citenamefont {Svrcek}\ and\ \citenamefont {Witten}(2006)}]{SW}%
  \BibitemOpen
  \bibfield  {author} {\bibinfo {author} {\bibfnamefont {P.}~\bibnamefont
  {Svrcek}}\ and\ \bibinfo {author} {\bibfnamefont {E.}~\bibnamefont
  {Witten}},\ }\bibfield  {title} {\enquote {\bibinfo {title} {{Axions in
  string theory}},}\ }\Doi {10.1088/1126-6708/2006/06/051} {\bibfield
  {journal} {\bibinfo  {journal} {JHEP},\ }\textbf {\bibinfo {volume} {06}},\
  \bibinfo {pages} {051} (\bibinfo {year} {2006})},\ \Eprint
  {http://arxiv.org/abs/hep-th/0605206} {arXiv:hep-th/0605206} \BibitemShut
  {NoStop}%
\bibitem [{\citenamefont {Arkani-Hamed}\ \emph {et~al.}(2012)\citenamefont
  {Arkani-Hamed}, \citenamefont {Gupta}, \citenamefont {Kaplan}, \citenamefont
  {Weiner},\ and\ \citenamefont {Zorawski}}]{SS}%
  \BibitemOpen
  \bibfield  {author} {\bibinfo {author} {\bibfnamefont {N.}~\bibnamefont
  {Arkani-Hamed}}, \bibinfo {author} {\bibfnamefont {A.}~\bibnamefont {Gupta}},
  \bibinfo {author} {\bibfnamefont {D.~E.}\ \bibnamefont {Kaplan}}, \bibinfo
  {author} {\bibfnamefont {N.}~\bibnamefont {Weiner}}, \ and\ \bibinfo {author}
  {\bibfnamefont {T.}~\bibnamefont {Zorawski}},\ }\bibfield  {title} {\enquote
  {\bibinfo {title} {{Simply unnatural supersymmetry}},}\ }\href@noop {} {
  (\bibinfo {year} {2012})},\ \Eprint {http://arxiv.org/abs/1212.6971}
  {arXiv:1212.6971} \BibitemShut {NoStop}%
\bibitem [{\citenamefont {Svrcek}(2006)}]{S}%
  \BibitemOpen
  \bibfield  {author} {\bibinfo {author} {\bibfnamefont {P.}~\bibnamefont
  {Svrcek}},\ }\bibfield  {title} {\enquote {\bibinfo {title} {{Cosmological
  constant and axions in string theory}},}\ }\href@noop {} {\bibfield
  {journal} {\bibinfo  {journal} {Submitted to: JHEP}} (\bibinfo {year}
  {2006})},\ \Eprint {http://arxiv.org/abs/hep-th/0607086}
  {arXiv:hep-th/0607086} \BibitemShut {NoStop}%
\bibitem [{\citenamefont {Preskill}\ \emph {et~al.}(1983)\citenamefont
  {Preskill}, \citenamefont {Wise},\ and\ \citenamefont {Wilczek}}]{PMW}%
  \BibitemOpen
  \bibfield  {author} {\bibinfo {author} {\bibfnamefont {J.}~\bibnamefont
  {Preskill}}, \bibinfo {author} {\bibfnamefont {M.~B.}\ \bibnamefont {Wise}},
  \ and\ \bibinfo {author} {\bibfnamefont {F.}~\bibnamefont {Wilczek}},\
  }\bibfield  {title} {\enquote {\bibinfo {title} {{Cosmology of the invisible
  axion}},}\ }\Doi {10.1016/0370-2693(83)90637-8} {\bibfield  {journal}
  {\bibinfo  {journal} {Phys. Lett.},\ }\textbf {\bibinfo {volume} {B120}},\
  \bibinfo {pages} {127--132} (\bibinfo {year} {1983})}\BibitemShut {NoStop}%
\bibitem [{\citenamefont {Abbott}\ and\ \citenamefont {Sikivie}(1983)}]{AS}%
  \BibitemOpen
  \bibfield  {author} {\bibinfo {author} {\bibfnamefont {L.~F.}\ \bibnamefont
  {Abbott}}\ and\ \bibinfo {author} {\bibfnamefont {P.}~\bibnamefont
  {Sikivie}},\ }\bibfield  {title} {\enquote {\bibinfo {title} {{A cosmological
  bound on the invisible axion}},}\ }\Doi {10.1016/0370-2693(83)90638-X}
  {\bibfield  {journal} {\bibinfo  {journal} {Phys. Lett.},\ }\textbf {\bibinfo
  {volume} {B120}},\ \bibinfo {pages} {133--136} (\bibinfo {year}
  {1983})}\BibitemShut {NoStop}%
\bibitem [{\citenamefont {Dine}\ and\ \citenamefont {Fischler}(1983)}]{DF}%
  \BibitemOpen
  \bibfield  {author} {\bibinfo {author} {\bibfnamefont {M.}~\bibnamefont
  {Dine}}\ and\ \bibinfo {author} {\bibfnamefont {W.}~\bibnamefont
  {Fischler}},\ }\bibfield  {title} {\enquote {\bibinfo {title} {{The not so
  harmless axion}},}\ }\Doi {10.1016/0370-2693(83)90639-1} {\bibfield
  {journal} {\bibinfo  {journal} {Phys. Lett.},\ }\textbf {\bibinfo {volume}
  {B120}},\ \bibinfo {pages} {137--141} (\bibinfo {year} {1983})}\BibitemShut
  {NoStop}%
\bibitem [{\citenamefont {Kim}\ and\ \citenamefont {Marsh}(2016)}]{KM}%
  \BibitemOpen
  \bibfield  {author} {\bibinfo {author} {\bibfnamefont {J.~E.}\ \bibnamefont
  {Kim}}\ and\ \bibinfo {author} {\bibfnamefont {D.J.E.}\ \bibnamefont
  {Marsh}},\ }\bibfield  {title} {\enquote {\bibinfo {title} {{An ultralight
  pseudoscalar boson}},}\ }\Doi {10.1103/PhysRevD.93.025027} {\bibfield
  {journal} {\bibinfo  {journal} {Phys. Rev.},\ }\textbf {\bibinfo {volume}
  {D93}},\ \bibinfo {pages} {025027} (\bibinfo {year} {2016})},\ \Eprint
  {http://arxiv.org/abs/1510.01701} {arXiv:1510.01701} \BibitemShut {NoStop}%
\bibitem [{\citenamefont {Marsh}\ \emph {et~al.}(2014)\citenamefont {Marsh},
  \citenamefont {Grin}, \citenamefont {Hlozek},\ and\ \citenamefont
  {Ferreira}}]{Marsh:2014qoa}%
  \BibitemOpen
  \bibfield  {author} {\bibinfo {author} {\bibfnamefont {D.J.E.}\ \bibnamefont
  {Marsh}}, \bibinfo {author} {\bibfnamefont {D.}~\bibnamefont {Grin}},
  \bibinfo {author} {\bibfnamefont {R.}~\bibnamefont {Hlozek}}, \ and\ \bibinfo
  {author} {\bibfnamefont {P.~G.}\ \bibnamefont {Ferreira}},\ }\bibfield
  {title} {\enquote {\bibinfo {title} {{Tensor interpretation of BICEP2 results
  severely constrains axion dark matter}},}\ }\Doi
  {10.1103/PhysRevLett.113.011801} {\bibfield  {journal} {\bibinfo  {journal}
  {Phys. Rev. Lett.},\ }\textbf {\bibinfo {volume} {113}},\ \bibinfo {pages}
  {011801} (\bibinfo {year} {2014})},\ \Eprint {http://arxiv.org/abs/1403.4216}
  {arXiv:1403.4216} \BibitemShut {NoStop}%
\bibitem [{\citenamefont {Graham}\ and\ \citenamefont
  {Rajendran}(2013)}]{CASPEr}%
  \BibitemOpen
  \bibfield  {author} {\bibinfo {author} {\bibfnamefont {P.~W.}\ \bibnamefont
  {Graham}}\ and\ \bibinfo {author} {\bibfnamefont {S.}~\bibnamefont
  {Rajendran}},\ }\bibfield  {title} {\enquote {\bibinfo {title} {{New
  observables for direct detection of axion dark matter}},}\ }\Doi
  {10.1103/PhysRevD.88.035023} {\bibfield  {journal} {\bibinfo  {journal}
  {Phys. Rev.},\ }\textbf {\bibinfo {volume} {D88}},\ \bibinfo {pages} {035023}
  (\bibinfo {year} {2013})},\ \Eprint {http://arxiv.org/abs/1306.6088}
  {arXiv:1306.6088} \BibitemShut {NoStop}%
\bibitem [{\citenamefont {Khmelnitsky}\ and\ \citenamefont
  {Rubakov}(2014)}]{kr14}%
  \BibitemOpen
  \bibfield  {author} {\bibinfo {author} {\bibfnamefont {A.}~\bibnamefont
  {Khmelnitsky}}\ and\ \bibinfo {author} {\bibfnamefont {V.}~\bibnamefont
  {Rubakov}},\ }\bibfield  {title} {\enquote {\bibinfo {title} {{Pulsar timing
  signal from ultralight scalar dark matter}},}\ }\Doi
  {10.1088/1475-7516/2014/02/019} {\bibfield  {journal} {\bibinfo  {journal}
  {JCAP},\ }\textbf {\bibinfo {volume} {1402}},\ \bibinfo {pages} {019}
  (\bibinfo {year} {2014})},\ \Eprint {http://arxiv.org/abs/1309.5888}
  {arXiv:1309.5888} \BibitemShut {NoStop}%
\bibitem [{\citenamefont {Guth}\ \emph {et~al.}(2015)\citenamefont {Guth},
  \citenamefont {Hertzberg},\ and\ \citenamefont
  {Prescod-Weinstein}}]{Guth2014}%
  \BibitemOpen
  \bibfield  {author} {\bibinfo {author} {\bibfnamefont {A.~H.}\ \bibnamefont
  {Guth}}, \bibinfo {author} {\bibfnamefont {M.~P.}\ \bibnamefont {Hertzberg}},
  \ and\ \bibinfo {author} {\bibfnamefont {C.}~\bibnamefont
  {Prescod-Weinstein}},\ }\bibfield  {title} {\enquote {\bibinfo {title} {{Do
  dark matter axions form a condensate with long-range correlation?}}}\ }\Doi
  {10.1103/PhysRevD.92.103513} {\bibfield  {journal} {\bibinfo  {journal}
  {Phys. Rev.},\ }\textbf {\bibinfo {volume} {D92}},\ \bibinfo {pages} {103513}
  (\bibinfo {year} {2015})},\ \Eprint {http://arxiv.org/abs/1412.5930}
  {arXiv:1412.5930} \BibitemShut {NoStop}%
\bibitem [{\citenamefont {Widrow}\ and\ \citenamefont
  {Kaiser}(1993)}]{WidrowKaiser}%
  \BibitemOpen
  \bibfield  {author} {\bibinfo {author} {\bibfnamefont {Lawrence~M.}\
  \bibnamefont {Widrow}}\ and\ \bibinfo {author} {\bibfnamefont {Nick}\
  \bibnamefont {Kaiser}},\ }\bibfield  {title} {\enquote {\bibinfo {title}
  {{Using the Schrodinger equation to simulate collisionless matter}},}\
  }\href@noop {} {\bibfield  {journal} {\bibinfo  {journal} {Astrophys. J.},\
  }\textbf {\bibinfo {volume} {416}},\ \bibinfo {pages} {L71--L74} (\bibinfo
  {year} {1993})}\BibitemShut {NoStop}%
\bibitem [{\citenamefont {Uhlemann}\ \emph {et~al.}(2014)\citenamefont
  {Uhlemann}, \citenamefont {Kopp},\ and\ \citenamefont {Haugg}}]{CU2014}%
  \BibitemOpen
  \bibfield  {author} {\bibinfo {author} {\bibfnamefont {Cora}\ \bibnamefont
  {Uhlemann}}, \bibinfo {author} {\bibfnamefont {Michael}\ \bibnamefont
  {Kopp}}, \ and\ \bibinfo {author} {\bibfnamefont {Thomas}\ \bibnamefont
  {Haugg}},\ }\bibfield  {title} {\enquote {\bibinfo {title} {{Schrödinger
  method as $N$-body double and UV completion of dust}},}\ }\Doi
  {10.1103/PhysRevD.90.023517} {\bibfield  {journal} {\bibinfo  {journal}
  {Phys. Rev.},\ }\textbf {\bibinfo {volume} {D90}},\ \bibinfo {pages} {023517}
  (\bibinfo {year} {2014})},\ \Eprint {http://arxiv.org/abs/1403.5567}
  {arXiv:1403.5567 [astro-ph.CO]} \BibitemShut {NoStop}%
\bibitem [{\citenamefont {Feynman}\ \emph {et~al.}(1963)\citenamefont
  {Feynman}, \citenamefont {Leighton},\ and\ \citenamefont {Sands}}]{Feynman}%
  \BibitemOpen
  \bibfield  {author} {\bibinfo {author} {\bibfnamefont {R.~P.}\ \bibnamefont
  {Feynman}}, \bibinfo {author} {\bibfnamefont {R.~B.}\ \bibnamefont
  {Leighton}}, \ and\ \bibinfo {author} {\bibfnamefont {M.}~\bibnamefont
  {Sands}},\ }\href {http://www.feynmanlectures.info/} {\emph {\bibinfo {title}
  {{The Feynman Lectures on Physics}}}}\ (\bibinfo  {publisher} {Addison Wesley
  Longman},\ \bibinfo {year} {1963})\ ISBN \bibinfo {isbn} {0201021153,
  9780201021158}\BibitemShut {NoStop}%
\bibitem [{\citenamefont {Spiegel}(1980)}]{spiegel80}%
  \BibitemOpen
  \bibfield  {author} {\bibinfo {author} {\bibfnamefont {E.~A.}\ \bibnamefont
  {Spiegel}},\ }\bibfield  {title} {\enquote {\bibinfo {title} {{Fluid
  dynamical form of the linear and nonlinear Schr\"odinger equations}},}\
  }\href@noop {} {\bibfield  {journal} {\bibinfo  {journal} {Physica},\
  }\textbf {\bibinfo {volume} {1D}},\ \bibinfo {pages} {236--240} (\bibinfo
  {year} {1980})}\BibitemShut {NoStop}%
\bibitem [{\citenamefont {{Chavanis}}(2011)}]{Chavanis2011}%
  \BibitemOpen
  \bibfield  {author} {\bibinfo {author} {\bibfnamefont {P.-H.}\ \bibnamefont
  {{Chavanis}}},\ }\bibfield  {title} {\enquote {\bibinfo {title} {{Mass-radius
  relation of Newtonian self-gravitating Bose-Einstein condensates with
  short-range interactions. I. Analytical results}},}\ }\Doi
  {10.1103/PhysRevD.84.043531} {\bibfield  {journal} {\bibinfo  {journal}
  {\prd},\ }\textbf {\bibinfo {volume} {84}},\ \bibinfo {eid} {043531}
  (\bibinfo {year} {2011})},\ \Eprint {http://arxiv.org/abs/1103.2050}
  {arXiv:1103.2050} \BibitemShut {NoStop}%
\bibitem [{\citenamefont {{Su{\'a}rez}}\ and\ \citenamefont
  {{Matos}}(2011)}]{suarez2011}%
  \BibitemOpen
  \bibfield  {author} {\bibinfo {author} {\bibfnamefont {A.}~\bibnamefont
  {{Su{\'a}rez}}}\ and\ \bibinfo {author} {\bibfnamefont {T.}~\bibnamefont
  {{Matos}}},\ }\bibfield  {title} {\enquote {\bibinfo {title} {{Structure
  formation with scalar-field dark matter: the fluid approach}},}\ }\Doi
  {10.1111/j.1365-2966.2011.19012.x} {\bibfield  {journal} {\bibinfo  {journal}
  {\mnras},\ }\textbf {\bibinfo {volume} {416}},\ \bibinfo {pages} {87--93}
  (\bibinfo {year} {2011})},\ \Eprint {http://arxiv.org/abs/1101.4039}
  {arXiv:1101.4039 [gr-qc]} \BibitemShut {NoStop}%
\bibitem [{\citenamefont {{Uhlemann}}\ \emph {et~al.}(2014)\citenamefont
  {{Uhlemann}}, \citenamefont {{Kopp}},\ and\ \citenamefont
  {{Haugg}}}]{uhlemann2014}%
  \BibitemOpen
  \bibfield  {author} {\bibinfo {author} {\bibfnamefont {C.}~\bibnamefont
  {{Uhlemann}}}, \bibinfo {author} {\bibfnamefont {M.}~\bibnamefont {{Kopp}}},
  \ and\ \bibinfo {author} {\bibfnamefont {T.}~\bibnamefont {{Haugg}}},\
  }\bibfield  {title} {\enquote {\bibinfo {title} {{Schr{\"o}dinger method as
  N-body double and UV completion of dust}},}\ }\Doi
  {10.1103/PhysRevD.90.023517} {\bibfield  {journal} {\bibinfo  {journal}
  {\prd},\ }\textbf {\bibinfo {volume} {90}},\ \bibinfo {eid} {023517}
  (\bibinfo {year} {2014})},\ \Eprint {http://arxiv.org/abs/1403.5567}
  {arXiv:1403.5567} \BibitemShut {NoStop}%
\bibitem [{\citenamefont {{Marsh}}(2015)}]{mar15a}%
  \BibitemOpen
  \bibfield  {author} {\bibinfo {author} {\bibfnamefont {D.J.E.}\ \bibnamefont
  {{Marsh}}},\ }\bibfield  {title} {\enquote {\bibinfo {title} {{Nonlinear
  hydrodynamics of axion dark matter: relative velocity effects and quantum
  forces}},}\ }\Doi {10.1103/PhysRevD.91.123520} {\bibfield  {journal}
  {\bibinfo  {journal} {\prd},\ }\textbf {\bibinfo {volume} {91}},\ \bibinfo
  {eid} {123520} (\bibinfo {year} {2015})},\ \Eprint
  {http://arxiv.org/abs/1504.00308} {arXiv:1504.00308} \BibitemShut {NoStop}%
\bibitem [{\citenamefont {{Schwabe}}\ \emph {et~al.}(2016)\citenamefont
  {{Schwabe}}, \citenamefont {{Niemeyer}},\ and\ \citenamefont
  {{Engels}}}]{schwabe16}%
  \BibitemOpen
  \bibfield  {author} {\bibinfo {author} {\bibfnamefont {B.}~\bibnamefont
  {{Schwabe}}}, \bibinfo {author} {\bibfnamefont {J.~C.}\ \bibnamefont
  {{Niemeyer}}}, \ and\ \bibinfo {author} {\bibfnamefont {J.~F.}\ \bibnamefont
  {{Engels}}},\ }\bibfield  {title} {\enquote {\bibinfo {title} {{Simulations
  of solitonic core mergers in ultralight axion dark matter cosmologies}},}\
  }\Doi {10.1103/PhysRevD.94.043513} {\bibfield  {journal} {\bibinfo  {journal}
  {\prd},\ }\textbf {\bibinfo {volume} {94}},\ \bibinfo {eid} {043513}
  (\bibinfo {year} {2016})},\ \Eprint {http://arxiv.org/abs/1606.05151}
  {arXiv:1606.05151} \BibitemShut {NoStop}%
\bibitem [{\citenamefont {Mocz}\ and\ \citenamefont {Succi}(2015)}]{Mocz}%
  \BibitemOpen
  \bibfield  {author} {\bibinfo {author} {\bibfnamefont {P.}~\bibnamefont
  {Mocz}}\ and\ \bibinfo {author} {\bibfnamefont {S.}~\bibnamefont {Succi}},\
  }\bibfield  {title} {\enquote {\bibinfo {title} {{Numerical solution of the
  nonlinear Schr\"odinger equation using smoothed-particle hydrodynamics}},}\
  }\Doi {10.1103/PhysRevE.91.053304} {\bibfield  {journal} {\bibinfo  {journal}
  {Phys. Rev.},\ }\textbf {\bibinfo {volume} {E91}},\ \bibinfo {pages} {053304}
  (\bibinfo {year} {2015})},\ \Eprint {http://arxiv.org/abs/1503.03869}
  {arXiv:1503.03869} \BibitemShut {NoStop}%
\bibitem [{\citenamefont {{Eby}}\ \emph {et~al.}(2016)\citenamefont {{Eby}},
  \citenamefont {{Suranyi}},\ and\ \citenamefont {{Wijewardhana}}}]{Eby2016}%
  \BibitemOpen
  \bibfield  {author} {\bibinfo {author} {\bibfnamefont {J.}~\bibnamefont
  {{Eby}}}, \bibinfo {author} {\bibfnamefont {P.}~\bibnamefont {{Suranyi}}}, \
  and\ \bibinfo {author} {\bibfnamefont {L.~C.~R.}\ \bibnamefont
  {{Wijewardhana}}},\ }\bibfield  {title} {\enquote {\bibinfo {title} {{The
  lifetime of axion stars}},}\ }\Doi {10.1142/S0217732316500905} {\bibfield
  {journal} {\bibinfo  {journal} {Modern Physics Letters A},\ }\textbf
  {\bibinfo {volume} {31}},\ \bibinfo {eid} {1650090} (\bibinfo {year}
  {2016})},\ \Eprint {http://arxiv.org/abs/1512.01709} {arXiv:1512.01709
  [hep-ph]} \BibitemShut {NoStop}%
\bibitem [{\citenamefont {McConnachie}(2012)}]{mc12}%
  \BibitemOpen
  \bibfield  {author} {\bibinfo {author} {\bibfnamefont {A.~W.}\ \bibnamefont
  {McConnachie}},\ }\bibfield  {title} {\enquote {\bibinfo {title} {{The
  observed properties of dwarf galaxies in and around the Local Group}},}\
  }\Doi {10.1088/0004-6256/144/1/4} {\bibfield  {journal} {\bibinfo  {journal}
  {Astron. J.},\ }\textbf {\bibinfo {volume} {144}},\ \bibinfo {pages} {4}
  (\bibinfo {year} {2012})},\ \Eprint {http://arxiv.org/abs/1204.1562}
  {arXiv:1204.1562} \BibitemShut {NoStop}%
\bibitem [{\citenamefont {Chen}\ \emph {et~al.}(2016)\citenamefont {Chen},
  \citenamefont {Schive},\ and\ \citenamefont {Chiueh}}]{chen16}%
  \BibitemOpen
  \bibfield  {author} {\bibinfo {author} {\bibfnamefont {S.-R.}\ \bibnamefont
  {Chen}}, \bibinfo {author} {\bibfnamefont {H.-Y.}\ \bibnamefont {Schive}}, \
  and\ \bibinfo {author} {\bibfnamefont {T.}~\bibnamefont {Chiueh}},\
  }\bibfield  {title} {\enquote {\bibinfo {title} {{Jeans analysis for dwarf
  spheroidal galaxies in wave dark matter}},}\ }\href@noop {} { (\bibinfo
  {year} {2016})},\ \Eprint {http://arxiv.org/abs/1606.09030}
  {arXiv:1606.09030} \BibitemShut {NoStop}%
\bibitem [{\citenamefont {Marsh}\ and\ \citenamefont {Pop}(2015)}]{mp15}%
  \BibitemOpen
  \bibfield  {author} {\bibinfo {author} {\bibfnamefont {D.J.E.}\ \bibnamefont
  {Marsh}}\ and\ \bibinfo {author} {\bibfnamefont {A.-R.}\ \bibnamefont
  {Pop}},\ }\bibfield  {title} {\enquote {\bibinfo {title} {{Axion dark matter,
  solitons and the cusp–-core problem}},}\ }\Doi {10.1093/mnras/stv1050}
  {\bibfield  {journal} {\bibinfo  {journal} {\mnras},\ }\textbf {\bibinfo
  {volume} {451}},\ \bibinfo {pages} {2479--2492} (\bibinfo {year} {2015})},\
  \Eprint {http://arxiv.org/abs/1502.03456} {arXiv:1502.03456} \BibitemShut
  {NoStop}%
\bibitem [{\citenamefont {{Calabrese}}\ and\ \citenamefont
  {{Spergel}}(2016)}]{cs16}%
  \BibitemOpen
  \bibfield  {author} {\bibinfo {author} {\bibfnamefont {E.}~\bibnamefont
  {{Calabrese}}}\ and\ \bibinfo {author} {\bibfnamefont {D.~N.}\ \bibnamefont
  {{Spergel}}},\ }\bibfield  {title} {\enquote {\bibinfo {title} {{Ultra-light
  dark matter in ultra-faint dwarf galaxies}},}\ }\Doi {10.1093/mnras/stw1256}
  {\bibfield  {journal} {\bibinfo  {journal} {\mnras},\ }\textbf {\bibinfo
  {volume} {460}},\ \bibinfo {pages} {4397--4402} (\bibinfo {year} {2016})},\
  \Eprint {http://arxiv.org/abs/1603.07321} {arXiv:1603.07321} \BibitemShut
  {NoStop}%
\bibitem [{\citenamefont {{Gonz{\'a}les-Morales}}\ \emph
  {et~al.}(2016)\citenamefont {{Gonz{\'a}les-Morales}}, \citenamefont
  {{Marsh}}, \citenamefont {{Pe{\~n}arrubia}},\ and\ \citenamefont
  {{Ure{\~n}a-L{\'o}pez}}}]{gm16}%
  \BibitemOpen
  \bibfield  {author} {\bibinfo {author} {\bibfnamefont {A.~X.}\ \bibnamefont
  {{Gonz{\'a}les-Morales}}}, \bibinfo {author} {\bibfnamefont {D.J.E.}\
  \bibnamefont {{Marsh}}}, \bibinfo {author} {\bibfnamefont {J.}~\bibnamefont
  {{Pe{\~n}arrubia}}}, \ and\ \bibinfo {author} {\bibfnamefont
  {L.}~\bibnamefont {{Ure{\~n}a-L{\'o}pez}}},\ }\bibfield  {title} {\enquote
  {\bibinfo {title} {{Unbiased constraints on ultralight axion mass from dwarf
  spheroidal galaxies}},}\ }\href@noop {} {\bibfield  {journal} {\bibinfo
  {journal} {ArXiv e-prints}} (\bibinfo {year} {2016})},\ \Eprint
  {http://arxiv.org/abs/1609.05856} {arXiv:1609.05856} \BibitemShut {NoStop}%
\bibitem [{\citenamefont {Schive}\ \emph
  {et~al.}(2014){\natexlab{b}}\citenamefont {Schive}, \citenamefont {Liao},
  \citenamefont {Woo}, \citenamefont {Wong}, \citenamefont {Chiueh},
  \citenamefont {Broadhurst},\ and\ \citenamefont {Hwang}}]{sch14}%
  \BibitemOpen
  \bibfield  {author} {\bibinfo {author} {\bibfnamefont {H.-Y.}\ \bibnamefont
  {Schive}}, \bibinfo {author} {\bibfnamefont {M.-H.}\ \bibnamefont {Liao}},
  \bibinfo {author} {\bibfnamefont {T.-P.}\ \bibnamefont {Woo}}, \bibinfo
  {author} {\bibfnamefont {S.-K.}\ \bibnamefont {Wong}}, \bibinfo {author}
  {\bibfnamefont {T.}~\bibnamefont {Chiueh}}, \bibinfo {author} {\bibfnamefont
  {T.}~\bibnamefont {Broadhurst}}, \ and\ \bibinfo {author} {\bibfnamefont
  {W.-Y.~Pauchy}\ \bibnamefont {Hwang}},\ }\bibfield  {title} {\enquote
  {\bibinfo {title} {{Understanding the core-halo relation of quantum wave dark
  matter from 3D simulations}},}\ }\Doi {10.1103/PhysRevLett.113.261302}
  {\bibfield  {journal} {\bibinfo  {journal} {Phys. Rev. Lett.},\ }\textbf
  {\bibinfo {volume} {113}},\ \bibinfo {pages} {261302} (\bibinfo {year}
  {2014}{\natexlab{b}})},\ \Eprint {http://arxiv.org/abs/1407.7762}
  {arXiv:1407.7762} \BibitemShut {NoStop}%
\bibitem [{\citenamefont {{Veltmaat}}\ and\ \citenamefont
  {{Niemeyer}}(2016)}]{velt2016}%
  \BibitemOpen
  \bibfield  {author} {\bibinfo {author} {\bibfnamefont {J.}~\bibnamefont
  {{Veltmaat}}}\ and\ \bibinfo {author} {\bibfnamefont {J.~C.}\ \bibnamefont
  {{Niemeyer}}},\ }\bibfield  {title} {\enquote {\bibinfo {title}
  {{Cosmological particle-in-cell simulations with ultra-light axion dark
  matter}},}\ }\href@noop {} {\bibfield  {journal} {\bibinfo  {journal} {ArXiv
  e-prints}} (\bibinfo {year} {2016})},\ \Eprint
  {http://arxiv.org/abs/1608.00802} {arXiv:1608.00802} \BibitemShut {NoStop}%
\bibitem [{\citenamefont {{Binney}}\ and\ \citenamefont
  {{Tremaine}}(2008)}]{BT}%
  \BibitemOpen
  \bibfield  {author} {\bibinfo {author} {\bibfnamefont {J.}~\bibnamefont
  {{Binney}}}\ and\ \bibinfo {author} {\bibfnamefont {S.}~\bibnamefont
  {{Tremaine}}},\ }\href@noop {} {\emph {\bibinfo {title} {{Galactic Dynamics,
  2nd ed.}}}}\ (\bibinfo  {publisher} {Princeton, NJ, Princeton University
  Press},\ \bibinfo {year} {2008})\BibitemShut {NoStop}%
\bibitem [{\citenamefont {Seidel}\ and\ \citenamefont {Suen}(1994)}]{ss94}%
  \BibitemOpen
  \bibfield  {author} {\bibinfo {author} {\bibfnamefont {E.}~\bibnamefont
  {Seidel}}\ and\ \bibinfo {author} {\bibfnamefont {W.-M.}\ \bibnamefont
  {Suen}},\ }\bibfield  {title} {\enquote {\bibinfo {title} {{Formation of
  solitonic stars through gravitational cooling}},}\ }\Doi
  {10.1103/PhysRevLett.72.2516} {\bibfield  {journal} {\bibinfo  {journal}
  {Phys. Rev. Lett.},\ }\textbf {\bibinfo {volume} {72}},\ \bibinfo {pages}
  {2516--2519} (\bibinfo {year} {1994})},\ \Eprint
  {http://arxiv.org/abs/gr-qc/9309015} {arXiv:gr-qc/9309015} \BibitemShut
  {NoStop}%
\bibitem [{\citenamefont {{Harrison}}\ \emph {et~al.}(2003)\citenamefont
  {{Harrison}}, \citenamefont {{Moroz}},\ and\ \citenamefont {{Tod}}}]{hmt03}%
  \BibitemOpen
  \bibfield  {author} {\bibinfo {author} {\bibfnamefont {R.}~\bibnamefont
  {{Harrison}}}, \bibinfo {author} {\bibfnamefont {I.}~\bibnamefont {{Moroz}}},
  \ and\ \bibinfo {author} {\bibfnamefont {K.~P.}\ \bibnamefont {{Tod}}},\
  }\bibfield  {title} {\enquote {\bibinfo {title} {{A numerical study of the
  Schr\"odinger-Newton equations}},}\ }\href@noop {} {\bibfield  {journal}
  {\bibinfo  {journal} {Nonlinearity},\ }\textbf {\bibinfo {volume} {16}},\
  \bibinfo {pages} {101--122} (\bibinfo {year} {2003})},\ \Eprint
  {http://arxiv.org/abs/math-ph/0208045} {arXiv:math-ph/0208045} \BibitemShut
  {NoStop}%
\bibitem [{\citenamefont {Guzman}\ and\ \citenamefont
  {Urena-Lopez}(2006)}]{gul06}%
  \BibitemOpen
  \bibfield  {author} {\bibinfo {author} {\bibfnamefont {F.~S.}\ \bibnamefont
  {Guzman}}\ and\ \bibinfo {author} {\bibfnamefont {L.~A.}\ \bibnamefont
  {Urena-Lopez}},\ }\bibfield  {title} {\enquote {\bibinfo {title}
  {{Gravitational cooling of self-gravitating Bose condensates}},}\ }\Doi
  {10.1086/504508} {\bibfield  {journal} {\bibinfo  {journal} {Astrophys. J.},\
  }\textbf {\bibinfo {volume} {645}},\ \bibinfo {pages} {814--819} (\bibinfo
  {year} {2006})},\ \Eprint {http://arxiv.org/abs/astro-ph/0603613}
  {arXiv:astro-ph/0603613} \BibitemShut {NoStop}%
\bibitem [{\citenamefont {{Lacey}}\ and\ \citenamefont
  {{Ostriker}}(1985)}]{lacey1985}%
  \BibitemOpen
  \bibfield  {author} {\bibinfo {author} {\bibfnamefont {C.~G.}\ \bibnamefont
  {{Lacey}}}\ and\ \bibinfo {author} {\bibfnamefont {J.~P.}\ \bibnamefont
  {{Ostriker}}},\ }\bibfield  {title} {\enquote {\bibinfo {title} {{Massive
  black holes in galactic halos?}}}\ }\Doi {10.1086/163729} {\bibfield
  {journal} {\bibinfo  {journal} {\apj},\ }\textbf {\bibinfo {volume} {299}},\
  \bibinfo {pages} {633--652} (\bibinfo {year} {1985})}\BibitemShut {NoStop}%
\bibitem [{\citenamefont {{Carr}}\ and\ \citenamefont
  {{Lacey}}(1987)}]{carr1987}%
  \BibitemOpen
  \bibfield  {author} {\bibinfo {author} {\bibfnamefont {B.~J.}\ \bibnamefont
  {{Carr}}}\ and\ \bibinfo {author} {\bibfnamefont {C.~G.}\ \bibnamefont
  {{Lacey}}},\ }\bibfield  {title} {\enquote {\bibinfo {title} {{Dark clusters
  in galactic halos?}}}\ }\Doi {10.1086/165176} {\bibfield  {journal} {\bibinfo
   {journal} {\apj},\ }\textbf {\bibinfo {volume} {316}},\ \bibinfo {pages}
  {23--35} (\bibinfo {year} {1987})}\BibitemShut {NoStop}%
\bibitem [{\citenamefont {{Begelman}}\ \emph {et~al.}(1980)\citenamefont
  {{Begelman}}, \citenamefont {{Blandford}},\ and\ \citenamefont
  {{Rees}}}]{BBR1980}%
  \BibitemOpen
  \bibfield  {author} {\bibinfo {author} {\bibfnamefont {M.~C.}\ \bibnamefont
  {{Begelman}}}, \bibinfo {author} {\bibfnamefont {R.~D.}\ \bibnamefont
  {{Blandford}}}, \ and\ \bibinfo {author} {\bibfnamefont {M.~J.}\ \bibnamefont
  {{Rees}}},\ }\bibfield  {title} {\enquote {\bibinfo {title} {{Massive black
  hole binaries in active galactic nuclei}},}\ }\Doi {10.1038/287307a0}
  {\bibfield  {journal} {\bibinfo  {journal} {\nat},\ }\textbf {\bibinfo
  {volume} {287}},\ \bibinfo {pages} {307--309} (\bibinfo {year}
  {1980})}\BibitemShut {NoStop}%
\bibitem [{\citenamefont {{Yu}}(2002)}]{Yu2002}%
  \BibitemOpen
  \bibfield  {author} {\bibinfo {author} {\bibfnamefont {Q.}~\bibnamefont
  {{Yu}}},\ }\bibfield  {title} {\enquote {\bibinfo {title} {{Evolution of
  massive binary black holes}},}\ }\Doi {10.1046/j.1365-8711.2002.05242.x}
  {\bibfield  {journal} {\bibinfo  {journal} {\mnras},\ }\textbf {\bibinfo
  {volume} {331}},\ \bibinfo {pages} {935--958} (\bibinfo {year} {2002})},\
  \Eprint {http://arxiv.org/abs/arXiv:astro-ph/0109530}
  {arXiv:astro-ph/0109530} \BibitemShut {NoStop}%
\bibitem [{\citenamefont {Navarro}\ \emph {et~al.}(1996)\citenamefont
  {Navarro}, \citenamefont {Eke},\ and\ \citenamefont {Frenk}}]{nef96}%
  \BibitemOpen
  \bibfield  {author} {\bibinfo {author} {\bibfnamefont {J.~F.}\ \bibnamefont
  {Navarro}}, \bibinfo {author} {\bibfnamefont {V.~R.}\ \bibnamefont {Eke}}, \
  and\ \bibinfo {author} {\bibfnamefont {C.~S.}\ \bibnamefont {Frenk}},\
  }\bibfield  {title} {\enquote {\bibinfo {title} {{The cores of dwarf galaxy
  halos}},}\ }\Doi {10.1093/mnras/283.3.72L, 10.1093/mnras/283.3.L72}
  {\bibfield  {journal} {\bibinfo  {journal} {\mnras},\ }\textbf {\bibinfo
  {volume} {283}},\ \bibinfo {pages} {L72--L78} (\bibinfo {year} {1996})},\
  \Eprint {http://arxiv.org/abs/astro-ph/9610187} {arXiv:astro-ph/9610187}
  \BibitemShut {NoStop}%
\bibitem [{\citenamefont {El-Zant}\ \emph {et~al.}(2001)\citenamefont
  {El-Zant}, \citenamefont {Shlosman},\ and\ \citenamefont {Hoffman}}]{esh01}%
  \BibitemOpen
  \bibfield  {author} {\bibinfo {author} {\bibfnamefont {A.}~\bibnamefont
  {El-Zant}}, \bibinfo {author} {\bibfnamefont {I.}~\bibnamefont {Shlosman}}, \
  and\ \bibinfo {author} {\bibfnamefont {Y.}~\bibnamefont {Hoffman}},\
  }\bibfield  {title} {\enquote {\bibinfo {title} {{Dark halos: the flattening
  of the density cusp by dynamical friction}},}\ }\Doi {10.1086/322516}
  {\bibfield  {journal} {\bibinfo  {journal} {Astrophys. J.},\ }\textbf
  {\bibinfo {volume} {560}},\ \bibinfo {pages} {636--643} (\bibinfo {year}
  {2001})},\ \Eprint {http://arxiv.org/abs/astro-ph/0103386}
  {arXiv:astro-ph/0103386} \BibitemShut {NoStop}%
\bibitem [{\citenamefont {Gnedin}\ and\ \citenamefont {Zhao}(2002)}]{gz02}%
  \BibitemOpen
  \bibfield  {author} {\bibinfo {author} {\bibfnamefont {O.~Y.}\ \bibnamefont
  {Gnedin}}\ and\ \bibinfo {author} {\bibfnamefont {H.}~\bibnamefont {Zhao}},\
  }\bibfield  {title} {\enquote {\bibinfo {title} {{Maximum feedback and dark
  matter profiles of dwarf galaxies}},}\ }\Doi
  {10.1046/j.1365-8711.2002.05361.x} {\bibfield  {journal} {\bibinfo  {journal}
  {\mnras},\ }\textbf {\bibinfo {volume} {333}},\ \bibinfo {pages} {299--306}
  (\bibinfo {year} {2002})},\ \Eprint {http://arxiv.org/abs/astro-ph/0108108}
  {arXiv:astro-ph/0108108} \BibitemShut {NoStop}%
\bibitem [{\citenamefont {Read}\ and\ \citenamefont {Gilmore}(2005)}]{rg05}%
  \BibitemOpen
  \bibfield  {author} {\bibinfo {author} {\bibfnamefont {J.~I.}\ \bibnamefont
  {Read}}\ and\ \bibinfo {author} {\bibfnamefont {G.}~\bibnamefont {Gilmore}},\
  }\bibfield  {title} {\enquote {\bibinfo {title} {{Mass loss from dwarf
  spheroidal galaxies: The origins of shallow dark matter cores and exponential
  surface brightness profiles}},}\ }\Doi {10.1111/j.1365-2966.2004.08424.x}
  {\bibfield  {journal} {\bibinfo  {journal} {\mnras},\ }\textbf {\bibinfo
  {volume} {356}},\ \bibinfo {pages} {107--124} (\bibinfo {year} {2005})},\
  \Eprint {http://arxiv.org/abs/astro-ph/0409565} {arXiv:astro-ph/0409565}
  \BibitemShut {NoStop}%
\bibitem [{\citenamefont {{Tonini}}\ \emph {et~al.}(2006)\citenamefont
  {{Tonini}}, \citenamefont {{Lapi}},\ and\ \citenamefont {{Salucci}}}]{tls06}%
  \BibitemOpen
  \bibfield  {author} {\bibinfo {author} {\bibfnamefont {C.}~\bibnamefont
  {{Tonini}}}, \bibinfo {author} {\bibfnamefont {A.}~\bibnamefont {{Lapi}}}, \
  and\ \bibinfo {author} {\bibfnamefont {P.}~\bibnamefont {{Salucci}}},\
  }\bibfield  {title} {\enquote {\bibinfo {title} {{Angular momentum transfer
  in dark matter halos: erasing the cusp}},}\ }\Doi {10.1086/506431} {\bibfield
   {journal} {\bibinfo  {journal} {\apj},\ }\textbf {\bibinfo {volume} {649}},\
  \bibinfo {pages} {591--598} (\bibinfo {year} {2006})},\ \Eprint
  {http://arxiv.org/abs/astro-ph/0603051} {arXiv:astro-ph/0603051} \BibitemShut
  {NoStop}%
\bibitem [{\citenamefont {Mashchenko}\ \emph {et~al.}(2008)\citenamefont
  {Mashchenko}, \citenamefont {Wadsley},\ and\ \citenamefont
  {Couchman}}]{mas08}%
  \BibitemOpen
  \bibfield  {author} {\bibinfo {author} {\bibfnamefont {S.}~\bibnamefont
  {Mashchenko}}, \bibinfo {author} {\bibfnamefont {J.}~\bibnamefont {Wadsley}},
  \ and\ \bibinfo {author} {\bibfnamefont {H.M.P.}\ \bibnamefont {Couchman}},\
  }\bibfield  {title} {\enquote {\bibinfo {title} {{Stellar feedback in dwarf
  galaxy formation}},}\ }\Doi {10.1126/science.1148666} {\bibfield  {journal}
  {\bibinfo  {journal} {Science},\ }\textbf {\bibinfo {volume} {319}},\
  \bibinfo {pages} {174--177} (\bibinfo {year} {2008})},\ \Eprint
  {http://arxiv.org/abs/0711.4803} {arXiv:0711.4803} \BibitemShut {NoStop}%
\bibitem [{\citenamefont {Romano-Diaz}\ \emph {et~al.}(2008)\citenamefont
  {Romano-Diaz}, \citenamefont {Shlosman}, \citenamefont {Hoffman},\ and\
  \citenamefont {Heller}}]{rd08}%
  \BibitemOpen
  \bibfield  {author} {\bibinfo {author} {\bibfnamefont {E.}~\bibnamefont
  {Romano-Diaz}}, \bibinfo {author} {\bibfnamefont {I.}~\bibnamefont
  {Shlosman}}, \bibinfo {author} {\bibfnamefont {Y.}~\bibnamefont {Hoffman}}, \
  and\ \bibinfo {author} {\bibfnamefont {C.}~\bibnamefont {Heller}},\
  }\bibfield  {title} {\enquote {\bibinfo {title} {{Erasing dark matter cusps
  in cosmological galactic halos with baryons}},}\ }\Doi {10.1086/592687}
  {\bibfield  {journal} {\bibinfo  {journal} {Astrophys. J.},\ }\textbf
  {\bibinfo {volume} {685}},\ \bibinfo {pages} {L105} (\bibinfo {year}
  {2008})},\ \Eprint {http://arxiv.org/abs/0808.0195} {arXiv:0808.0195}
  \BibitemShut {NoStop}%
\bibitem [{\citenamefont {Goerdt}\ \emph {et~al.}(2010)\citenamefont {Goerdt},
  \citenamefont {Read}, \citenamefont {Moore},\ and\ \citenamefont
  {Stadel}}]{goe10}%
  \BibitemOpen
  \bibfield  {author} {\bibinfo {author} {\bibfnamefont {T.}~\bibnamefont
  {Goerdt}}, \bibinfo {author} {\bibfnamefont {J.~I.}\ \bibnamefont {Read}},
  \bibinfo {author} {\bibfnamefont {B.}~\bibnamefont {Moore}}, \ and\ \bibinfo
  {author} {\bibfnamefont {J.}~\bibnamefont {Stadel}},\ }\bibfield  {title}
  {\enquote {\bibinfo {title} {{Core creation in galaxies and haloes via
  sinking massive objects}},}\ }\Doi {10.1088/0004-637X/725/2/1707} {\bibfield
  {journal} {\bibinfo  {journal} {Astrophys. J.},\ }\textbf {\bibinfo {volume}
  {725}},\ \bibinfo {pages} {1707--1716} (\bibinfo {year} {2010})},\ \Eprint
  {http://arxiv.org/abs/0806.1951} {arXiv:0806.1951} \BibitemShut {NoStop}%
\bibitem [{\citenamefont {{Governato}}\ \emph {et~al.}(2010)\citenamefont
  {{Governato}}, \citenamefont {{Brook}}, \citenamefont {{Mayer}},
  \citenamefont {{Brooks}}, \citenamefont {{Rhee}}, \citenamefont {{Wadsley}},
  \citenamefont {{Jonsson}}, \citenamefont {{Willman}}, \citenamefont
  {{Stinson}}, \citenamefont {{Quinn}},\ and\ \citenamefont {{Madau}}}]{gov10}%
  \BibitemOpen
  \bibfield  {author} {\bibinfo {author} {\bibfnamefont {F.}~\bibnamefont
  {{Governato}}}, \bibinfo {author} {\bibfnamefont {C.}~\bibnamefont
  {{Brook}}}, \bibinfo {author} {\bibfnamefont {L.}~\bibnamefont {{Mayer}}},
  \bibinfo {author} {\bibfnamefont {A.}~\bibnamefont {{Brooks}}}, \bibinfo
  {author} {\bibfnamefont {G.}~\bibnamefont {{Rhee}}}, \bibinfo {author}
  {\bibfnamefont {J.}~\bibnamefont {{Wadsley}}}, \bibinfo {author}
  {\bibfnamefont {P.}~\bibnamefont {{Jonsson}}}, \bibinfo {author}
  {\bibfnamefont {B.}~\bibnamefont {{Willman}}}, \bibinfo {author}
  {\bibfnamefont {G.}~\bibnamefont {{Stinson}}}, \bibinfo {author}
  {\bibfnamefont {T.}~\bibnamefont {{Quinn}}}, \ and\ \bibinfo {author}
  {\bibfnamefont {P.}~\bibnamefont {{Madau}}},\ }\bibfield  {title} {\enquote
  {\bibinfo {title} {{Bulgeless dwarf galaxies and dark matter cores from
  supernova-driven outflows}},}\ }\Doi {10.1038/nature08640} {\bibfield
  {journal} {\bibinfo  {journal} {\nat},\ }\textbf {\bibinfo {volume} {463}},\
  \bibinfo {pages} {203--206} (\bibinfo {year} {2010})},\ \Eprint
  {http://arxiv.org/abs/0911.2237} {arXiv:0911.2237} \BibitemShut {NoStop}%
\bibitem [{\citenamefont {Cole}\ \emph {et~al.}(2011)\citenamefont {Cole},
  \citenamefont {Dehnen},\ and\ \citenamefont {Wilkinson}}]{cole11}%
  \BibitemOpen
  \bibfield  {author} {\bibinfo {author} {\bibfnamefont {D.}~\bibnamefont
  {Cole}}, \bibinfo {author} {\bibfnamefont {W.}~\bibnamefont {Dehnen}}, \ and\
  \bibinfo {author} {\bibfnamefont {M.}~\bibnamefont {Wilkinson}},\ }\bibfield
  {title} {\enquote {\bibinfo {title} {{Weakening dark-matter cusps by clumpy
  baryonic infall}},}\ }\Doi {10.1111/j.1365-2966.2011.19110.x} {\bibfield
  {journal} {\bibinfo  {journal} {\mnras},\ }\textbf {\bibinfo {volume}
  {416}},\ \bibinfo {pages} {1118--1134} (\bibinfo {year} {2011})},\ \Eprint
  {http://arxiv.org/abs/1105.4050} {arXiv:1105.4050} \BibitemShut {NoStop}%
\bibitem [{\citenamefont {Pontzen}\ and\ \citenamefont
  {Governato}(2012)}]{pg12}%
  \BibitemOpen
  \bibfield  {author} {\bibinfo {author} {\bibfnamefont {A.}~\bibnamefont
  {Pontzen}}\ and\ \bibinfo {author} {\bibfnamefont {F.}~\bibnamefont
  {Governato}},\ }\bibfield  {title} {\enquote {\bibinfo {title} {{How
  supernova feedback turns dark matter cusps into cores}},}\ }\Doi
  {10.1111/j.1365-2966.2012.20571.x} {\bibfield  {journal} {\bibinfo  {journal}
  {\mnras},\ }\textbf {\bibinfo {volume} {421}},\ \bibinfo {pages} {3464--3471}
  (\bibinfo {year} {2012})},\ \Eprint {http://arxiv.org/abs/1106.0499}
  {arXiv:1106.0499} \BibitemShut {NoStop}%
\bibitem [{\citenamefont {{Kleyna}}\ \emph {et~al.}(2002)\citenamefont
  {{Kleyna}}, \citenamefont {{Wilkinson}}, \citenamefont {{Evans}},
  \citenamefont {{Gilmore}},\ and\ \citenamefont {{Frayn}}}]{kle02}%
  \BibitemOpen
  \bibfield  {author} {\bibinfo {author} {\bibfnamefont {J.}~\bibnamefont
  {{Kleyna}}}, \bibinfo {author} {\bibfnamefont {M.~I.}\ \bibnamefont
  {{Wilkinson}}}, \bibinfo {author} {\bibfnamefont {N.~W.}\ \bibnamefont
  {{Evans}}}, \bibinfo {author} {\bibfnamefont {G.}~\bibnamefont {{Gilmore}}},
  \ and\ \bibinfo {author} {\bibfnamefont {C.}~\bibnamefont {{Frayn}}},\
  }\bibfield  {title} {\enquote {\bibinfo {title} {{Dark matter in dwarf
  spheroidals - II. Observations and modelling of Draco}},}\ }\Doi
  {10.1046/j.1365-8711.2002.05155.x} {\bibfield  {journal} {\bibinfo  {journal}
  {\mnras},\ }\textbf {\bibinfo {volume} {330}},\ \bibinfo {pages} {792--806}
  (\bibinfo {year} {2002})},\ \Eprint {http://arxiv.org/abs/astro-ph/0109450}
  {arXiv:astro-ph/0109450} \BibitemShut {NoStop}%
\bibitem [{\citenamefont {Goerdt}\ \emph {et~al.}(2006)\citenamefont {Goerdt},
  \citenamefont {Moore}, \citenamefont {Read}, \citenamefont {Stadel},\ and\
  \citenamefont {Zemp}}]{goe06}%
  \BibitemOpen
  \bibfield  {author} {\bibinfo {author} {\bibfnamefont {T.}~\bibnamefont
  {Goerdt}}, \bibinfo {author} {\bibfnamefont {B.}~\bibnamefont {Moore}},
  \bibinfo {author} {\bibfnamefont {J.~I.}\ \bibnamefont {Read}}, \bibinfo
  {author} {\bibfnamefont {J.}~\bibnamefont {Stadel}}, \ and\ \bibinfo {author}
  {\bibfnamefont {M.}~\bibnamefont {Zemp}},\ }\bibfield  {title} {\enquote
  {\bibinfo {title} {{Does the Fornax dwarf spheroidal have a central cusp or
  core?}}}\ }\Doi {10.1111/j.1365-2966.2006.10182.x} {\bibfield  {journal}
  {\bibinfo  {journal} {\mnras},\ }\textbf {\bibinfo {volume} {368}},\ \bibinfo
  {pages} {1073--1077} (\bibinfo {year} {2006})},\ \Eprint
  {http://arxiv.org/abs/astro-ph/0601404} {arXiv:astro-ph/0601404} \BibitemShut
  {NoStop}%
\bibitem [{\citenamefont {Wu}(2007)}]{wu07}%
  \BibitemOpen
  \bibfield  {author} {\bibinfo {author} {\bibfnamefont {X.}~\bibnamefont
  {Wu}},\ }\bibfield  {title} {\enquote {\bibinfo {title} {{The mass
  distribution of dwarf spheroidal galaxies from stellar kinematics: Draco,
  Ursa Minor and Fornax}},}\ }\href@noop {} { (\bibinfo {year} {2007})},\
  \Eprint {http://arxiv.org/abs/astro-ph/0702233} {arXiv:astro-ph/0702233}
  \BibitemShut {NoStop}%
\bibitem [{\citenamefont {Battaglia}\ \emph {et~al.}(2008)\citenamefont
  {Battaglia}, \citenamefont {Helmi}, \citenamefont {Tolstoy}, \citenamefont
  {Irwin}, \citenamefont {Hill},\ and\ \citenamefont {Jablonka}}]{bat08}%
  \BibitemOpen
  \bibfield  {author} {\bibinfo {author} {\bibfnamefont {G.}~\bibnamefont
  {Battaglia}}, \bibinfo {author} {\bibfnamefont {A.}~\bibnamefont {Helmi}},
  \bibinfo {author} {\bibfnamefont {E.}~\bibnamefont {Tolstoy}}, \bibinfo
  {author} {\bibfnamefont {M.}~\bibnamefont {Irwin}}, \bibinfo {author}
  {\bibfnamefont {V.}~\bibnamefont {Hill}}, \ and\ \bibinfo {author}
  {\bibfnamefont {P.}~\bibnamefont {Jablonka}},\ }\bibfield  {title} {\enquote
  {\bibinfo {title} {{The kinematic status and mass content of the Sculptor
  dwarf spheroidal galaxy}},}\ }\Doi {10.1086/590179} {\bibfield  {journal}
  {\bibinfo  {journal} {Astrophys. J.},\ }\textbf {\bibinfo {volume} {681}},\
  \bibinfo {pages} {L13} (\bibinfo {year} {2008})},\ \Eprint
  {http://arxiv.org/abs/0802.4220} {arXiv:0802.4220} \BibitemShut {NoStop}%
\bibitem [{\citenamefont {Walker}\ \emph {et~al.}(2009)\citenamefont {Walker},
  \citenamefont {Mateo}, \citenamefont {Olszewski}, \citenamefont {Penarrubia},
  \citenamefont {Evans},\ and\ \citenamefont {Gilmore}}]{wal09}%
  \BibitemOpen
  \bibfield  {author} {\bibinfo {author} {\bibfnamefont {M.~G.}\ \bibnamefont
  {Walker}}, \bibinfo {author} {\bibfnamefont {M.}~\bibnamefont {Mateo}},
  \bibinfo {author} {\bibfnamefont {E.~W.}\ \bibnamefont {Olszewski}}, \bibinfo
  {author} {\bibfnamefont {J.}~\bibnamefont {Penarrubia}}, \bibinfo {author}
  {\bibfnamefont {N.~W.}\ \bibnamefont {Evans}}, \ and\ \bibinfo {author}
  {\bibfnamefont {G.}~\bibnamefont {Gilmore}},\ }\bibfield  {title} {\enquote
  {\bibinfo {title} {{A universal mass profile for dwarf spheroidal
  galaxies}},}\ }\Doi {10.1088/0004-637X/704/2/1274,
  10.1088/0004-637X/710/1/886} {\bibfield  {journal} {\bibinfo  {journal}
  {Astrophys. J.},\ }\textbf {\bibinfo {volume} {704}},\ \bibinfo {pages}
  {1274--1287} (\bibinfo {year} {2009})},\ \bibinfo {note} {[Erratum:
  Astrophys. J.710,886(2010)]},\ \Eprint {http://arxiv.org/abs/0906.0341}
  {arXiv:0906.0341} \BibitemShut {NoStop}%
\bibitem [{\citenamefont {Strigari}\ \emph {et~al.}(2010)\citenamefont
  {Strigari}, \citenamefont {Frenk},\ and\ \citenamefont {White}}]{str10}%
  \BibitemOpen
  \bibfield  {author} {\bibinfo {author} {\bibfnamefont {L.~E.}\ \bibnamefont
  {Strigari}}, \bibinfo {author} {\bibfnamefont {C.~S.}\ \bibnamefont {Frenk}},
  \ and\ \bibinfo {author} {\bibfnamefont {S.D.M.}\ \bibnamefont {White}},\
  }\bibfield  {title} {\enquote {\bibinfo {title} {{Kinematics of Milky Way
  satellites in a Lambda Cold Dark Matter universe}},}\ }\Doi
  {10.1111/j.1365-2966.2010.17287.x} {\bibfield  {journal} {\bibinfo  {journal}
  {\mnras},\ }\textbf {\bibinfo {volume} {408}},\ \bibinfo {pages} {2364--2372}
  (\bibinfo {year} {2010})},\ \Eprint {http://arxiv.org/abs/1003.4268}
  {arXiv:1003.4268} \BibitemShut {NoStop}%
\bibitem [{\citenamefont {Walker}\ and\ \citenamefont
  {Penarrubia}(2011)}]{wal11}%
  \BibitemOpen
  \bibfield  {author} {\bibinfo {author} {\bibfnamefont {M.~G.}\ \bibnamefont
  {Walker}}\ and\ \bibinfo {author} {\bibfnamefont {J.}~\bibnamefont
  {Penarrubia}},\ }\bibfield  {title} {\enquote {\bibinfo {title} {{A method
  for measuring (slopes of) the mass profiles of dwarf spheroidal galaxies}},}\
  }\Doi {10.1088/0004-637X/742/1/20} {\bibfield  {journal} {\bibinfo  {journal}
  {Astrophys. J.},\ }\textbf {\bibinfo {volume} {742}},\ \bibinfo {pages} {20}
  (\bibinfo {year} {2011})},\ \Eprint {http://arxiv.org/abs/1108.2404}
  {arXiv:1108.2404} \BibitemShut {NoStop}%
\bibitem [{\citenamefont {Amorisco}\ and\ \citenamefont {Evans}(2012)}]{ae12}%
  \BibitemOpen
  \bibfield  {author} {\bibinfo {author} {\bibfnamefont {N.~C.}\ \bibnamefont
  {Amorisco}}\ and\ \bibinfo {author} {\bibfnamefont {N.~W.}\ \bibnamefont
  {Evans}},\ }\bibfield  {title} {\enquote {\bibinfo {title} {{Dark matter
  cores and cusps: the case of multiple stellar populations in dwarf
  spheroidals}},}\ }\Doi {10.1111/j.1365-2966.2011.19684.x} {\bibfield
  {journal} {\bibinfo  {journal} {\mnras},\ }\textbf {\bibinfo {volume}
  {419}},\ \bibinfo {pages} {184--196} (\bibinfo {year} {2012})},\ \Eprint
  {http://arxiv.org/abs/1106.1062} {arXiv:1106.1062} \BibitemShut {NoStop}%
\bibitem [{\citenamefont {Amorisco}\ \emph {et~al.}(2013)\citenamefont
  {Amorisco}, \citenamefont {Agnello},\ and\ \citenamefont {Evans}}]{aae13}%
  \BibitemOpen
  \bibfield  {author} {\bibinfo {author} {\bibfnamefont {N.~C.}\ \bibnamefont
  {Amorisco}}, \bibinfo {author} {\bibfnamefont {A.}~\bibnamefont {Agnello}}, \
  and\ \bibinfo {author} {\bibfnamefont {N.~W.}\ \bibnamefont {Evans}},\
  }\bibfield  {title} {\enquote {\bibinfo {title} {{The core size of the Fornax
  dwarf Spheroidal}},}\ }\Doi {10.1093/mnrasl/sls031} {\bibfield  {journal}
  {\bibinfo  {journal} {\mnras},\ }\textbf {\bibinfo {volume} {429}},\ \bibinfo
  {pages} {L89--L93} (\bibinfo {year} {2013})},\ \Eprint
  {http://arxiv.org/abs/1210.3157} {arXiv:1210.3157} \BibitemShut {NoStop}%
\bibitem [{\citenamefont {Breddels}\ and\ \citenamefont
  {Helmi}(2013)}]{bred13}%
  \BibitemOpen
  \bibfield  {author} {\bibinfo {author} {\bibfnamefont {M.~A.}\ \bibnamefont
  {Breddels}}\ and\ \bibinfo {author} {\bibfnamefont {A.}~\bibnamefont
  {Helmi}},\ }\bibfield  {title} {\enquote {\bibinfo {title} {{Model comparison
  of the dark matter profiles of Fornax, Sculptor, Carina and Sextans}},}\
  }\Doi {10.1051/0004-6361/201321606} {\bibfield  {journal} {\bibinfo
  {journal} {Astron. Astrophys.},\ }\textbf {\bibinfo {volume} {558}},\
  \bibinfo {pages} {A35} (\bibinfo {year} {2013})},\ \Eprint
  {http://arxiv.org/abs/1304.2976} {arXiv:1304.2976} \BibitemShut {NoStop}%
\bibitem [{\citenamefont {Jardel}\ and\ \citenamefont {Gebhardt}(2013)}]{jg13}%
  \BibitemOpen
  \bibfield  {author} {\bibinfo {author} {\bibfnamefont {J.~R.}\ \bibnamefont
  {Jardel}}\ and\ \bibinfo {author} {\bibfnamefont {K.}~\bibnamefont
  {Gebhardt}},\ }\bibfield  {title} {\enquote {\bibinfo {title} {{Variations in
  a universal dark matter profile for dwarf spheroidals}},}\ }\Doi
  {10.1088/2041-8205/775/1/L30} {\bibfield  {journal} {\bibinfo  {journal}
  {Astrophys. J.},\ }\textbf {\bibinfo {volume} {775}},\ \bibinfo {pages} {L30}
  (\bibinfo {year} {2013})}\BibitemShut {NoStop}%
\bibitem [{\citenamefont {Strigari}\ \emph {et~al.}(2014)\citenamefont
  {Strigari}, \citenamefont {Frenk},\ and\ \citenamefont {White}}]{str14}%
  \BibitemOpen
  \bibfield  {author} {\bibinfo {author} {\bibfnamefont {L.~E.}\ \bibnamefont
  {Strigari}}, \bibinfo {author} {\bibfnamefont {C.~S.}\ \bibnamefont {Frenk}},
  \ and\ \bibinfo {author} {\bibfnamefont {S.D.M.}\ \bibnamefont {White}},\
  }\bibfield  {title} {\enquote {\bibinfo {title} {{Dynamical models for the
  Sculptor dwarf spheroidal in a Lambda CDM universe}},}\ }\href@noop {} {
  (\bibinfo {year} {2014})},\ \Eprint {http://arxiv.org/abs/1406.6079}
  {arXiv:1406.6079} \BibitemShut {NoStop}%
\bibitem [{\citenamefont {Fattahi}\ \emph {et~al.}(2016)\citenamefont
  {Fattahi}, \citenamefont {Navarro}, \citenamefont {Sawala}, \citenamefont
  {Frenk}, \citenamefont {Sales}, \citenamefont {Oman}, \citenamefont
  {Schaller},\ and\ \citenamefont {Wang}}]{fat16}%
  \BibitemOpen
  \bibfield  {author} {\bibinfo {author} {\bibfnamefont {A.}~\bibnamefont
  {Fattahi}}, \bibinfo {author} {\bibfnamefont {J.~F.}\ \bibnamefont
  {Navarro}}, \bibinfo {author} {\bibfnamefont {T.}~\bibnamefont {Sawala}},
  \bibinfo {author} {\bibfnamefont {C.~S.}\ \bibnamefont {Frenk}}, \bibinfo
  {author} {\bibfnamefont {L.~V.}\ \bibnamefont {Sales}}, \bibinfo {author}
  {\bibfnamefont {K.}~\bibnamefont {Oman}}, \bibinfo {author} {\bibfnamefont
  {M.}~\bibnamefont {Schaller}}, \ and\ \bibinfo {author} {\bibfnamefont
  {J.}~\bibnamefont {Wang}},\ }\bibfield  {title} {\enquote {\bibinfo {title}
  {{The cold dark matter content of Galactic dwarf spheroidals: no cores, no
  failures, no problem}},}\ }\href@noop {} { (\bibinfo {year} {2016})},\
  \Eprint {http://arxiv.org/abs/1607.06479} {arXiv:1607.06479} \BibitemShut
  {NoStop}%
\bibitem [{\citenamefont {de~Blok}\ \emph {et~al.}(2001)\citenamefont
  {de~Blok}, \citenamefont {McGaugh}, \citenamefont {Bosma},\ and\
  \citenamefont {Rubin}}]{deblok01}%
  \BibitemOpen
  \bibfield  {author} {\bibinfo {author} {\bibfnamefont {W.J.G.}\ \bibnamefont
  {de~Blok}}, \bibinfo {author} {\bibfnamefont {S.~S.}\ \bibnamefont
  {McGaugh}}, \bibinfo {author} {\bibfnamefont {A.}~\bibnamefont {Bosma}}, \
  and\ \bibinfo {author} {\bibfnamefont {V.~C.}\ \bibnamefont {Rubin}},\
  }\bibfield  {title} {\enquote {\bibinfo {title} {{Mass density profiles of
  LSB galaxies}},}\ }\Doi {10.1086/320262} {\bibfield  {journal} {\bibinfo
  {journal} {Astrophys. J.},\ }\textbf {\bibinfo {volume} {552}},\ \bibinfo
  {pages} {L23--L26} (\bibinfo {year} {2001})},\ \Eprint
  {http://arxiv.org/abs/astro-ph/0103102} {arXiv:astro-ph/0103102} \BibitemShut
  {NoStop}%
\bibitem [{\citenamefont {de~Blok}(2005)}]{deblok05}%
  \BibitemOpen
  \bibfield  {author} {\bibinfo {author} {\bibfnamefont {W.J.G.}\ \bibnamefont
  {de~Blok}},\ }\bibfield  {title} {\enquote {\bibinfo {title} {{Halo mass
  profiles and low surface brightness galaxies rotation curves}},}\ }\Doi
  {10.1086/496912} {\bibfield  {journal} {\bibinfo  {journal} {Astrophys. J.},\
  }\textbf {\bibinfo {volume} {634}},\ \bibinfo {pages} {227--238} (\bibinfo
  {year} {2005})},\ \Eprint {http://arxiv.org/abs/astro-ph/0506753}
  {arXiv:astro-ph/0506753} \BibitemShut {NoStop}%
\bibitem [{\citenamefont {Oh}\ \emph {et~al.}(2008)\citenamefont {Oh},
  \citenamefont {de~Blok}, \citenamefont {Walter}, \citenamefont {Brinks},\
  and\ \citenamefont {Kennicutt}}]{oh08}%
  \BibitemOpen
  \bibfield  {author} {\bibinfo {author} {\bibfnamefont {S.-H.}\ \bibnamefont
  {Oh}}, \bibinfo {author} {\bibfnamefont {W.J.G.}\ \bibnamefont {de~Blok}},
  \bibinfo {author} {\bibfnamefont {F.}~\bibnamefont {Walter}}, \bibinfo
  {author} {\bibfnamefont {E.}~\bibnamefont {Brinks}}, \ and\ \bibinfo {author}
  {\bibfnamefont {R.~C.}\ \bibnamefont {Kennicutt}},\ }\bibfield  {title}
  {\enquote {\bibinfo {title} {{High-resolution dark matter density profiles of
  THINGS dwarf galaxies: Correcting for non-circular motions}},}\ }\Doi
  {10.1088/0004-6256/136/6/2761} {\bibfield  {journal} {\bibinfo  {journal}
  {Astron. J.},\ }\textbf {\bibinfo {volume} {136}},\ \bibinfo {pages}
  {2761--2781} (\bibinfo {year} {2008})},\ \Eprint
  {http://arxiv.org/abs/0810.2119} {arXiv:0810.2119} \BibitemShut {NoStop}%
\bibitem [{\citenamefont {de~Blok}(2010)}]{deblok09}%
  \BibitemOpen
  \bibfield  {author} {\bibinfo {author} {\bibfnamefont {W.J.G.}\ \bibnamefont
  {de~Blok}},\ }\bibfield  {title} {\enquote {\bibinfo {title} {{The core-cusp
  problem}},}\ }\Doi {10.1155/2010/789293} {\bibfield  {journal} {\bibinfo
  {journal} {Adv. Astron.},\ }\textbf {\bibinfo {volume} {2010}},\ \bibinfo
  {pages} {789293} (\bibinfo {year} {2010})},\ \Eprint
  {http://arxiv.org/abs/0910.3538} {arXiv:0910.3538} \BibitemShut {NoStop}%
\bibitem [{\citenamefont {Khlopov}\ \emph {et~al.}(1985)\citenamefont
  {Khlopov}, \citenamefont {Malomed},\ and\ \citenamefont {Zeldovich}}]{kmz85}%
  \BibitemOpen
  \bibfield  {author} {\bibinfo {author} {\bibfnamefont {M.}~\bibnamefont
  {Khlopov}}, \bibinfo {author} {\bibfnamefont {B.~A.}\ \bibnamefont
  {Malomed}}, \ and\ \bibinfo {author} {\bibfnamefont {Ia.~B.}\ \bibnamefont
  {Zeldovich}},\ }\bibfield  {title} {\enquote {\bibinfo {title}
  {{Gravitational instability of scalar fields and formation of primordial
  black holes}},}\ }\href@noop {} {\bibfield  {journal} {\bibinfo  {journal}
  {\mnras},\ }\textbf {\bibinfo {volume} {215}},\ \bibinfo {pages} {575--589}
  (\bibinfo {year} {1985})}\BibitemShut {NoStop}%
\bibitem [{\citenamefont {{Bianchi}}\ \emph {et~al.}(1990)\citenamefont
  {{Bianchi}}, \citenamefont {{Grasso}},\ and\ \citenamefont
  {{Ruffini}}}]{bianchi_1990}%
  \BibitemOpen
  \bibfield  {author} {\bibinfo {author} {\bibfnamefont {M.}~\bibnamefont
  {{Bianchi}}}, \bibinfo {author} {\bibfnamefont {D.}~\bibnamefont {{Grasso}}},
  \ and\ \bibinfo {author} {\bibfnamefont {R.}~\bibnamefont {{Ruffini}}},\
  }\bibfield  {title} {\enquote {\bibinfo {title} {{Jeans mass of a
  cosmological coherent scalar field}},}\ }\href@noop {} {\bibfield  {journal}
  {\bibinfo  {journal} {\aap},\ }\textbf {\bibinfo {volume} {231}},\ \bibinfo
  {pages} {301--308} (\bibinfo {year} {1990})}\BibitemShut {NoStop}%
\bibitem [{\citenamefont {{Johnson}}\ and\ \citenamefont
  {{Kamionkowski}}(2008)}]{jk2008}%
  \BibitemOpen
  \bibfield  {author} {\bibinfo {author} {\bibfnamefont {M.~C.}\ \bibnamefont
  {{Johnson}}}\ and\ \bibinfo {author} {\bibfnamefont {M.}~\bibnamefont
  {{Kamionkowski}}},\ }\bibfield  {title} {\enquote {\bibinfo {title}
  {{Dynamical and gravitational instability of an oscillating-field dark energy
  and dark matter}},}\ }\Doi {10.1103/PhysRevD.78.063010} {\bibfield  {journal}
  {\bibinfo  {journal} {\prd},\ }\textbf {\bibinfo {volume} {78}},\ \bibinfo
  {eid} {063010} (\bibinfo {year} {2008})},\ \Eprint
  {http://arxiv.org/abs/0805.1748} {arXiv:0805.1748} \BibitemShut {NoStop}%
\bibitem [{\citenamefont {{Lee}}\ and\ \citenamefont {{Lim}}(2010)}]{Lee2010}%
  \BibitemOpen
  \bibfield  {author} {\bibinfo {author} {\bibfnamefont {J.-W.}\ \bibnamefont
  {{Lee}}}\ and\ \bibinfo {author} {\bibfnamefont {S.}~\bibnamefont {{Lim}}},\
  }\bibfield  {title} {\enquote {\bibinfo {title} {{Minimum mass of galaxies
  from BEC or scalar field dark matter}},}\ }\Doi
  {10.1088/1475-7516/2010/01/007} {\bibfield  {journal} {\bibinfo  {journal}
  {JCAP},\ }\textbf {\bibinfo {volume} {1}},\ \bibinfo {eid} {007} (\bibinfo
  {year} {2010})},\ \Eprint {http://arxiv.org/abs/0812.1342} {arXiv:0812.1342}
  \BibitemShut {NoStop}%
\bibitem [{\citenamefont {Kaup}(1968)}]{kaup68}%
  \BibitemOpen
  \bibfield  {author} {\bibinfo {author} {\bibfnamefont {D.~J.}\ \bibnamefont
  {Kaup}},\ }\bibfield  {title} {\enquote {\bibinfo {title} {Klein--{G}ordon
  geon},}\ }\Doi {10.1103/PhysRev.172.1331} {\bibfield  {journal} {\bibinfo
  {journal} {Phys. Rev.},\ }\textbf {\bibinfo {volume} {172}},\ \bibinfo
  {pages} {1331--1342} (\bibinfo {year} {1968})}\BibitemShut {NoStop}%
\bibitem [{\citenamefont {Ruffini}\ and\ \citenamefont
  {Bonazzola}(1969)}]{rb69}%
  \BibitemOpen
  \bibfield  {author} {\bibinfo {author} {\bibfnamefont {R.}~\bibnamefont
  {Ruffini}}\ and\ \bibinfo {author} {\bibfnamefont {S.}~\bibnamefont
  {Bonazzola}},\ }\bibfield  {title} {\enquote {\bibinfo {title} {{Systems of
  self-gravitating particles in general relativity and the concept of an
  equation of state}},}\ }\Doi {10.1103/PhysRev.187.1767} {\bibfield  {journal}
  {\bibinfo  {journal} {Phys. Rev.},\ }\textbf {\bibinfo {volume} {187}},\
  \bibinfo {pages} {1767--1783} (\bibinfo {year} {1969})}\BibitemShut {NoStop}%
\bibitem [{\citenamefont {{Helfer}}\ \emph {et~al.}(2016)\citenamefont
  {{Helfer}}, \citenamefont {{Marsh}}, \citenamefont {{Clough}}, \citenamefont
  {{Fairbairn}}, \citenamefont {{Lim}},\ and\ \citenamefont
  {{Becerril}}}]{helfer2016}%
  \BibitemOpen
  \bibfield  {author} {\bibinfo {author} {\bibfnamefont {T.}~\bibnamefont
  {{Helfer}}}, \bibinfo {author} {\bibfnamefont {D.J.E.}\ \bibnamefont
  {{Marsh}}}, \bibinfo {author} {\bibfnamefont {K.}~\bibnamefont {{Clough}}},
  \bibinfo {author} {\bibfnamefont {M.}~\bibnamefont {{Fairbairn}}}, \bibinfo
  {author} {\bibfnamefont {E.~A.}\ \bibnamefont {{Lim}}}, \ and\ \bibinfo
  {author} {\bibfnamefont {R.}~\bibnamefont {{Becerril}}},\ }\bibfield  {title}
  {\enquote {\bibinfo {title} {{Black hole formation from axion stars}},}\
  }\href@noop {} {\bibfield  {journal} {\bibinfo  {journal} {ArXiv e-prints}}
  (\bibinfo {year} {2016})},\ \Eprint {http://arxiv.org/abs/1609.04724}
  {arXiv:1609.04724} \BibitemShut {NoStop}%
\bibitem [{\citenamefont {Eby}\ \emph {et~al.}(2016){\natexlab{a}}\citenamefont
  {Eby}, \citenamefont {Kouvaris}, \citenamefont {Nielsen},\ and\ \citenamefont
  {Wijewardhana}}]{Eby1}%
  \BibitemOpen
  \bibfield  {author} {\bibinfo {author} {\bibfnamefont {J.}~\bibnamefont
  {Eby}}, \bibinfo {author} {\bibfnamefont {C.}~\bibnamefont {Kouvaris}},
  \bibinfo {author} {\bibfnamefont {N.~G.}\ \bibnamefont {Nielsen}}, \ and\
  \bibinfo {author} {\bibfnamefont {L.C.R.}\ \bibnamefont {Wijewardhana}},\
  }\bibfield  {title} {\enquote {\bibinfo {title} {{Boson stars from
  self-interacting dark matter}},}\ }\Doi {10.1007/JHEP02(2016)028} {\bibfield
  {journal} {\bibinfo  {journal} {JHEP},\ }\textbf {\bibinfo {volume} {02}},\
  \bibinfo {pages} {028} (\bibinfo {year} {2016}{\natexlab{a}})},\ \Eprint
  {http://arxiv.org/abs/1511.04474} {arXiv:1511.04474} \BibitemShut {NoStop}%
\bibitem [{\citenamefont {Eby}\ \emph {et~al.}(2016){\natexlab{b}}\citenamefont
  {Eby}, \citenamefont {Suranyi},\ and\ \citenamefont {Wijewardhana}}]{Eby2}%
  \BibitemOpen
  \bibfield  {author} {\bibinfo {author} {\bibfnamefont {J.}~\bibnamefont
  {Eby}}, \bibinfo {author} {\bibfnamefont {P.}~\bibnamefont {Suranyi}}, \ and\
  \bibinfo {author} {\bibfnamefont {L.C.R.}\ \bibnamefont {Wijewardhana}},\
  }\bibfield  {title} {\enquote {\bibinfo {title} {{The lifetime of axion
  stars}},}\ }\Doi {10.1142/S0217732316500905} {\bibfield  {journal} {\bibinfo
  {journal} {Mod. Phys. Lett.},\ }\textbf {\bibinfo {volume} {A31}},\ \bibinfo
  {pages} {1650090} (\bibinfo {year} {2016}{\natexlab{b}})},\ \Eprint
  {http://arxiv.org/abs/1512.01709} {arXiv:1512.01709} \BibitemShut {NoStop}%
\bibitem [{\citenamefont {Klypin}\ \emph {et~al.}(1999)\citenamefont {Klypin},
  \citenamefont {Kravtsov}, \citenamefont {Valenzuela},\ and\ \citenamefont
  {Prada}}]{kly99}%
  \BibitemOpen
  \bibfield  {author} {\bibinfo {author} {\bibfnamefont {A.~A.}\ \bibnamefont
  {Klypin}}, \bibinfo {author} {\bibfnamefont {A.~V.}\ \bibnamefont
  {Kravtsov}}, \bibinfo {author} {\bibfnamefont {O.}~\bibnamefont
  {Valenzuela}}, \ and\ \bibinfo {author} {\bibfnamefont {F.}~\bibnamefont
  {Prada}},\ }\bibfield  {title} {\enquote {\bibinfo {title} {{Where are the
  missing Galactic satellites?}}}\ }\Doi {10.1086/307643} {\bibfield  {journal}
  {\bibinfo  {journal} {Astrophys. J.},\ }\textbf {\bibinfo {volume} {522}},\
  \bibinfo {pages} {82--92} (\bibinfo {year} {1999})},\ \Eprint
  {http://arxiv.org/abs/astro-ph/9901240} {arXiv:astro-ph/9901240} \BibitemShut
  {NoStop}%
\bibitem [{\citenamefont {Moore}\ \emph {et~al.}(1999)\citenamefont {Moore},
  \citenamefont {Ghigna}, \citenamefont {Governato}, \citenamefont {Lake},
  \citenamefont {Quinn}, \citenamefont {Stadel},\ and\ \citenamefont
  {Tozzi}}]{moore99}%
  \BibitemOpen
  \bibfield  {author} {\bibinfo {author} {\bibfnamefont {B.}~\bibnamefont
  {Moore}}, \bibinfo {author} {\bibfnamefont {S.}~\bibnamefont {Ghigna}},
  \bibinfo {author} {\bibfnamefont {F.}~\bibnamefont {Governato}}, \bibinfo
  {author} {\bibfnamefont {G.}~\bibnamefont {Lake}}, \bibinfo {author}
  {\bibfnamefont {Thomas~R.}\ \bibnamefont {Quinn}}, \bibinfo {author}
  {\bibfnamefont {J.}~\bibnamefont {Stadel}}, \ and\ \bibinfo {author}
  {\bibfnamefont {P.}~\bibnamefont {Tozzi}},\ }\bibfield  {title} {\enquote
  {\bibinfo {title} {{Dark matter substructure within galactic halos}},}\ }\Doi
  {10.1086/312287} {\bibfield  {journal} {\bibinfo  {journal} {Astrophys. J.},\
  }\textbf {\bibinfo {volume} {524}},\ \bibinfo {pages} {L19--L22} (\bibinfo
  {year} {1999})},\ \Eprint {http://arxiv.org/abs/astro-ph/9907411}
  {arXiv:astro-ph/9907411} \BibitemShut {NoStop}%
\bibitem [{\citenamefont {{Wetzel}}\ \emph {et~al.}(2016)\citenamefont
  {{Wetzel}}, \citenamefont {{Hopkins}}, \citenamefont {{Kim}}, \citenamefont
  {{Faucher-Gigu{\`e}re}}, \citenamefont {{Kere{\v s}}},\ and\ \citenamefont
  {{Quataert}}}]{wetzel2016}%
  \BibitemOpen
  \bibfield  {author} {\bibinfo {author} {\bibfnamefont {A.~R.}\ \bibnamefont
  {{Wetzel}}}, \bibinfo {author} {\bibfnamefont {P.~F.}\ \bibnamefont
  {{Hopkins}}}, \bibinfo {author} {\bibfnamefont {J.-H.}\ \bibnamefont
  {{Kim}}}, \bibinfo {author} {\bibfnamefont {C.-A.}\ \bibnamefont
  {{Faucher-Gigu{\`e}re}}}, \bibinfo {author} {\bibfnamefont {D.}~\bibnamefont
  {{Kere{\v s}}}}, \ and\ \bibinfo {author} {\bibfnamefont {E.}~\bibnamefont
  {{Quataert}}},\ }\bibfield  {title} {\enquote {\bibinfo {title} {{Reconciling
  dwarf galaxies with {$\Lambda$}CDM cosmology: simulating a realistic
  population of satellites around a Milky Way--mass galaxy}},}\ }\Doi
  {10.3847/2041-8205/827/2/L23} {\bibfield  {journal} {\bibinfo  {journal}
  {\apjl},\ }\textbf {\bibinfo {volume} {827}},\ \bibinfo {eid} {L23} (\bibinfo
  {year} {2016})},\ \Eprint {http://arxiv.org/abs/1602.05957}
  {arXiv:1602.05957} \BibitemShut {NoStop}%
\bibitem [{\citenamefont {Nierenberg}\ \emph {et~al.}(2016)\citenamefont
  {Nierenberg}, \citenamefont {Treu}, \citenamefont {Menci}, \citenamefont
  {Lu}, \citenamefont {Torrey},\ and\ \citenamefont {Vogelsberger}}]{nier16}%
  \BibitemOpen
  \bibfield  {author} {\bibinfo {author} {\bibfnamefont {A.~M.}\ \bibnamefont
  {Nierenberg}}, \bibinfo {author} {\bibfnamefont {T.}~\bibnamefont {Treu}},
  \bibinfo {author} {\bibfnamefont {N.}~\bibnamefont {Menci}}, \bibinfo
  {author} {\bibfnamefont {Y.}~\bibnamefont {Lu}}, \bibinfo {author}
  {\bibfnamefont {P.}~\bibnamefont {Torrey}}, \ and\ \bibinfo {author}
  {\bibfnamefont {M.}~\bibnamefont {Vogelsberger}},\ }\bibfield  {title}
  {\enquote {\bibinfo {title} {{The missing satellite problem in 3D}},}\
  }\href@noop {} { (\bibinfo {year} {2016})},\ \Eprint
  {http://arxiv.org/abs/1603.01614} {arXiv:1603.01614} \BibitemShut {NoStop}%
\bibitem [{\citenamefont {Mao}\ and\ \citenamefont {Schneider}(1998)}]{ms98}%
  \BibitemOpen
  \bibfield  {author} {\bibinfo {author} {\bibfnamefont {S.}~\bibnamefont
  {Mao}}\ and\ \bibinfo {author} {\bibfnamefont {P.}~\bibnamefont
  {Schneider}},\ }\bibfield  {title} {\enquote {\bibinfo {title} {{Evidence for
  substructure in lens galaxies?}}}\ }\Doi {10.1046/j.1365-8711.1998.01319.x}
  {\bibfield  {journal} {\bibinfo  {journal} {\mnras},\ }\textbf {\bibinfo
  {volume} {295}},\ \bibinfo {pages} {587--594} (\bibinfo {year} {1998})},\
  \Eprint {http://arxiv.org/abs/astro-ph/9707187} {arXiv:astro-ph/9707187}
  \BibitemShut {NoStop}%
\bibitem [{\citenamefont {{Li}}\ \emph {et~al.}(2016)\citenamefont {{Li}},
  \citenamefont {{Frenk}}, \citenamefont {{Cole}}, \citenamefont {{Gao}},
  \citenamefont {{Bose}},\ and\ \citenamefont {{Hellwing}}}]{li16}%
  \BibitemOpen
  \bibfield  {author} {\bibinfo {author} {\bibfnamefont {R.}~\bibnamefont
  {{Li}}}, \bibinfo {author} {\bibfnamefont {C.~S.}\ \bibnamefont {{Frenk}}},
  \bibinfo {author} {\bibfnamefont {S.}~\bibnamefont {{Cole}}}, \bibinfo
  {author} {\bibfnamefont {L.}~\bibnamefont {{Gao}}}, \bibinfo {author}
  {\bibfnamefont {S.}~\bibnamefont {{Bose}}}, \ and\ \bibinfo {author}
  {\bibfnamefont {W.~A.}\ \bibnamefont {{Hellwing}}},\ }\bibfield  {title}
  {\enquote {\bibinfo {title} {{Constraints on the identity of the dark matter
  from strong gravitational lenses}},}\ }\Doi {10.1093/mnras/stw939} {\bibfield
   {journal} {\bibinfo  {journal} {\mnras},\ }\textbf {\bibinfo {volume}
  {460}},\ \bibinfo {pages} {363--372} (\bibinfo {year} {2016})},\ \Eprint
  {http://arxiv.org/abs/1512.06507} {arXiv:1512.06507} \BibitemShut {NoStop}%
\bibitem [{\citenamefont {Hezaveh}\ \emph {et~al.}(2016)\citenamefont {Hezaveh}
  \emph {et~al.}}]{hez16}%
  \BibitemOpen
  \bibfield  {author} {\bibinfo {author} {\bibfnamefont {Y.~D.}\ \bibnamefont
  {Hezaveh}} \emph {et~al.},\ }\bibfield  {title} {\enquote {\bibinfo {title}
  {{Detection of lensing substructure using ALMA observations of the dusty
  galaxy SDP.81}},}\ }\Doi {10.3847/0004-637X/823/1/37} {\bibfield  {journal}
  {\bibinfo  {journal} {Astrophys. J.},\ }\textbf {\bibinfo {volume} {823}},\
  \bibinfo {pages} {37} (\bibinfo {year} {2016})},\ \Eprint
  {http://arxiv.org/abs/1601.01388} {arXiv:1601.01388} \BibitemShut {NoStop}%
\bibitem [{\citenamefont {{Read}}\ \emph {et~al.}(2006)\citenamefont {{Read}},
  \citenamefont {{Wilkinson}}, \citenamefont {{Evans}}, \citenamefont
  {{Gilmore}},\ and\ \citenamefont {{Kleyna}}}]{read2006}%
  \BibitemOpen
  \bibfield  {author} {\bibinfo {author} {\bibfnamefont {J.~I.}\ \bibnamefont
  {{Read}}}, \bibinfo {author} {\bibfnamefont {M.~I.}\ \bibnamefont
  {{Wilkinson}}}, \bibinfo {author} {\bibfnamefont {N.~W.}\ \bibnamefont
  {{Evans}}}, \bibinfo {author} {\bibfnamefont {G.}~\bibnamefont {{Gilmore}}},
  \ and\ \bibinfo {author} {\bibfnamefont {J.~T.}\ \bibnamefont {{Kleyna}}},\
  }\bibfield  {title} {\enquote {\bibinfo {title} {{The importance of tides for
  the Local Group dwarf spheroidals}},}\ }\Doi
  {10.1111/j.1365-2966.2005.09959.x} {\bibfield  {journal} {\bibinfo  {journal}
  {\mnras},\ }\textbf {\bibinfo {volume} {367}},\ \bibinfo {pages} {387--399}
  (\bibinfo {year} {2006})},\ \Eprint {http://arxiv.org/abs/astro-ph/0511759}
  {arXiv:astro-ph/0511759} \BibitemShut {NoStop}%
\bibitem [{\citenamefont {{Boylan-Kolchin}}\ \emph {et~al.}(2011)\citenamefont
  {{Boylan-Kolchin}}, \citenamefont {{Bullock}},\ and\ \citenamefont
  {{Kaplinghat}}}]{boylin2011}%
  \BibitemOpen
  \bibfield  {author} {\bibinfo {author} {\bibfnamefont {M.}~\bibnamefont
  {{Boylan-Kolchin}}}, \bibinfo {author} {\bibfnamefont {J.~S.}\ \bibnamefont
  {{Bullock}}}, \ and\ \bibinfo {author} {\bibfnamefont {M.}~\bibnamefont
  {{Kaplinghat}}},\ }\bibfield  {title} {\enquote {\bibinfo {title} {{Too big
  to fail? The puzzling darkness of massive Milky Way subhaloes}},}\ }\Doi
  {10.1111/j.1745-3933.2011.01074.x} {\bibfield  {journal} {\bibinfo  {journal}
  {\mnras},\ }\textbf {\bibinfo {volume} {415}},\ \bibinfo {pages} {L40--L44}
  (\bibinfo {year} {2011})},\ \Eprint {http://arxiv.org/abs/1103.0007}
  {arXiv:1103.0007} \BibitemShut {NoStop}%
\bibitem [{\citenamefont {{Marsh}}\ and\ \citenamefont
  {{Silk}}(2014)}]{ms2014}%
  \BibitemOpen
  \bibfield  {author} {\bibinfo {author} {\bibfnamefont {D.J.E.}\ \bibnamefont
  {{Marsh}}}\ and\ \bibinfo {author} {\bibfnamefont {J.}~\bibnamefont
  {{Silk}}},\ }\bibfield  {title} {\enquote {\bibinfo {title} {{A model for
  halo formation with axion mixed dark matter}},}\ }\Doi
  {10.1093/mnras/stt2079} {\bibfield  {journal} {\bibinfo  {journal} {\mnras},\
  }\textbf {\bibinfo {volume} {437}},\ \bibinfo {pages} {2652--2663} (\bibinfo
  {year} {2014})},\ \Eprint {http://arxiv.org/abs/1307.1705} {arXiv:1307.1705}
  \BibitemShut {NoStop}%
\bibitem [{\citenamefont {{Renzini}}(2008)}]{renzini2008}%
  \BibitemOpen
  \bibfield  {author} {\bibinfo {author} {\bibfnamefont {A.}~\bibnamefont
  {{Renzini}}},\ }\bibfield  {title} {\enquote {\bibinfo {title} {{Origin of
  multiple stellar populations in globular clusters and their helium
  enrichment}},}\ }\Doi {10.1111/j.1365-2966.2008.13892.x} {\bibfield
  {journal} {\bibinfo  {journal} {\mnras},\ }\textbf {\bibinfo {volume}
  {391}},\ \bibinfo {pages} {354--362} (\bibinfo {year} {2008})},\ \Eprint
  {http://arxiv.org/abs/0808.4095} {arXiv:0808.4095} \BibitemShut {NoStop}%
\bibitem [{\citenamefont {{Peebles}}\ and\ \citenamefont
  {{Dicke}}(1968)}]{pd1968}%
  \BibitemOpen
  \bibfield  {author} {\bibinfo {author} {\bibfnamefont {P.J.E.}\ \bibnamefont
  {{Peebles}}}\ and\ \bibinfo {author} {\bibfnamefont {R.~H.}\ \bibnamefont
  {{Dicke}}},\ }\bibfield  {title} {\enquote {\bibinfo {title} {{Origin of the
  globular star clusters}},}\ }\Doi {10.1086/149811} {\bibfield  {journal}
  {\bibinfo  {journal} {\apj},\ }\textbf {\bibinfo {volume} {154}},\ \bibinfo
  {pages} {891} (\bibinfo {year} {1968})}\BibitemShut {NoStop}%
\bibitem [{\citenamefont {{Peebles}}(1984)}]{peebles1984}%
  \BibitemOpen
  \bibfield  {author} {\bibinfo {author} {\bibfnamefont {P.J.E.}\ \bibnamefont
  {{Peebles}}},\ }\bibfield  {title} {\enquote {\bibinfo {title} {{Dark matter
  and the origin of galaxies and globular star clusters}},}\ }\Doi
  {10.1086/161714} {\bibfield  {journal} {\bibinfo  {journal} {\apj},\ }\textbf
  {\bibinfo {volume} {277}},\ \bibinfo {pages} {470--477} (\bibinfo {year}
  {1984})}\BibitemShut {NoStop}%
\bibitem [{\citenamefont {{Conroy}}\ \emph {et~al.}(2011)\citenamefont
  {{Conroy}}, \citenamefont {{Loeb}},\ and\ \citenamefont
  {{Spergel}}}]{conroy2011}%
  \BibitemOpen
  \bibfield  {author} {\bibinfo {author} {\bibfnamefont {C.}~\bibnamefont
  {{Conroy}}}, \bibinfo {author} {\bibfnamefont {A.}~\bibnamefont {{Loeb}}}, \
  and\ \bibinfo {author} {\bibfnamefont {D.~N.}\ \bibnamefont {{Spergel}}},\
  }\bibfield  {title} {\enquote {\bibinfo {title} {{Evidence against dark
  matter halos surrounding the globular clusters MGC1 and NGC 2419}},}\ }\Doi
  {10.1088/0004-637X/741/2/72} {\bibfield  {journal} {\bibinfo  {journal}
  {\apj},\ }\textbf {\bibinfo {volume} {741}},\ \bibinfo {eid} {72} (\bibinfo
  {year} {2011})},\ \Eprint {http://arxiv.org/abs/1010.5783} {arXiv:1010.5783}
  \BibitemShut {NoStop}%
\bibitem [{\citenamefont {{Ibata}}\ \emph {et~al.}(2013)\citenamefont
  {{Ibata}}, \citenamefont {{Nipoti}}, \citenamefont {{Sollima}}, \citenamefont
  {{Bellazzini}}, \citenamefont {{Chapman}},\ and\ \citenamefont
  {{Dalessandro}}}]{ibata2013}%
  \BibitemOpen
  \bibfield  {author} {\bibinfo {author} {\bibfnamefont {R.}~\bibnamefont
  {{Ibata}}}, \bibinfo {author} {\bibfnamefont {C.}~\bibnamefont {{Nipoti}}},
  \bibinfo {author} {\bibfnamefont {A.}~\bibnamefont {{Sollima}}}, \bibinfo
  {author} {\bibfnamefont {M.}~\bibnamefont {{Bellazzini}}}, \bibinfo {author}
  {\bibfnamefont {S.~C.}\ \bibnamefont {{Chapman}}}, \ and\ \bibinfo {author}
  {\bibfnamefont {E.}~\bibnamefont {{Dalessandro}}},\ }\bibfield  {title}
  {\enquote {\bibinfo {title} {{Do globular clusters possess dark matter
  haloes? A case study in NGC 2419}},}\ }\Doi {10.1093/mnras/sts302} {\bibfield
   {journal} {\bibinfo  {journal} {\mnras},\ }\textbf {\bibinfo {volume}
  {428}},\ \bibinfo {pages} {3648--3659} (\bibinfo {year} {2013})},\ \Eprint
  {http://arxiv.org/abs/1210.7787} {arXiv:1210.7787} \BibitemShut {NoStop}%
\bibitem [{\citenamefont {{Diakogiannis}}\ \emph {et~al.}(2014)\citenamefont
  {{Diakogiannis}}, \citenamefont {{Lewis}},\ and\ \citenamefont
  {{Ibata}}}]{dia2014}%
  \BibitemOpen
  \bibfield  {author} {\bibinfo {author} {\bibfnamefont {F.~I.}\ \bibnamefont
  {{Diakogiannis}}}, \bibinfo {author} {\bibfnamefont {G.~F.}\ \bibnamefont
  {{Lewis}}}, \ and\ \bibinfo {author} {\bibfnamefont {R.~A.}\ \bibnamefont
  {{Ibata}}},\ }\bibfield  {title} {\enquote {\bibinfo {title} {{Dynamical
  modelling of NGC 6809: selecting the best model using Bayesian inference}},}\
  }\Doi {10.1093/mnras/stt2093} {\bibfield  {journal} {\bibinfo  {journal}
  {\mnras},\ }\textbf {\bibinfo {volume} {437}},\ \bibinfo {pages} {3172--3182}
  (\bibinfo {year} {2014})},\ \Eprint {http://arxiv.org/abs/1310.8096}
  {arXiv:1310.8096} \BibitemShut {NoStop}%
\bibitem [{\citenamefont {{Johnston}}\ and\ \citenamefont
  {{Carlberg}}(2016)}]{jc16}%
  \BibitemOpen
  \bibfield  {author} {\bibinfo {author} {\bibfnamefont {K.~V.}\ \bibnamefont
  {{Johnston}}}\ and\ \bibinfo {author} {\bibfnamefont {R.~G.}\ \bibnamefont
  {{Carlberg}}},\ }\bibfield  {title} {\enquote {\bibinfo {title} {{Tidal
  debris as a dark matter probe}},}\ }in\ \Doi {10.1007/978-3-319-19336-6_7}
  {\emph {\bibinfo {booktitle} {Astrophysics and Space Science Library}}},\
  Vol.\ \bibinfo {volume} {420},\ \bibinfo {editor} {edited by\ \bibinfo
  {editor} {\bibfnamefont {H.~J.}\ \bibnamefont {{Newberg}}}\ and\ \bibinfo
  {editor} {\bibfnamefont {J.~L.}\ \bibnamefont {{Carlin}}}}\ (\bibinfo {year}
  {2016})\ p.\ \bibinfo {pages} {169},\ \Eprint
  {http://arxiv.org/abs/1603.06602} {arXiv:1603.06602} \BibitemShut {NoStop}%
\bibitem [{\citenamefont {{Chandrasekhar}}(1944)}]{chandra1944}%
  \BibitemOpen
  \bibfield  {author} {\bibinfo {author} {\bibfnamefont {S.}~\bibnamefont
  {{Chandrasekhar}}},\ }\bibfield  {title} {\enquote {\bibinfo {title} {{On the
  stability of binary systems}},}\ }\Doi {10.1086/144589} {\bibfield  {journal}
  {\bibinfo  {journal} {\apj},\ }\textbf {\bibinfo {volume} {99}},\ \bibinfo
  {pages} {54--58} (\bibinfo {year} {1944})}\BibitemShut {NoStop}%
\bibitem [{\citenamefont {{Bahcall}}\ \emph {et~al.}(1985)\citenamefont
  {{Bahcall}}, \citenamefont {{Hut}},\ and\ \citenamefont
  {{Tremaine}}}]{bht1985}%
  \BibitemOpen
  \bibfield  {author} {\bibinfo {author} {\bibfnamefont {J.~N.}\ \bibnamefont
  {{Bahcall}}}, \bibinfo {author} {\bibfnamefont {P.}~\bibnamefont {{Hut}}}, \
  and\ \bibinfo {author} {\bibfnamefont {S.}~\bibnamefont {{Tremaine}}},\
  }\bibfield  {title} {\enquote {\bibinfo {title} {{Maximum mass of objects
  that constitute unseen disk material}},}\ }\Doi {10.1086/162953} {\bibfield
  {journal} {\bibinfo  {journal} {\apj},\ }\textbf {\bibinfo {volume} {290}},\
  \bibinfo {pages} {15--20} (\bibinfo {year} {1985})}\BibitemShut {NoStop}%
\bibitem [{\citenamefont {{Johnston}}\ \emph {et~al.}(2002)\citenamefont
  {{Johnston}}, \citenamefont {{Spergel}},\ and\ \citenamefont
  {{Haydn}}}]{jsh2002}%
  \BibitemOpen
  \bibfield  {author} {\bibinfo {author} {\bibfnamefont {K.~V.}\ \bibnamefont
  {{Johnston}}}, \bibinfo {author} {\bibfnamefont {D.~N.}\ \bibnamefont
  {{Spergel}}}, \ and\ \bibinfo {author} {\bibfnamefont {C.}~\bibnamefont
  {{Haydn}}},\ }\bibfield  {title} {\enquote {\bibinfo {title} {{How lumpy Is
  the Milky Way's dark matter halo?}}}\ }\Doi {10.1086/339791} {\bibfield
  {journal} {\bibinfo  {journal} {\apj},\ }\textbf {\bibinfo {volume} {570}},\
  \bibinfo {pages} {656--664} (\bibinfo {year} {2002})},\ \Eprint
  {http://arxiv.org/abs/arXiv:astro-ph/0111196} {arXiv:astro-ph/0111196}
  \BibitemShut {NoStop}%
\bibitem [{\citenamefont {Yoon}\ \emph {et~al.}(2011)\citenamefont {Yoon},
  \citenamefont {Johnston},\ and\ \citenamefont {Hogg}}]{yoon11}%
  \BibitemOpen
  \bibfield  {author} {\bibinfo {author} {\bibfnamefont {J.~H.}\ \bibnamefont
  {Yoon}}, \bibinfo {author} {\bibfnamefont {K.~V.}\ \bibnamefont {Johnston}},
  \ and\ \bibinfo {author} {\bibfnamefont {D.~W.}\ \bibnamefont {Hogg}},\
  }\bibfield  {title} {\enquote {\bibinfo {title} {{Clumpy streams from clumpy
  halos: detecting missing satellites with cold stellar structures}},}\ }\Doi
  {10.1088/0004-637X/731/1/58} {\bibfield  {journal} {\bibinfo  {journal}
  {Astrophys. J.},\ }\textbf {\bibinfo {volume} {731}},\ \bibinfo {pages} {58}
  (\bibinfo {year} {2011})},\ \Eprint {http://arxiv.org/abs/1012.2884}
  {arXiv:1012.2884} \BibitemShut {NoStop}%
\bibitem [{\citenamefont {{Sanders}}\ \emph {et~al.}(2016)\citenamefont
  {{Sanders}}, \citenamefont {{Bovy}},\ and\ \citenamefont {{Erkal}}}]{sbe16}%
  \BibitemOpen
  \bibfield  {author} {\bibinfo {author} {\bibfnamefont {J.~L.}\ \bibnamefont
  {{Sanders}}}, \bibinfo {author} {\bibfnamefont {J.}~\bibnamefont {{Bovy}}}, \
  and\ \bibinfo {author} {\bibfnamefont {D.}~\bibnamefont {{Erkal}}},\
  }\bibfield  {title} {\enquote {\bibinfo {title} {{Dynamics of stream-subhalo
  interactions}},}\ }\Doi {10.1093/mnras/stw232} {\bibfield  {journal}
  {\bibinfo  {journal} {\mnras},\ }\textbf {\bibinfo {volume} {457}},\ \bibinfo
  {pages} {3817--3835} (\bibinfo {year} {2016})},\ \Eprint
  {http://arxiv.org/abs/1510.03426} {arXiv:1510.03426} \BibitemShut {NoStop}%
\bibitem [{\citenamefont {{Erkal}}\ \emph {et~al.}(2016)\citenamefont
  {{Erkal}}, \citenamefont {{Belokurov}}, \citenamefont {{Bovy}},\ and\
  \citenamefont {{Sanders}}}]{erkal16}%
  \BibitemOpen
  \bibfield  {author} {\bibinfo {author} {\bibfnamefont {D.}~\bibnamefont
  {{Erkal}}}, \bibinfo {author} {\bibfnamefont {V.}~\bibnamefont
  {{Belokurov}}}, \bibinfo {author} {\bibfnamefont {J.}~\bibnamefont {{Bovy}}},
  \ and\ \bibinfo {author} {\bibfnamefont {J.~L.}\ \bibnamefont {{Sanders}}},\
  }\bibfield  {title} {\enquote {\bibinfo {title} {{The number and size of
  subhalo-induced gaps in stellar streams}},}\ }\Doi {10.1093/mnras/stw1957}
  {\bibfield  {journal} {\bibinfo  {journal} {\mnras},\ }\textbf {\bibinfo
  {volume} {463}},\ \bibinfo {pages} {102--119} (\bibinfo {year} {2016})},\
  \Eprint {http://arxiv.org/abs/1606.04946} {arXiv:1606.04946} \BibitemShut
  {NoStop}%
\bibitem [{\citenamefont {Carlberg}(2009)}]{carl09}%
  \BibitemOpen
  \bibfield  {author} {\bibinfo {author} {\bibfnamefont {R.~G.}\ \bibnamefont
  {Carlberg}},\ }\bibfield  {title} {\enquote {\bibinfo {title} {{Star stream
  folding by dark galactic sub-halos}},}\ }\Doi {10.1088/0004-637X/705/2/L223}
  {\bibfield  {journal} {\bibinfo  {journal} {Astrophys. J.},\ }\textbf
  {\bibinfo {volume} {705}},\ \bibinfo {pages} {L223--L226} (\bibinfo {year}
  {2009})},\ \Eprint {http://arxiv.org/abs/0908.4345} {arXiv:0908.4345}
  \BibitemShut {NoStop}%
\bibitem [{\citenamefont {{Erkal}}\ and\ \citenamefont
  {{Belokurov}}(2015)}]{erkal15}%
  \BibitemOpen
  \bibfield  {author} {\bibinfo {author} {\bibfnamefont {D.}~\bibnamefont
  {{Erkal}}}\ and\ \bibinfo {author} {\bibfnamefont {V.}~\bibnamefont
  {{Belokurov}}},\ }\bibfield  {title} {\enquote {\bibinfo {title} {{Forensics
  of subhalo-stream encounters: the three phases of gap growth}},}\ }\Doi
  {10.1093/mnras/stv655} {\bibfield  {journal} {\bibinfo  {journal} {\mnras},\
  }\textbf {\bibinfo {volume} {450}},\ \bibinfo {pages} {1136--1149} (\bibinfo
  {year} {2015})},\ \Eprint {http://arxiv.org/abs/1412.6035} {arXiv:1412.6035}
  \BibitemShut {NoStop}%
\bibitem [{\citenamefont {Amorisco}\ \emph {et~al.}(2016)\citenamefont
  {Amorisco}, \citenamefont {G\`omez}, \citenamefont {Vegetti},\ and\
  \citenamefont {White}}]{agvw16}%
  \BibitemOpen
  \bibfield  {author} {\bibinfo {author} {\bibfnamefont {N.~C.}\ \bibnamefont
  {Amorisco}}, \bibinfo {author} {\bibfnamefont {F.~A.}\ \bibnamefont
  {G\`omez}}, \bibinfo {author} {\bibfnamefont {S.}~\bibnamefont {Vegetti}}, \
  and\ \bibinfo {author} {\bibfnamefont {S.D.M.}\ \bibnamefont {White}},\
  }\bibfield  {title} {\enquote {\bibinfo {title} {{Gaps in globular cluster
  streams: giant molecular clouds can cause them too}},}\ }\href@noop {} {
  (\bibinfo {year} {2016})},\ \Eprint {http://arxiv.org/abs/1606.02715}
  {arXiv:1606.02715} \BibitemShut {NoStop}%
\bibitem [{\citenamefont {{Diemand}}\ \emph {et~al.}(2007)\citenamefont
  {{Diemand}}, \citenamefont {{Kuhlen}},\ and\ \citenamefont
  {{Madau}}}]{diemand07}%
  \BibitemOpen
  \bibfield  {author} {\bibinfo {author} {\bibfnamefont {J.}~\bibnamefont
  {{Diemand}}}, \bibinfo {author} {\bibfnamefont {M.}~\bibnamefont {{Kuhlen}}},
  \ and\ \bibinfo {author} {\bibfnamefont {P.}~\bibnamefont {{Madau}}},\
  }\bibfield  {title} {\enquote {\bibinfo {title} {{Dark matter substructure
  and gamma-ray annihilation in the Milky Way halo}},}\ }\Doi {10.1086/510736}
  {\bibfield  {journal} {\bibinfo  {journal} {\apj},\ }\textbf {\bibinfo
  {volume} {657}},\ \bibinfo {pages} {262--270} (\bibinfo {year} {2007})},\
  \Eprint {http://arxiv.org/abs/arXiv:astro-ph/0611370}
  {arXiv:astro-ph/0611370} \BibitemShut {NoStop}%
\bibitem [{\citenamefont {Ngan}\ \emph {et~al.}(2016)\citenamefont {Ngan},
  \citenamefont {Carlberg}, \citenamefont {Bozek}, \citenamefont {Wyse},
  \citenamefont {Szalay},\ and\ \citenamefont {Madau}}]{ngan16}%
  \BibitemOpen
  \bibfield  {author} {\bibinfo {author} {\bibfnamefont {W.}~\bibnamefont
  {Ngan}}, \bibinfo {author} {\bibfnamefont {R.~G.}\ \bibnamefont {Carlberg}},
  \bibinfo {author} {\bibfnamefont {B.}~\bibnamefont {Bozek}}, \bibinfo
  {author} {\bibfnamefont {R.F.G.}\ \bibnamefont {Wyse}}, \bibinfo {author}
  {\bibfnamefont {A.~S.}\ \bibnamefont {Szalay}}, \ and\ \bibinfo {author}
  {\bibfnamefont {P.}~\bibnamefont {Madau}},\ }\bibfield  {title} {\enquote
  {\bibinfo {title} {{Dispersal of tidal debris in a Milky-Way-sized dark
  matter halo}},}\ }\Doi {10.3847/0004-637X/818/2/194} {\bibfield  {journal}
  {\bibinfo  {journal} {Astrophys. J.},\ }\textbf {\bibinfo {volume} {818}},\
  \bibinfo {pages} {194} (\bibinfo {year} {2016})},\ \Eprint
  {http://arxiv.org/abs/1601.04681} {arXiv:1601.04681} \BibitemShut {NoStop}%
\bibitem [{\citenamefont {Bovy}\ and\ \citenamefont {Rix}(2013)}]{br13}%
  \BibitemOpen
  \bibfield  {author} {\bibinfo {author} {\bibfnamefont {J.}~\bibnamefont
  {Bovy}}\ and\ \bibinfo {author} {\bibfnamefont {H.-W.}\ \bibnamefont {Rix}},\
  }\bibfield  {title} {\enquote {\bibinfo {title} {{A direct dynamical
  measurement of the Milky Way's disk surface density profile, disk scale
  length, and dark matter profile at $\kpc\lesssim R \lesssim 9\kpc$}},}\ }\Doi
  {10.1088/0004-637X/779/2/115} {\bibfield  {journal} {\bibinfo  {journal}
  {Astrophys. J.},\ }\textbf {\bibinfo {volume} {779}},\ \bibinfo {pages} {115}
  (\bibinfo {year} {2013})},\ \Eprint {http://arxiv.org/abs/1309.0809}
  {arXiv:1309.0809} \BibitemShut {NoStop}%
\bibitem [{\citenamefont {Holmberg}\ and\ \citenamefont {Flynn}(2000)}]{hf00}%
  \BibitemOpen
  \bibfield  {author} {\bibinfo {author} {\bibfnamefont {J.}~\bibnamefont
  {Holmberg}}\ and\ \bibinfo {author} {\bibfnamefont {C.}~\bibnamefont
  {Flynn}},\ }\bibfield  {title} {\enquote {\bibinfo {title} {{The local
  density of matter mapped by Hipparcos}},}\ }\Doi
  {10.1046/j.1365-8711.2000.02905.x} {\bibfield  {journal} {\bibinfo  {journal}
  {\mnras},\ }\textbf {\bibinfo {volume} {313}},\ \bibinfo {pages} {209--216}
  (\bibinfo {year} {2000})},\ \Eprint {http://arxiv.org/abs/astro-ph/9812404}
  {arXiv:astro-ph/9812404} \BibitemShut {NoStop}%
\bibitem [{\citenamefont {Bovy}\ and\ \citenamefont {Tremaine}(2012)}]{bt12}%
  \BibitemOpen
  \bibfield  {author} {\bibinfo {author} {\bibfnamefont {J.}~\bibnamefont
  {Bovy}}\ and\ \bibinfo {author} {\bibfnamefont {S.}~\bibnamefont
  {Tremaine}},\ }\bibfield  {title} {\enquote {\bibinfo {title} {{On the local
  dark matter density}},}\ }\Doi {10.1088/0004-637X/756/1/89} {\bibfield
  {journal} {\bibinfo  {journal} {Astrophys. J.},\ }\textbf {\bibinfo {volume}
  {756}},\ \bibinfo {pages} {89} (\bibinfo {year} {2012})},\ \Eprint
  {http://arxiv.org/abs/1205.4033} {arXiv:1205.4033} \BibitemShut {NoStop}%
\bibitem [{\citenamefont {Randall}\ and\ \citenamefont {Reece}(2014)}]{rr14}%
  \BibitemOpen
  \bibfield  {author} {\bibinfo {author} {\bibfnamefont {L.}~\bibnamefont
  {Randall}}\ and\ \bibinfo {author} {\bibfnamefont {M.}~\bibnamefont
  {Reece}},\ }\bibfield  {title} {\enquote {\bibinfo {title} {{Dark matter as a
  trigger for periodic comet impacts}},}\ }\Doi
  {10.1103/PhysRevLett.112.161301} {\bibfield  {journal} {\bibinfo  {journal}
  {Phys. Rev. Lett.},\ }\textbf {\bibinfo {volume} {112}},\ \bibinfo {pages}
  {161301} (\bibinfo {year} {2014})},\ \Eprint {http://arxiv.org/abs/1403.0576}
  {arXiv:1403.0576} \BibitemShut {NoStop}%
\bibitem [{\citenamefont {Chandrasekhar}(1943)}]{chandra43}%
  \BibitemOpen
  \bibfield  {author} {\bibinfo {author} {\bibfnamefont {S.}~\bibnamefont
  {Chandrasekhar}},\ }\bibfield  {title} {\enquote {\bibinfo {title}
  {{Dynamical friction. I. General considerations: the coefficient of dynamical
  friction}},}\ }\Doi {10.1086/144517} {\bibfield  {journal} {\bibinfo
  {journal} {Astrophys. J.},\ }\textbf {\bibinfo {volume} {97}},\ \bibinfo
  {pages} {255--262} (\bibinfo {year} {1943})}\BibitemShut {NoStop}%
\bibitem [{\citenamefont {{Tremaine}}(1976)}]{tre76}%
  \BibitemOpen
  \bibfield  {author} {\bibinfo {author} {\bibfnamefont {S.~D.}\ \bibnamefont
  {{Tremaine}}},\ }\bibfield  {title} {\enquote {\bibinfo {title} {{The
  formation of the nuclei of galaxies. II - The Local Group}},}\ }\Doi
  {10.1086/154085} {\bibfield  {journal} {\bibinfo  {journal} {\apj},\ }\textbf
  {\bibinfo {volume} {203}},\ \bibinfo {pages} {345--351} (\bibinfo {year}
  {1976})}\BibitemShut {NoStop}%
\bibitem [{\citenamefont {Lotz}\ \emph {et~al.}(2001)\citenamefont {Lotz},
  \citenamefont {Telford}, \citenamefont {Ferguson}, \citenamefont {Miller},
  \citenamefont {Stiavelli},\ and\ \citenamefont {Mack}}]{lotz01}%
  \BibitemOpen
  \bibfield  {author} {\bibinfo {author} {\bibfnamefont {J.~M.}\ \bibnamefont
  {Lotz}}, \bibinfo {author} {\bibfnamefont {R.}~\bibnamefont {Telford}},
  \bibinfo {author} {\bibfnamefont {H.~C.}\ \bibnamefont {Ferguson}}, \bibinfo
  {author} {\bibfnamefont {B.~W.}\ \bibnamefont {Miller}}, \bibinfo {author}
  {\bibfnamefont {M.}~\bibnamefont {Stiavelli}}, \ and\ \bibinfo {author}
  {\bibfnamefont {J.}~\bibnamefont {Mack}},\ }\bibfield  {title} {\enquote
  {\bibinfo {title} {{Dynamical friction in dE globular cluster systems}},}\
  }\Doi {10.1086/320545} {\bibfield  {journal} {\bibinfo  {journal} {Astrophys.
  J.},\ }\textbf {\bibinfo {volume} {552}},\ \bibinfo {pages} {572--581}
  (\bibinfo {year} {2001})},\ \Eprint {http://arxiv.org/abs/astro-ph/0102079}
  {arXiv:astro-ph/0102079} \BibitemShut {NoStop}%
\bibitem [{\citenamefont {{Oh}}\ \emph {et~al.}(2000)\citenamefont {{Oh}},
  \citenamefont {{Lin}},\ and\ \citenamefont {{Richer}}}]{oh2000}%
  \BibitemOpen
  \bibfield  {author} {\bibinfo {author} {\bibfnamefont {K.~S.}\ \bibnamefont
  {{Oh}}}, \bibinfo {author} {\bibfnamefont {D.N.C.}\ \bibnamefont {{Lin}}}, \
  and\ \bibinfo {author} {\bibfnamefont {H.~B.}\ \bibnamefont {{Richer}}},\
  }\bibfield  {title} {\enquote {\bibinfo {title} {{Globular clusters in the
  Fornax dwarf spheroidal galaxy}},}\ }\Doi {10.1086/308477} {\bibfield
  {journal} {\bibinfo  {journal} {\apj},\ }\textbf {\bibinfo {volume} {531}},\
  \bibinfo {pages} {727--738} (\bibinfo {year} {2000})}\BibitemShut {NoStop}%
\bibitem [{\citenamefont {Inoue}(2011)}]{in11}%
  \BibitemOpen
  \bibfield  {author} {\bibinfo {author} {\bibfnamefont {S.}~\bibnamefont
  {Inoue}},\ }\bibfield  {title} {\enquote {\bibinfo {title} {{Corrective
  effect of many-body interactions in dynamical friction}},}\ }\Doi
  {10.1111/j.1365-2966.2011.19122.x} {\bibfield  {journal} {\bibinfo  {journal}
  {\mnras},\ }\textbf {\bibinfo {volume} {416}},\ \bibinfo {pages} {1181--1190}
  (\bibinfo {year} {2011})},\ \Eprint {http://arxiv.org/abs/0912.2409}
  {arXiv:0912.2409} \BibitemShut {NoStop}%
\bibitem [{\citenamefont {Cole}\ \emph {et~al.}(2012)\citenamefont {Cole},
  \citenamefont {Dehnen}, \citenamefont {Read},\ and\ \citenamefont
  {Wilkinson}}]{cole12}%
  \BibitemOpen
  \bibfield  {author} {\bibinfo {author} {\bibfnamefont {D.~R.}\ \bibnamefont
  {Cole}}, \bibinfo {author} {\bibfnamefont {W.}~\bibnamefont {Dehnen}},
  \bibinfo {author} {\bibfnamefont {J.~I.}\ \bibnamefont {Read}}, \ and\
  \bibinfo {author} {\bibfnamefont {M.~I.}\ \bibnamefont {Wilkinson}},\
  }\bibfield  {title} {\enquote {\bibinfo {title} {{The mass distribution of
  the Fornax dSph: constraints from its globular cluster distribution}},}\
  }\Doi {10.1111/j.1365-2966.2012.21885.x} {\bibfield  {journal} {\bibinfo
  {journal} {\mnras},\ }\textbf {\bibinfo {volume} {426}},\ \bibinfo {pages}
  {601--613} (\bibinfo {year} {2012})},\ \Eprint
  {http://arxiv.org/abs/1205.6327} {arXiv:1205.6327} \BibitemShut {NoStop}%
\bibitem [{\citenamefont {{Tremaine}}\ and\ \citenamefont
  {{Weinberg}}(1984)}]{tw84}%
  \BibitemOpen
  \bibfield  {author} {\bibinfo {author} {\bibfnamefont {S.}~\bibnamefont
  {{Tremaine}}}\ and\ \bibinfo {author} {\bibfnamefont {M.~D.}\ \bibnamefont
  {{Weinberg}}},\ }\bibfield  {title} {\enquote {\bibinfo {title} {{Dynamical
  friction in spherical systems}},}\ }\Doi {10.1093/mnras/209.4.729} {\bibfield
   {journal} {\bibinfo  {journal} {\mnras},\ }\textbf {\bibinfo {volume}
  {209}},\ \bibinfo {pages} {729--757} (\bibinfo {year} {1984})}\BibitemShut
  {NoStop}%
\bibitem [{\citenamefont {Read}\ \emph {et~al.}(2006)\citenamefont {Read},
  \citenamefont {Goerdt}, \citenamefont {Moore}, \citenamefont {Pontzen},
  \citenamefont {Stadel},\ and\ \citenamefont {Lake}}]{rgm06}%
  \BibitemOpen
  \bibfield  {author} {\bibinfo {author} {\bibfnamefont {J.~I.}\ \bibnamefont
  {Read}}, \bibinfo {author} {\bibfnamefont {T.}~\bibnamefont {Goerdt}},
  \bibinfo {author} {\bibfnamefont {B.}~\bibnamefont {Moore}}, \bibinfo
  {author} {\bibfnamefont {A.~P.}\ \bibnamefont {Pontzen}}, \bibinfo {author}
  {\bibfnamefont {J.}~\bibnamefont {Stadel}}, \ and\ \bibinfo {author}
  {\bibfnamefont {G.}~\bibnamefont {Lake}},\ }\bibfield  {title} {\enquote
  {\bibinfo {title} {{Dynamical friction in constant density cores: a failure
  of the Chandrasekhar formula}},}\ }\Doi {10.1111/j.1365-2966.2006.11022.x}
  {\bibfield  {journal} {\bibinfo  {journal} {\mnras},\ }\textbf {\bibinfo
  {volume} {373}},\ \bibinfo {pages} {1451--1460} (\bibinfo {year} {2006})},\
  \Eprint {http://arxiv.org/abs/astro-ph/0606636} {arXiv:astro-ph/0606636}
  \BibitemShut {NoStop}%
\bibitem [{\citenamefont {Kochanek}\ and\ \citenamefont {White}(2000)}]{kw00}%
  \BibitemOpen
  \bibfield  {author} {\bibinfo {author} {\bibfnamefont {C.~S.}\ \bibnamefont
  {Kochanek}}\ and\ \bibinfo {author} {\bibfnamefont {M.~J.}\ \bibnamefont
  {White}},\ }\bibfield  {title} {\enquote {\bibinfo {title} {{A quantitative
  study of interacting dark matter in halos}},}\ }\Doi {10.1086/317149}
  {\bibfield  {journal} {\bibinfo  {journal} {Astrophys. J.},\ }\textbf
  {\bibinfo {volume} {543}},\ \bibinfo {pages} {514--520} (\bibinfo {year}
  {2000})},\ \Eprint {http://arxiv.org/abs/astro-ph/0003483}
  {arXiv:astro-ph/0003483} \BibitemShut {NoStop}%
\bibitem [{\citenamefont {Dalcanton}\ and\ \citenamefont {Hogan}(2001)}]{dh01}%
  \BibitemOpen
  \bibfield  {author} {\bibinfo {author} {\bibfnamefont {J.~J.}\ \bibnamefont
  {Dalcanton}}\ and\ \bibinfo {author} {\bibfnamefont {C.~J.}\ \bibnamefont
  {Hogan}},\ }\bibfield  {title} {\enquote {\bibinfo {title} {{Halo cores and
  phase space densities: Observational constraints on dark matter physics and
  structure formation}},}\ }\Doi {10.1086/323207} {\bibfield  {journal}
  {\bibinfo  {journal} {Astrophys. J.},\ }\textbf {\bibinfo {volume} {561}},\
  \bibinfo {pages} {35--45} (\bibinfo {year} {2001})},\ \Eprint
  {http://arxiv.org/abs/astro-ph/0004381} {arXiv:astro-ph/0004381} \BibitemShut
  {NoStop}%
\bibitem [{\citenamefont {Sanchez-Salcedo}\ \emph {et~al.}(2006)\citenamefont
  {Sanchez-Salcedo}, \citenamefont {Reyes-Iturbide},\ and\ \citenamefont
  {Hernandez}}]{sal06}%
  \BibitemOpen
  \bibfield  {author} {\bibinfo {author} {\bibfnamefont {F.~J.}\ \bibnamefont
  {Sanchez-Salcedo}}, \bibinfo {author} {\bibfnamefont {J.}~\bibnamefont
  {Reyes-Iturbide}}, \ and\ \bibinfo {author} {\bibfnamefont {X.}~\bibnamefont
  {Hernandez}},\ }\bibfield  {title} {\enquote {\bibinfo {title} {{An extensive
  study of dynamical friction in dwarf galaxies: the role of stars, dark
  matter, halo profiles and mond}},}\ }\Doi {10.1111/j.1365-2966.2006.10602.x}
  {\bibfield  {journal} {\bibinfo  {journal} {\mnras},\ }\textbf {\bibinfo
  {volume} {370}},\ \bibinfo {pages} {1829--1840} (\bibinfo {year} {2006})},\
  \Eprint {http://arxiv.org/abs/astro-ph/0601490} {arXiv:astro-ph/0601490}
  \BibitemShut {NoStop}%
\bibitem [{\citenamefont {Villaescusa-Navarro}\ and\ \citenamefont
  {Dalal}(2011)}]{vnd11}%
  \BibitemOpen
  \bibfield  {author} {\bibinfo {author} {\bibfnamefont {F.}~\bibnamefont
  {Villaescusa-Navarro}}\ and\ \bibinfo {author} {\bibfnamefont
  {N.}~\bibnamefont {Dalal}},\ }\bibfield  {title} {\enquote {\bibinfo {title}
  {{Cores and cusps in warm dark matter halos}},}\ }\Doi
  {10.1088/1475-7516/2011/03/024} {\bibfield  {journal} {\bibinfo  {journal}
  {JCAP},\ }\textbf {\bibinfo {volume} {1103}},\ \bibinfo {pages} {024}
  (\bibinfo {year} {2011})},\ \Eprint {http://arxiv.org/abs/1010.3008}
  {arXiv:1010.3008} \BibitemShut {NoStop}%
\bibitem [{\citenamefont {Maccio}\ \emph {et~al.}(2012)\citenamefont {Maccio},
  \citenamefont {Paduroiu}, \citenamefont {Anderhalden}, \citenamefont
  {Schneider},\ and\ \citenamefont {Moore}}]{mac12}%
  \BibitemOpen
  \bibfield  {author} {\bibinfo {author} {\bibfnamefont {A.~V.}\ \bibnamefont
  {Maccio}}, \bibinfo {author} {\bibfnamefont {S.}~\bibnamefont {Paduroiu}},
  \bibinfo {author} {\bibfnamefont {D.}~\bibnamefont {Anderhalden}}, \bibinfo
  {author} {\bibfnamefont {A.}~\bibnamefont {Schneider}}, \ and\ \bibinfo
  {author} {\bibfnamefont {B.}~\bibnamefont {Moore}},\ }\bibfield  {title}
  {\enquote {\bibinfo {title} {{Cores in warm dark matter haloes: a Catch 22
  problem}},}\ }\Doi {10.1111/j.1365-2966.2012.21284.x} {\bibfield  {journal}
  {\bibinfo  {journal} {\mnras},\ }\textbf {\bibinfo {volume} {424}},\ \bibinfo
  {pages} {1105--1112} (\bibinfo {year} {2012})},\ \Eprint
  {http://arxiv.org/abs/1202.1282} {arXiv:1202.1282} \BibitemShut {NoStop}%
\bibitem [{\citenamefont {Berezhiani}\ and\ \citenamefont
  {Khoury}(2015)}]{bk15}%
  \BibitemOpen
  \bibfield  {author} {\bibinfo {author} {\bibfnamefont {L.}~\bibnamefont
  {Berezhiani}}\ and\ \bibinfo {author} {\bibfnamefont {J.}~\bibnamefont
  {Khoury}},\ }\bibfield  {title} {\enquote {\bibinfo {title} {{Theory of dark
  matter superfluidity}},}\ }\Doi {10.1103/PhysRevD.92.103510} {\bibfield
  {journal} {\bibinfo  {journal} {Phys. Rev.},\ }\textbf {\bibinfo {volume}
  {D92}},\ \bibinfo {pages} {103510} (\bibinfo {year} {2015})},\ \Eprint
  {http://arxiv.org/abs/1507.01019} {arXiv:1507.01019} \BibitemShut {NoStop}%
\bibitem [{\citenamefont {Lora}\ \emph {et~al.}(2012)\citenamefont {Lora},
  \citenamefont {Magana}, \citenamefont {Bernal}, \citenamefont
  {Sanchez-Salcedo},\ and\ \citenamefont {Grebel}}]{lora12}%
  \BibitemOpen
  \bibfield  {author} {\bibinfo {author} {\bibfnamefont {V.}~\bibnamefont
  {Lora}}, \bibinfo {author} {\bibfnamefont {J.}~\bibnamefont {Magana}},
  \bibinfo {author} {\bibfnamefont {A.}~\bibnamefont {Bernal}}, \bibinfo
  {author} {\bibfnamefont {F.~J.}\ \bibnamefont {Sanchez-Salcedo}}, \ and\
  \bibinfo {author} {\bibfnamefont {E.~K.}\ \bibnamefont {Grebel}},\ }\bibfield
   {title} {\enquote {\bibinfo {title} {{On the mass of ultra-light bosonic
  dark matter from galactic dynamics}},}\ }\Doi {10.1088/1475-7516/2012/02/011}
  {\bibfield  {journal} {\bibinfo  {journal} {JCAP},\ }\textbf {\bibinfo
  {volume} {1202}},\ \bibinfo {pages} {011} (\bibinfo {year} {2012})},\ \Eprint
  {http://arxiv.org/abs/1110.2684} {arXiv:1110.2684} \BibitemShut {NoStop}%
\bibitem [{\citenamefont {{Sellwood}}(1980)}]{sell80}%
  \BibitemOpen
  \bibfield  {author} {\bibinfo {author} {\bibfnamefont {J.~A.}\ \bibnamefont
  {{Sellwood}}},\ }\bibfield  {title} {\enquote {\bibinfo {title} {{Galaxy
  models with live halos}},}\ }\href@noop {} {\bibfield  {journal} {\bibinfo
  {journal} {\aap},\ }\textbf {\bibinfo {volume} {89}},\ \bibinfo {pages}
  {296--307} (\bibinfo {year} {1980})}\BibitemShut {NoStop}%
\bibitem [{\citenamefont {{Weinberg}}(1985)}]{wein85}%
  \BibitemOpen
  \bibfield  {author} {\bibinfo {author} {\bibfnamefont {M.~D.}\ \bibnamefont
  {{Weinberg}}},\ }\bibfield  {title} {\enquote {\bibinfo {title} {{Evolution
  of barred galaxies by dynamical friction}},}\ }\Doi {10.1093/mnras/213.3.451}
  {\bibfield  {journal} {\bibinfo  {journal} {\mnras},\ }\textbf {\bibinfo
  {volume} {213}},\ \bibinfo {pages} {451--471} (\bibinfo {year}
  {1985})}\BibitemShut {NoStop}%
\bibitem [{\citenamefont {Debattista}\ and\ \citenamefont
  {Sellwood}(2000)}]{ds00}%
  \BibitemOpen
  \bibfield  {author} {\bibinfo {author} {\bibfnamefont {V.~P.}\ \bibnamefont
  {Debattista}}\ and\ \bibinfo {author} {\bibfnamefont {J.~A.}\ \bibnamefont
  {Sellwood}},\ }\bibfield  {title} {\enquote {\bibinfo {title} {{Constraints
  from dynamical friction on the dark matter content of barred galaxies}},}\
  }\Doi {10.1086/317148} {\bibfield  {journal} {\bibinfo  {journal} {Astrophys.
  J.},\ }\textbf {\bibinfo {volume} {543}},\ \bibinfo {pages} {704--721}
  (\bibinfo {year} {2000})},\ \Eprint {http://arxiv.org/abs/astro-ph/0006275}
  {arXiv:astro-ph/0006275} \BibitemShut {NoStop}%
\bibitem [{\citenamefont {{Weinberg}}\ and\ \citenamefont
  {{Katz}}(2002)}]{wk02}%
  \BibitemOpen
  \bibfield  {author} {\bibinfo {author} {\bibfnamefont {M.~D.}\ \bibnamefont
  {{Weinberg}}}\ and\ \bibinfo {author} {\bibfnamefont {N.}~\bibnamefont
  {{Katz}}},\ }\bibfield  {title} {\enquote {\bibinfo {title} {{Bar-driven dark
  halo evolution: a resolution of the cusp-core controversy}},}\ }\Doi
  {10.1086/343847} {\bibfield  {journal} {\bibinfo  {journal} {\apj},\ }\textbf
  {\bibinfo {volume} {580}},\ \bibinfo {pages} {627--633} (\bibinfo {year}
  {2002})},\ \Eprint {http://arxiv.org/abs/arXiv:astro-ph/0110632}
  {arXiv:astro-ph/0110632} \BibitemShut {NoStop}%
\bibitem [{\citenamefont {Dubinski}\ \emph {et~al.}(2009)\citenamefont
  {Dubinski}, \citenamefont {Berentzen},\ and\ \citenamefont
  {Shlosman}}]{dbs09}%
  \BibitemOpen
  \bibfield  {author} {\bibinfo {author} {\bibfnamefont {J.}~\bibnamefont
  {Dubinski}}, \bibinfo {author} {\bibfnamefont {I.}~\bibnamefont {Berentzen}},
  \ and\ \bibinfo {author} {\bibfnamefont {I.}~\bibnamefont {Shlosman}},\
  }\bibfield  {title} {\enquote {\bibinfo {title} {{Anatomy of the bar
  instability in cuspy dark matter halos}},}\ }\Doi
  {10.1088/0004-637X/697/1/293} {\bibfield  {journal} {\bibinfo  {journal}
  {Astrophys. J.},\ }\textbf {\bibinfo {volume} {697}},\ \bibinfo {pages}
  {293--310} (\bibinfo {year} {2009})},\ \Eprint
  {http://arxiv.org/abs/0810.4925} {arXiv:0810.4925} \BibitemShut {NoStop}%
\bibitem [{\citenamefont {{Athanassoula}}(2014)}]{ath14}%
  \BibitemOpen
  \bibfield  {author} {\bibinfo {author} {\bibfnamefont {E.}~\bibnamefont
  {{Athanassoula}}},\ }\bibfield  {title} {\enquote {\bibinfo {title} {{Bar
  slowdown and the distribution of dark matter in barred galaxies}},}\ }\Doi
  {10.1093/mnrasl/slt163} {\bibfield  {journal} {\bibinfo  {journal} {\mnras},\
  }\textbf {\bibinfo {volume} {438}},\ \bibinfo {pages} {L81--L85} (\bibinfo
  {year} {2014})},\ \Eprint {http://arxiv.org/abs/1312.1690} {arXiv:1312.1690}
  \BibitemShut {NoStop}%
\bibitem [{\citenamefont {Sellwood}(2014)}]{sell14}%
  \BibitemOpen
  \bibfield  {author} {\bibinfo {author} {\bibfnamefont {J.~A.}\ \bibnamefont
  {Sellwood}},\ }\bibfield  {title} {\enquote {\bibinfo {title} {{Secular
  evolution in disk galaxies}},}\ }\Doi {10.1103/RevModPhys.86.1} {\bibfield
  {journal} {\bibinfo  {journal} {Rev. Mod. Phys.},\ }\textbf {\bibinfo
  {volume} {86}},\ \bibinfo {pages} {1} (\bibinfo {year} {2014})},\ \Eprint
  {http://arxiv.org/abs/1310.0403} {arXiv:1310.0403} \BibitemShut {NoStop}%
\bibitem [{\citenamefont {Aguerri}\ \emph {et~al.}(2015)\citenamefont {Aguerri}
  \emph {et~al.}}]{agu15}%
  \BibitemOpen
  \bibfield  {author} {\bibinfo {author} {\bibfnamefont {J.A.L.}\ \bibnamefont
  {Aguerri}} \emph {et~al.},\ }\bibfield  {title} {\enquote {\bibinfo {title}
  {{Bar pattern speeds in CALIFA galaxies: I. Fast bars across the Hubble
  sequence}},}\ }\Doi {10.1051/0004-6361/201423383} {\bibfield  {journal}
  {\bibinfo  {journal} {Astron. Astrophys.},\ }\textbf {\bibinfo {volume}
  {576}},\ \bibinfo {pages} {A102} (\bibinfo {year} {2015})},\ \Eprint
  {http://arxiv.org/abs/1501.05498} {arXiv:1501.05498} \BibitemShut {NoStop}%
\bibitem [{\citenamefont {Petersen}\ \emph {et~al.}(2016)\citenamefont
  {Petersen}, \citenamefont {Weinberg},\ and\ \citenamefont {Katz}}]{pwk16}%
  \BibitemOpen
  \bibfield  {author} {\bibinfo {author} {\bibfnamefont {M.~S.}\ \bibnamefont
  {Petersen}}, \bibinfo {author} {\bibfnamefont {M.~D.}\ \bibnamefont
  {Weinberg}}, \ and\ \bibinfo {author} {\bibfnamefont {N.}~\bibnamefont
  {Katz}},\ }\bibfield  {title} {\enquote {\bibinfo {title} {{Dark matter
  trapping by stellar bars: the shadow bar}},}\ }\href@noop {} { (\bibinfo
  {year} {2016})},\ \Eprint {http://arxiv.org/abs/1602.04826}
  {arXiv:1602.04826} \BibitemShut {NoStop}%
\bibitem [{\citenamefont {{Gebhardt}}\ \emph {et~al.}(2011)\citenamefont
  {{Gebhardt}}, \citenamefont {{Adams}}, \citenamefont {{Richstone}},
  \citenamefont {{Lauer}}, \citenamefont {{Faber}}, \citenamefont
  {{G{\"u}ltekin}}, \citenamefont {{Murphy}},\ and\ \citenamefont
  {{Tremaine}}}]{geb11}%
  \BibitemOpen
  \bibfield  {author} {\bibinfo {author} {\bibfnamefont {K.}~\bibnamefont
  {{Gebhardt}}}, \bibinfo {author} {\bibfnamefont {J.}~\bibnamefont {{Adams}}},
  \bibinfo {author} {\bibfnamefont {D.}~\bibnamefont {{Richstone}}}, \bibinfo
  {author} {\bibfnamefont {T.~R.}\ \bibnamefont {{Lauer}}}, \bibinfo {author}
  {\bibfnamefont {S.~M.}\ \bibnamefont {{Faber}}}, \bibinfo {author}
  {\bibfnamefont {K.}~\bibnamefont {{G{\"u}ltekin}}}, \bibinfo {author}
  {\bibfnamefont {J.}~\bibnamefont {{Murphy}}}, \ and\ \bibinfo {author}
  {\bibfnamefont {S.}~\bibnamefont {{Tremaine}}},\ }\bibfield  {title}
  {\enquote {\bibinfo {title} {{The black hole mass in M87 from Gemini/NIFS
  adaptive optics observations}},}\ }\Doi {10.1088/0004-637X/729/2/119}
  {\bibfield  {journal} {\bibinfo  {journal} {\apj},\ }\textbf {\bibinfo
  {volume} {729}},\ \bibinfo {eid} {119} (\bibinfo {year} {2011})},\ \Eprint
  {http://arxiv.org/abs/1101.1954} {arXiv:1101.1954} \BibitemShut {NoStop}%
\bibitem [{\citenamefont {{McConnell}}\ \emph {et~al.}(2011)\citenamefont
  {{McConnell}}, \citenamefont {{Ma}}, \citenamefont {{Gebhardt}},
  \citenamefont {{Wright}}, \citenamefont {{Murphy}}, \citenamefont {{Lauer}},
  \citenamefont {{Graham}},\ and\ \citenamefont {{Richstone}}}]{mcconnell11}%
  \BibitemOpen
  \bibfield  {author} {\bibinfo {author} {\bibfnamefont {N.~J.}\ \bibnamefont
  {{McConnell}}}, \bibinfo {author} {\bibfnamefont {C.-P.}\ \bibnamefont
  {{Ma}}}, \bibinfo {author} {\bibfnamefont {K.}~\bibnamefont {{Gebhardt}}},
  \bibinfo {author} {\bibfnamefont {S.~A.}\ \bibnamefont {{Wright}}}, \bibinfo
  {author} {\bibfnamefont {J.~D.}\ \bibnamefont {{Murphy}}}, \bibinfo {author}
  {\bibfnamefont {T.~R.}\ \bibnamefont {{Lauer}}}, \bibinfo {author}
  {\bibfnamefont {J.~R.}\ \bibnamefont {{Graham}}}, \ and\ \bibinfo {author}
  {\bibfnamefont {D.~O.}\ \bibnamefont {{Richstone}}},\ }\bibfield  {title}
  {\enquote {\bibinfo {title} {{Two ten-billion-solar-mass black holes at the
  centres of giant elliptical galaxies}},}\ }\Doi {10.1038/nature10636}
  {\bibfield  {journal} {\bibinfo  {journal} {\nat},\ }\textbf {\bibinfo
  {volume} {480}},\ \bibinfo {pages} {215--218} (\bibinfo {year} {2011})},\
  \Eprint {http://arxiv.org/abs/1112.1078} {arXiv:1112.1078 [astro-ph.CO]}
  \BibitemShut {NoStop}%
\bibitem [{\citenamefont {{Kormendy}}\ and\ \citenamefont
  {{Ho}}(2013)}]{kormendy2013}%
  \BibitemOpen
  \bibfield  {author} {\bibinfo {author} {\bibfnamefont {J.}~\bibnamefont
  {{Kormendy}}}\ and\ \bibinfo {author} {\bibfnamefont {L.~C.}\ \bibnamefont
  {{Ho}}},\ }\bibfield  {title} {\enquote {\bibinfo {title} {{Coevolution (or
  not) of supermassive black holes and host galaxies}},}\ }\Doi
  {10.1146/annurev-astro-082708-101811} {\bibfield  {journal} {\bibinfo
  {journal} {Ann.\ Rev.\ Astron.\ Astrophys.},\ }\textbf {\bibinfo {volume}
  {51}},\ \bibinfo {pages} {511--653} (\bibinfo {year} {2013})},\ \Eprint
  {http://arxiv.org/abs/1304.7762} {arXiv:1304.7762} \BibitemShut {NoStop}%
\bibitem [{\citenamefont {{Binggeli}}\ \emph {et~al.}(1987)\citenamefont
  {{Binggeli}}, \citenamefont {{Tammann}},\ and\ \citenamefont
  {{Sandage}}}]{bing87}%
  \BibitemOpen
  \bibfield  {author} {\bibinfo {author} {\bibfnamefont {B.}~\bibnamefont
  {{Binggeli}}}, \bibinfo {author} {\bibfnamefont {G.~A.}\ \bibnamefont
  {{Tammann}}}, \ and\ \bibinfo {author} {\bibfnamefont {A.}~\bibnamefont
  {{Sandage}}},\ }\bibfield  {title} {\enquote {\bibinfo {title} {{Studies of
  the Virgo cluster. VI - Morphological and kinematical structure of the Virgo
  cluster}},}\ }\Doi {10.1086/114467} {\bibfield  {journal} {\bibinfo
  {journal} {Astron.\ J.},\ }\textbf {\bibinfo {volume} {94}},\ \bibinfo
  {pages} {251--277} (\bibinfo {year} {1987})}\BibitemShut {NoStop}%
\bibitem [{\citenamefont {{Fitchett}}\ and\ \citenamefont
  {{Webster}}(1987)}]{fitchett87}%
  \BibitemOpen
  \bibfield  {author} {\bibinfo {author} {\bibfnamefont {M.}~\bibnamefont
  {{Fitchett}}}\ and\ \bibinfo {author} {\bibfnamefont {R.}~\bibnamefont
  {{Webster}}},\ }\bibfield  {title} {\enquote {\bibinfo {title} {{Substructure
  in the Coma cluster}},}\ }\Doi {10.1086/165310} {\bibfield  {journal}
  {\bibinfo  {journal} {\apj},\ }\textbf {\bibinfo {volume} {317}},\ \bibinfo
  {pages} {653--667} (\bibinfo {year} {1987})}\BibitemShut {NoStop}%
\bibitem [{\citenamefont {{Unruh}}(1976)}]{unruh1976}%
  \BibitemOpen
  \bibfield  {author} {\bibinfo {author} {\bibfnamefont {W.~G.}\ \bibnamefont
  {{Unruh}}},\ }\bibfield  {title} {\enquote {\bibinfo {title} {{Absorption
  cross section of small black holes}},}\ }\Doi {10.1103/PhysRevD.14.3251}
  {\bibfield  {journal} {\bibinfo  {journal} {\prd},\ }\textbf {\bibinfo
  {volume} {14}},\ \bibinfo {pages} {3251--3259} (\bibinfo {year}
  {1976})}\BibitemShut {NoStop}%
\bibitem [{\citenamefont {Schive}\ \emph {et~al.}(2016)\citenamefont {Schive},
  \citenamefont {Chiueh}, \citenamefont {Broadhurst},\ and\ \citenamefont
  {Huang}}]{sch16}%
  \BibitemOpen
  \bibfield  {author} {\bibinfo {author} {\bibfnamefont {H.-Y.}\ \bibnamefont
  {Schive}}, \bibinfo {author} {\bibfnamefont {T.}~\bibnamefont {Chiueh}},
  \bibinfo {author} {\bibfnamefont {T.}~\bibnamefont {Broadhurst}}, \ and\
  \bibinfo {author} {\bibfnamefont {K.-W.}\ \bibnamefont {Huang}},\ }\bibfield
  {title} {\enquote {\bibinfo {title} {{Contrasting galaxy formation from
  quantum wave dark matter, $\psi$DM, with $\Lambda$CDM, using Planck and
  Hubble data}},}\ }\Doi {10.3847/0004-637X/818/1/89} {\bibfield  {journal}
  {\bibinfo  {journal} {Astrophys. J.},\ }\textbf {\bibinfo {volume} {818}},\
  \bibinfo {pages} {89} (\bibinfo {year} {2016})},\ \Eprint
  {http://arxiv.org/abs/1508.04621} {arXiv:1508.04621} \BibitemShut {NoStop}%
\bibitem [{\citenamefont {Adam}\ \emph {et~al.}(2016)\citenamefont {Adam} \emph
  {et~al.}}]{planck16}%
  \BibitemOpen
  \bibfield  {author} {\bibinfo {author} {\bibfnamefont {R.}~\bibnamefont
  {Adam}} \emph {et~al.} (\bibinfo {collaboration} {Planck}),\ }\bibfield
  {title} {\enquote {\bibinfo {title} {{Planck intermediate results. XLVII.
  Planck constraints on reionization history}},}\ }\href@noop {} { (\bibinfo
  {year} {2016})},\ \Eprint {http://arxiv.org/abs/1605.03507}
  {arXiv:1605.03507} \BibitemShut {NoStop}%
\bibitem [{\citenamefont {Corasaniti}\ \emph {et~al.}(2016)\citenamefont
  {Corasaniti}, \citenamefont {Agarwal}, \citenamefont {Marsh},\ and\
  \citenamefont {Das}}]{cora16}%
  \BibitemOpen
  \bibfield  {author} {\bibinfo {author} {\bibfnamefont {P.\~S.}\ \bibnamefont
  {Corasaniti}}, \bibinfo {author} {\bibfnamefont {S.}~\bibnamefont {Agarwal}},
  \bibinfo {author} {\bibfnamefont {D.J.E.}\ \bibnamefont {Marsh}}, \ and\
  \bibinfo {author} {\bibfnamefont {S.}~\bibnamefont {Das}},\ }\bibfield
  {title} {\enquote {\bibinfo {title} {{Constraints on dark matter scenarios
  from measurements of the galaxy luminosity function at high redshifts}},}\
  }\href@noop {} { (\bibinfo {year} {2016})},\ \Eprint
  {http://arxiv.org/abs/arXiv:1611.05892} {arXiv:arXiv:1611.05892} \BibitemShut
  {NoStop}%
\bibitem [{\citenamefont {Bozek}\ \emph {et~al.}(2015)\citenamefont {Bozek},
  \citenamefont {Marsh}, \citenamefont {Silk},\ and\ \citenamefont
  {Wyse}}]{bozek15}%
  \BibitemOpen
  \bibfield  {author} {\bibinfo {author} {\bibfnamefont {B.}~\bibnamefont
  {Bozek}}, \bibinfo {author} {\bibfnamefont {D.J.E.}\ \bibnamefont {Marsh}},
  \bibinfo {author} {\bibfnamefont {J.}~\bibnamefont {Silk}}, \ and\ \bibinfo
  {author} {\bibfnamefont {R.F.G.}\ \bibnamefont {Wyse}},\ }\bibfield  {title}
  {\enquote {\bibinfo {title} {{Galaxy UV-luminosity function and reionization
  constraints on axion dark matter}},}\ }\Doi {10.1093/mnras/stv624} {\bibfield
   {journal} {\bibinfo  {journal} {\mnras},\ }\textbf {\bibinfo {volume}
  {450}},\ \bibinfo {pages} {209--222} (\bibinfo {year} {2015})},\ \Eprint
  {http://arxiv.org/abs/1409.3544} {arXiv:1409.3544} \BibitemShut {NoStop}%
\bibitem [{\citenamefont {Cen}\ \emph {et~al.}(1994)\citenamefont {Cen},
  \citenamefont {Miralda-Escude}, \citenamefont {Ostriker},\ and\ \citenamefont
  {Rauch}}]{Cen:1994da}%
  \BibitemOpen
  \bibfield  {author} {\bibinfo {author} {\bibfnamefont {R.}~\bibnamefont
  {Cen}}, \bibinfo {author} {\bibfnamefont {J.}~\bibnamefont {Miralda-Escude}},
  \bibinfo {author} {\bibfnamefont {J.~P.}\ \bibnamefont {Ostriker}}, \ and\
  \bibinfo {author} {\bibfnamefont {M.}~\bibnamefont {Rauch}},\ }\bibfield
  {title} {\enquote {\bibinfo {title} {{Gravitational collapse of small scale
  structure as the origin of the Lyman-$\alpha$ forest}},}\ }\Doi
  {10.1086/187670} {\bibfield  {journal} {\bibinfo  {journal} {Astrophys. J.},\
  }\textbf {\bibinfo {volume} {437}},\ \bibinfo {pages} {L9--L12} (\bibinfo
  {year} {1994})},\ \Eprint {http://arxiv.org/abs/astro-ph/9409017}
  {arXiv:astro-ph/9409017} \BibitemShut {NoStop}%
\bibitem [{\citenamefont {Hernquist}\ \emph {et~al.}(1996)\citenamefont
  {Hernquist}, \citenamefont {Katz}, \citenamefont {Weinberg},\ and\
  \citenamefont {Miralda-Escude}}]{Hernquist:1995uma}%
  \BibitemOpen
  \bibfield  {author} {\bibinfo {author} {\bibfnamefont {L.}~\bibnamefont
  {Hernquist}}, \bibinfo {author} {\bibfnamefont {N.}~\bibnamefont {Katz}},
  \bibinfo {author} {\bibfnamefont {D.~H.}\ \bibnamefont {Weinberg}}, \ and\
  \bibinfo {author} {\bibfnamefont {J.}~\bibnamefont {Miralda-Escude}},\
  }\bibfield  {title} {\enquote {\bibinfo {title} {{The Lyman-$\alpha$ forest
  in the cold dark matter model}},}\ }\Doi {10.1086/309899} {\bibfield
  {journal} {\bibinfo  {journal} {Astrophys. J.},\ }\textbf {\bibinfo {volume}
  {457}},\ \bibinfo {pages} {L51--L55} (\bibinfo {year} {1996})},\ \Eprint
  {http://arxiv.org/abs/astro-ph/9509105} {arXiv:astro-ph/9509105} \BibitemShut
  {NoStop}%
\bibitem [{\citenamefont {Zhang}\ \emph {et~al.}(1995)\citenamefont {Zhang},
  \citenamefont {Anninos},\ and\ \citenamefont {Norman}}]{Zhang:1995zh}%
  \BibitemOpen
  \bibfield  {author} {\bibinfo {author} {\bibfnamefont {Y.}~\bibnamefont
  {Zhang}}, \bibinfo {author} {\bibfnamefont {P.}~\bibnamefont {Anninos}}, \
  and\ \bibinfo {author} {\bibfnamefont {M.~L.}\ \bibnamefont {Norman}},\
  }\bibfield  {title} {\enquote {\bibinfo {title} {{A multi-species model for
  hydrogen and helium absorbers in Lyman-$\alpha$ forest clouds}},}\ }\Doi
  {10.1086/309752} {\bibfield  {journal} {\bibinfo  {journal} {Astrophys. J.},\
  }\textbf {\bibinfo {volume} {453}},\ \bibinfo {pages} {L57--L60} (\bibinfo
  {year} {1995})},\ \Eprint {http://arxiv.org/abs/astro-ph/9508133}
  {arXiv:astro-ph/9508133} \BibitemShut {NoStop}%
\bibitem [{\citenamefont {Hui}\ \emph {et~al.}(1997)\citenamefont {Hui},
  \citenamefont {Gnedin},\ and\ \citenamefont {Zhang}}]{HGZ97}%
  \BibitemOpen
  \bibfield  {author} {\bibinfo {author} {\bibfnamefont {L.}~\bibnamefont
  {Hui}}, \bibinfo {author} {\bibfnamefont {N.~Y.}\ \bibnamefont {Gnedin}}, \
  and\ \bibinfo {author} {\bibfnamefont {Y.}~\bibnamefont {Zhang}},\ }\bibfield
   {title} {\enquote {\bibinfo {title} {{The statistics of density peaks and
  the column density distribution of the Lyman-$\alpha$ forest}},}\ }\Doi
  {10.1086/304539} {\bibfield  {journal} {\bibinfo  {journal} {Astrophys. J.},\
  }\textbf {\bibinfo {volume} {486}},\ \bibinfo {pages} {599--622} (\bibinfo
  {year} {1997})},\ \Eprint {http://arxiv.org/abs/astro-ph/9608157}
  {arXiv:astro-ph/9608157} \BibitemShut {NoStop}%
\bibitem [{\citenamefont {Croft}\ \emph {et~al.}(1998)\citenamefont {Croft},
  \citenamefont {Weinberg}, \citenamefont {Katz},\ and\ \citenamefont
  {Hernquist}}]{Croft:1997jf}%
  \BibitemOpen
  \bibfield  {author} {\bibinfo {author} {\bibfnamefont {R.A.C.}\ \bibnamefont
  {Croft}}, \bibinfo {author} {\bibfnamefont {D.~H.}\ \bibnamefont {Weinberg}},
  \bibinfo {author} {\bibfnamefont {N.}~\bibnamefont {Katz}}, \ and\ \bibinfo
  {author} {\bibfnamefont {L.}~\bibnamefont {Hernquist}},\ }\bibfield  {title}
  {\enquote {\bibinfo {title} {{Recovery of the power spectrum of mass
  fluctuations from observations of the Lyman-$\alpha$ forest}},}\ }\Doi
  {10.1086/305289} {\bibfield  {journal} {\bibinfo  {journal} {Astrophys. J.},\
  }\textbf {\bibinfo {volume} {495}},\ \bibinfo {pages} {44--62} (\bibinfo
  {year} {1998})},\ \Eprint {http://arxiv.org/abs/astro-ph/9708018}
  {arXiv:astro-ph/9708018} \BibitemShut {NoStop}%
\bibitem [{\citenamefont {Gnedin}\ and\ \citenamefont {Hui}(1998)}]{GH97}%
  \BibitemOpen
  \bibfield  {author} {\bibinfo {author} {\bibfnamefont {N.~Y.}\ \bibnamefont
  {Gnedin}}\ and\ \bibinfo {author} {\bibfnamefont {L.}~\bibnamefont {Hui}},\
  }\bibfield  {title} {\enquote {\bibinfo {title} {{Probing the universe with
  the Lyman-$\alpha$ forest: 1. Hydrodynamics of the low density IGM}},}\ }\Doi
  {10.1046/j.1365-8711.1998.01249.x} {\bibfield  {journal} {\bibinfo  {journal}
  {\mnras},\ }\textbf {\bibinfo {volume} {296}},\ \bibinfo {pages} {44--55}
  (\bibinfo {year} {1998})},\ \Eprint {http://arxiv.org/abs/astro-ph/9706219}
  {arXiv:astro-ph/9706219} \BibitemShut {NoStop}%
\bibitem [{\citenamefont {Croft}\ \emph {et~al.}(1999)\citenamefont {Croft},
  \citenamefont {Weinberg}, \citenamefont {Pettini}, \citenamefont
  {Hernquist},\ and\ \citenamefont {Katz}}]{Croft:1998pe}%
  \BibitemOpen
  \bibfield  {author} {\bibinfo {author} {\bibfnamefont {R.A.C.}\ \bibnamefont
  {Croft}}, \bibinfo {author} {\bibfnamefont {D.~H.}\ \bibnamefont {Weinberg}},
  \bibinfo {author} {\bibfnamefont {M.}~\bibnamefont {Pettini}}, \bibinfo
  {author} {\bibfnamefont {L.}~\bibnamefont {Hernquist}}, \ and\ \bibinfo
  {author} {\bibfnamefont {N.}~\bibnamefont {Katz}},\ }\bibfield  {title}
  {\enquote {\bibinfo {title} {{The power spectrum of mass fluctuations
  measured from the Lyman-$\alpha$ forest at redshift $z=2.5$}},}\ }\Doi
  {10.1086/307438} {\bibfield  {journal} {\bibinfo  {journal} {Astrophys. J.},\
  }\textbf {\bibinfo {volume} {520}},\ \bibinfo {pages} {1--23} (\bibinfo
  {year} {1999})},\ \Eprint {http://arxiv.org/abs/astro-ph/9809401}
  {arXiv:astro-ph/9809401} \BibitemShut {NoStop}%
\bibitem [{\citenamefont {Hui}(1999)}]{H98}%
  \BibitemOpen
  \bibfield  {author} {\bibinfo {author} {\bibfnamefont {L.}~\bibnamefont
  {Hui}},\ }\bibfield  {title} {\enquote {\bibinfo {title} {{Recovery of the
  shape of the mass power spectrum from the Lyman-$\alpha$ forest}},}\ }\Doi
  {10.1086/307134} {\bibfield  {journal} {\bibinfo  {journal} {Astrophys. J.},\
  }\textbf {\bibinfo {volume} {516}},\ \bibinfo {pages} {519--526} (\bibinfo
  {year} {1999})},\ \Eprint {http://arxiv.org/abs/astro-ph/9807068}
  {arXiv:astro-ph/9807068} \BibitemShut {NoStop}%
\bibitem [{\citenamefont {Hui}\ and\ \citenamefont {Gnedin}(1997)}]{HG97}%
  \BibitemOpen
  \bibfield  {author} {\bibinfo {author} {\bibfnamefont {L.}~\bibnamefont
  {Hui}}\ and\ \bibinfo {author} {\bibfnamefont {N.~Y.}\ \bibnamefont
  {Gnedin}},\ }\bibfield  {title} {\enquote {\bibinfo {title} {{Equation of
  state of the photoionized intergalactic medium}},}\ }\Doi
  {10.1093/mnras/292.1.27} {\bibfield  {journal} {\bibinfo  {journal}
  {\mnras},\ }\textbf {\bibinfo {volume} {292}},\ \bibinfo {pages} {27--42}
  (\bibinfo {year} {1997})},\ \Eprint {http://arxiv.org/abs/astro-ph/9612232}
  {arXiv:astro-ph/9612232} \BibitemShut {NoStop}%
\bibitem [{\citenamefont {McDonald}\ \emph {et~al.}(2005)\citenamefont
  {McDonald}, \citenamefont {Seljak}, \citenamefont {Cen}, \citenamefont
  {Bode},\ and\ \citenamefont {Ostriker}}]{McDonald04}%
  \BibitemOpen
  \bibfield  {author} {\bibinfo {author} {\bibfnamefont {P.}~\bibnamefont
  {McDonald}}, \bibinfo {author} {\bibfnamefont {U.}~\bibnamefont {Seljak}},
  \bibinfo {author} {\bibfnamefont {R.}~\bibnamefont {Cen}}, \bibinfo {author}
  {\bibfnamefont {P.}~\bibnamefont {Bode}}, \ and\ \bibinfo {author}
  {\bibfnamefont {J.~P.}\ \bibnamefont {Ostriker}},\ }\bibfield  {title}
  {\enquote {\bibinfo {title} {{Physical effects on the Ly--$\alpha$ forest
  flux power spectrum: Damping wings, ionizing radiation fluctuations, and
  galactic winds}},}\ }\Doi {10.1111/j.1365-2966.2005.09141.x} {\bibfield
  {journal} {\bibinfo  {journal} {Mon. Not. Roy. Astron. Soc.},\ }\textbf
  {\bibinfo {volume} {360}},\ \bibinfo {pages} {1471--1482} (\bibinfo {year}
  {2005})},\ \Eprint {http://arxiv.org/abs/astro-ph/0407378}
  {arXiv:astro-ph/0407378} \BibitemShut {NoStop}%
\bibitem [{\citenamefont {Croft}(2004)}]{Croft2003}%
  \BibitemOpen
  \bibfield  {author} {\bibinfo {author} {\bibfnamefont {R.A.C.}\ \bibnamefont
  {Croft}},\ }\bibfield  {title} {\enquote {\bibinfo {title} {{Ionizing
  radiation fluctuations and large scale structure in the Lyman-$\alpha$
  forest}},}\ }\Doi {10.1086/421839} {\bibfield  {journal} {\bibinfo  {journal}
  {Astrophys. J.},\ }\textbf {\bibinfo {volume} {610}},\ \bibinfo {pages}
  {642--662} (\bibinfo {year} {2004})},\ \Eprint
  {http://arxiv.org/abs/astro-ph/0310890} {arXiv:astro-ph/0310890} \BibitemShut
  {NoStop}%
\bibitem [{\citenamefont {Meiksin}\ and\ \citenamefont
  {White}(2004)}]{MeiksinWhite}%
  \BibitemOpen
  \bibfield  {author} {\bibinfo {author} {\bibfnamefont {A.}~\bibnamefont
  {Meiksin}}\ and\ \bibinfo {author} {\bibfnamefont {M.~J.}\ \bibnamefont
  {White}},\ }\bibfield  {title} {\enquote {\bibinfo {title} {{The effects of
  UV background correlations on Ly-$\alpha$ forest flux statistics}},}\ }\Doi
  {10.1111/j.1365-2966.2004.07724.x} {\bibfield  {journal} {\bibinfo  {journal}
  {Mon. Not. Roy. Astron. Soc.},\ }\textbf {\bibinfo {volume} {350}},\ \bibinfo
  {pages} {1107} (\bibinfo {year} {2004})},\ \Eprint
  {http://arxiv.org/abs/astro-ph/0307289} {arXiv:astro-ph/0307289} \BibitemShut
  {NoStop}%
\bibitem [{\citenamefont {Hui}\ and\ \citenamefont {Haiman}(2003)}]{HH03}%
  \BibitemOpen
  \bibfield  {author} {\bibinfo {author} {\bibfnamefont {L.}~\bibnamefont
  {Hui}}\ and\ \bibinfo {author} {\bibfnamefont {Z.}~\bibnamefont {Haiman}},\
  }\bibfield  {title} {\enquote {\bibinfo {title} {{The thermal memory of
  reionization history}},}\ }\Doi {10.1086/377229} {\bibfield  {journal}
  {\bibinfo  {journal} {Astrophys. J.},\ }\textbf {\bibinfo {volume} {596}},\
  \bibinfo {pages} {9--18} (\bibinfo {year} {2003})},\ \Eprint
  {http://arxiv.org/abs/astro-ph/0302439} {arXiv:astro-ph/0302439} \BibitemShut
  {NoStop}%
\bibitem [{\citenamefont {D'Aloisio}\ \emph {et~al.}(2015)\citenamefont
  {D'Aloisio}, \citenamefont {McQuinn},\ and\ \citenamefont
  {Trac}}]{Aloisio2015}%
  \BibitemOpen
  \bibfield  {author} {\bibinfo {author} {\bibfnamefont {A.}~\bibnamefont
  {D'Aloisio}}, \bibinfo {author} {\bibfnamefont {M.}~\bibnamefont {McQuinn}},
  \ and\ \bibinfo {author} {\bibfnamefont {H.}~\bibnamefont {Trac}},\
  }\bibfield  {title} {\enquote {\bibinfo {title} {{Large opacity variations in
  the high-redshift Ly--$\alpha$ forest: the signature of relic temperature
  fluctuations from patchy reionization}},}\ }\Doi
  {10.1088/2041-8205/813/2/L38} {\bibfield  {journal} {\bibinfo  {journal}
  {Astrophys. J.},\ }\textbf {\bibinfo {volume} {813}},\ \bibinfo {pages} {L38}
  (\bibinfo {year} {2015})},\ \Eprint {http://arxiv.org/abs/1509.02523}
  {arXiv:1509.02523} \BibitemShut {NoStop}%
\bibitem [{\citenamefont {Cen}\ \emph {et~al.}(2009)\citenamefont {Cen},
  \citenamefont {McDonald}, \citenamefont {Trac},\ and\ \citenamefont
  {Loeb}}]{Cen2009}%
  \BibitemOpen
  \bibfield  {author} {\bibinfo {author} {\bibfnamefont {R.}~\bibnamefont
  {Cen}}, \bibinfo {author} {\bibfnamefont {P.}~\bibnamefont {McDonald}},
  \bibinfo {author} {\bibfnamefont {H.}~\bibnamefont {Trac}}, \ and\ \bibinfo
  {author} {\bibfnamefont {A.}~\bibnamefont {Loeb}},\ }\bibfield  {title}
  {\enquote {\bibinfo {title} {{Probing the epoch of reionization with the
  Lyman-$\alpha$ forest at $z\sim4$--5}},}\ }\Doi
  {10.1088/0004-637X/706/1/L164} {\bibfield  {journal} {\bibinfo  {journal}
  {Astrophys. J.},\ }\textbf {\bibinfo {volume} {706}},\ \bibinfo {pages}
  {L164--L167} (\bibinfo {year} {2009})},\ \Eprint
  {http://arxiv.org/abs/0907.0735} {arXiv:0907.0735} \BibitemShut {NoStop}%
\bibitem [{\citenamefont {Palanque-Delabrouille}\ \emph
  {et~al.}(2013)\citenamefont {Palanque-Delabrouille} \emph
  {et~al.}}]{Palanque2013}%
  \BibitemOpen
  \bibfield  {author} {\bibinfo {author} {\bibfnamefont {N.}~\bibnamefont
  {Palanque-Delabrouille}} \emph {et~al.},\ }\bibfield  {title} {\enquote
  {\bibinfo {title} {{The one-dimensional Ly-$\alpha$ forest power spectrum
  from BOSS}},}\ }\Doi {10.1051/0004-6361/201322130} {\bibfield  {journal}
  {\bibinfo  {journal} {Astron. Astrophys.},\ }\textbf {\bibinfo {volume}
  {559}},\ \bibinfo {pages} {A85} (\bibinfo {year} {2013})},\ \Eprint
  {http://arxiv.org/abs/1306.5896} {arXiv:1306.5896} \BibitemShut {NoStop}%
\bibitem [{\citenamefont {Zaldarriaga}\ \emph {et~al.}(2001)\citenamefont
  {Zaldarriaga}, \citenamefont {Hui},\ and\ \citenamefont {Tegmark}}]{ZHT2001}%
  \BibitemOpen
  \bibfield  {author} {\bibinfo {author} {\bibfnamefont {M.}~\bibnamefont
  {Zaldarriaga}}, \bibinfo {author} {\bibfnamefont {L.}~\bibnamefont {Hui}}, \
  and\ \bibinfo {author} {\bibfnamefont {M.}~\bibnamefont {Tegmark}},\
  }\bibfield  {title} {\enquote {\bibinfo {title} {{Constraints from the
  Lyman-$\alpha$ forest power spectrum}},}\ }\Doi {10.1086/321652} {\bibfield
  {journal} {\bibinfo  {journal} {Astrophys. J.},\ }\textbf {\bibinfo {volume}
  {557}},\ \bibinfo {pages} {519--526} (\bibinfo {year} {2001})},\ \Eprint
  {http://arxiv.org/abs/astro-ph/0011559} {arXiv:astro-ph/0011559} \BibitemShut
  {NoStop}%
\bibitem [{\citenamefont {Zaldarriaga}\ \emph {et~al.}(2003)\citenamefont
  {Zaldarriaga}, \citenamefont {Scoccimarro},\ and\ \citenamefont
  {Hui}}]{ZSH2003}%
  \BibitemOpen
  \bibfield  {author} {\bibinfo {author} {\bibfnamefont {M.}~\bibnamefont
  {Zaldarriaga}}, \bibinfo {author} {\bibfnamefont {R.}~\bibnamefont
  {Scoccimarro}}, \ and\ \bibinfo {author} {\bibfnamefont {L.}~\bibnamefont
  {Hui}},\ }\bibfield  {title} {\enquote {\bibinfo {title} {{Inferring the
  linear power spectrum from the Lyman-$\alpha$ forest}},}\ }\Doi
  {10.1086/374407} {\bibfield  {journal} {\bibinfo  {journal} {Astrophys. J.},\
  }\textbf {\bibinfo {volume} {590}},\ \bibinfo {pages} {1--7} (\bibinfo {year}
  {2003})},\ \Eprint {http://arxiv.org/abs/astro-ph/0111230}
  {arXiv:astro-ph/0111230} \BibitemShut {NoStop}%
\bibitem [{\citenamefont {Garzilli}\ \emph {et~al.}(2015)\citenamefont
  {Garzilli}, \citenamefont {Boyarsky},\ and\ \citenamefont
  {Ruchayskiy}}]{Garzilli2015}%
  \BibitemOpen
  \bibfield  {author} {\bibinfo {author} {\bibfnamefont {A.}~\bibnamefont
  {Garzilli}}, \bibinfo {author} {\bibfnamefont {A.}~\bibnamefont {Boyarsky}},
  \ and\ \bibinfo {author} {\bibfnamefont {O.}~\bibnamefont {Ruchayskiy}},\
  }\bibfield  {title} {\enquote {\bibinfo {title} {{Cutoff in the Lyman alpha
  forest power spectrum: warm IGM or warm dark matter?}}}\ }\href@noop {} {
  (\bibinfo {year} {2015})},\ \Eprint {http://arxiv.org/abs/1510.07006}
  {arXiv:1510.07006} \BibitemShut {NoStop}%
\bibitem [{\citenamefont {Seljak}\ \emph {et~al.}(2006)\citenamefont {Seljak},
  \citenamefont {Makarov}, \citenamefont {McDonald},\ and\ \citenamefont
  {Trac}}]{Seljak2006}%
  \BibitemOpen
  \bibfield  {author} {\bibinfo {author} {\bibfnamefont {U.}~\bibnamefont
  {Seljak}}, \bibinfo {author} {\bibfnamefont {A.}~\bibnamefont {Makarov}},
  \bibinfo {author} {\bibfnamefont {P.}~\bibnamefont {McDonald}}, \ and\
  \bibinfo {author} {\bibfnamefont {H.}~\bibnamefont {Trac}},\ }\bibfield
  {title} {\enquote {\bibinfo {title} {{Can sterile neutrinos be the dark
  matter?}}}\ }\Doi {10.1103/PhysRevLett.97.191303} {\bibfield  {journal}
  {\bibinfo  {journal} {Phys. Rev. Lett.},\ }\textbf {\bibinfo {volume} {97}},\
  \bibinfo {pages} {191303} (\bibinfo {year} {2006})},\ \Eprint
  {http://arxiv.org/abs/arXiv:astro-ph/0602430} {arXiv:astro-ph/0602430}
  \BibitemShut {NoStop}%
\bibitem [{\citenamefont {Baur}\ \emph {et~al.}(2015)\citenamefont {Baur},
  \citenamefont {Palanque-Delabrouille}, \citenamefont {Yèche}, \citenamefont
  {Magneville},\ and\ \citenamefont {Viel}}]{Baur2015}%
  \BibitemOpen
  \bibfield  {author} {\bibinfo {author} {\bibfnamefont {J.}~\bibnamefont
  {Baur}}, \bibinfo {author} {\bibfnamefont {N.}~\bibnamefont
  {Palanque-Delabrouille}}, \bibinfo {author} {\bibfnamefont {C.}~\bibnamefont
  {Yèche}}, \bibinfo {author} {\bibfnamefont {C.}~\bibnamefont {Magneville}},
  \ and\ \bibinfo {author} {\bibfnamefont {M.}~\bibnamefont {Viel}},\
  }\bibfield  {title} {\enquote {\bibinfo {title} {{Lyman-$\alpha$ forests cool
  warm dark matter}},}\ }in\ \href
  {https://inspirehep.net/record/1408473/files/arXiv:1512.01981.pdf} {\emph
  {\bibinfo {booktitle} {{SDSS-IV Collaboration Meeting, July 20-23, 2015}}}}\
  (\bibinfo {year} {2015})\ \Eprint {http://arxiv.org/abs/1512.01981}
  {arXiv:1512.01981} \BibitemShut {NoStop}%
\bibitem [{\citenamefont {Viel}\ \emph
  {et~al.}(2013){\natexlab{b}}\citenamefont {Viel}, \citenamefont {Schaye},\
  and\ \citenamefont {Booth}}]{Viel2012}%
  \BibitemOpen
  \bibfield  {author} {\bibinfo {author} {\bibfnamefont {M.}~\bibnamefont
  {Viel}}, \bibinfo {author} {\bibfnamefont {J.}~\bibnamefont {Schaye}}, \ and\
  \bibinfo {author} {\bibfnamefont {C.~M.}\ \bibnamefont {Booth}},\ }\bibfield
  {title} {\enquote {\bibinfo {title} {{The impact of feedback from galaxy
  formation on the Lyman-$\alpha$ transmitted flux}},}\ }\Doi
  {10.1093/mnras/sts465} {\bibfield  {journal} {\bibinfo  {journal} {\mnras},\
  }\textbf {\bibinfo {volume} {429}},\ \bibinfo {pages} {1734--1746} (\bibinfo
  {year} {2013}{\natexlab{b}})},\ \Eprint {http://arxiv.org/abs/1207.6567}
  {arXiv:1207.6567} \BibitemShut {NoStop}%
\bibitem [{\citenamefont {Adelberger}\ \emph {et~al.}(2003)\citenamefont
  {Adelberger}, \citenamefont {Steidel}, \citenamefont {Shapley},\ and\
  \citenamefont {Pettini}}]{Adelberger2003}%
  \BibitemOpen
  \bibfield  {author} {\bibinfo {author} {\bibfnamefont {K.~L.}\ \bibnamefont
  {Adelberger}}, \bibinfo {author} {\bibfnamefont {C.~C.}\ \bibnamefont
  {Steidel}}, \bibinfo {author} {\bibfnamefont {A.~E.}\ \bibnamefont
  {Shapley}}, \ and\ \bibinfo {author} {\bibfnamefont {M.}~\bibnamefont
  {Pettini}},\ }\bibfield  {title} {\enquote {\bibinfo {title} {{Galaxies and
  intergalactic matter at redshift $z\sim 3$: overview}},}\ }\Doi
  {10.1086/345660} {\bibfield  {journal} {\bibinfo  {journal} {Astrophys. J.},\
  }\textbf {\bibinfo {volume} {584}},\ \bibinfo {pages} {45--75} (\bibinfo
  {year} {2003})},\ \Eprint {http://arxiv.org/abs/astro-ph/0210314}
  {arXiv:astro-ph/0210314} \BibitemShut {NoStop}%
\bibitem [{\citenamefont {{van Dokkum}}\ \emph {et~al.}(2016)\citenamefont
  {{van Dokkum}}, \citenamefont {{Abraham}}, \citenamefont {{Brodie}},
  \citenamefont {{Conroy}}, \citenamefont {{Danieli}}, \citenamefont
  {{Merritt}}, \citenamefont {{Mowla}}, \citenamefont {{Romanowsky}},\ and\
  \citenamefont {{Zhang}}}]{vandokkum2016}%
  \BibitemOpen
  \bibfield  {author} {\bibinfo {author} {\bibfnamefont {P.}~\bibnamefont {{van
  Dokkum}}}, \bibinfo {author} {\bibfnamefont {R.}~\bibnamefont {{Abraham}}},
  \bibinfo {author} {\bibfnamefont {J.}~\bibnamefont {{Brodie}}}, \bibinfo
  {author} {\bibfnamefont {C.}~\bibnamefont {{Conroy}}}, \bibinfo {author}
  {\bibfnamefont {S.}~\bibnamefont {{Danieli}}}, \bibinfo {author}
  {\bibfnamefont {A.}~\bibnamefont {{Merritt}}}, \bibinfo {author}
  {\bibfnamefont {L.}~\bibnamefont {{Mowla}}}, \bibinfo {author} {\bibfnamefont
  {A.}~\bibnamefont {{Romanowsky}}}, \ and\ \bibinfo {author} {\bibfnamefont
  {J.}~\bibnamefont {{Zhang}}},\ }\bibfield  {title} {\enquote {\bibinfo
  {title} {{A high stellar velocity dispersion and $\sim 100$ globular clusters
  for the ultra-diffuse galaxy Dragonfly 44}},}\ }\Doi
  {10.3847/2041-8205/828/1/L6} {\bibfield  {journal} {\bibinfo  {journal}
  {\apjl},\ }\textbf {\bibinfo {volume} {828}},\ \bibinfo {eid} {L6} (\bibinfo
  {year} {2016})},\ \Eprint {http://arxiv.org/abs/1606.06291}
  {arXiv:1606.06291} \BibitemShut {NoStop}%
\bibitem [{\citenamefont {Bernstein}\ \emph {et~al.}(1998)\citenamefont
  {Bernstein}, \citenamefont {Giladi},\ and\ \citenamefont {Jones}}]{bgj98}%
  \BibitemOpen
  \bibfield  {author} {\bibinfo {author} {\bibfnamefont {D.~H.}\ \bibnamefont
  {Bernstein}}, \bibinfo {author} {\bibfnamefont {E.}~\bibnamefont {Giladi}}, \
  and\ \bibinfo {author} {\bibfnamefont {K.R.W.}\ \bibnamefont {Jones}},\
  }\bibfield  {title} {\enquote {\bibinfo {title} {{Eigenstates of the general
  Schr\"odinger equation}},}\ }\Doi {10.1103/PhysRev.187.1767} {\bibfield
  {journal} {\bibinfo  {journal} {Mod. Phys. Lett. A},\ }\textbf {\bibinfo
  {volume} {13}},\ \bibinfo {pages} {2327--2336} (\bibinfo {year}
  {1998})}\BibitemShut {NoStop}%
\bibitem [{\citenamefont {{Guzm{\'a}n}}\ and\ \citenamefont
  {{Ure{\~n}a-L{\'o}pez}}(2004)}]{guzman2004}%
  \BibitemOpen
  \bibfield  {author} {\bibinfo {author} {\bibfnamefont {F.~S.}\ \bibnamefont
  {{Guzm{\'a}n}}}\ and\ \bibinfo {author} {\bibfnamefont {L.~A.}\ \bibnamefont
  {{Ure{\~n}a-L{\'o}pez}}},\ }\bibfield  {title} {\enquote {\bibinfo {title}
  {{Evolution of the Schr{\"o}dinger-Newton system for a self-gravitating
  scalar field}},}\ }\Doi {10.1103/PhysRevD.69.124033} {\bibfield  {journal}
  {\bibinfo  {journal} {\prd},\ }\textbf {\bibinfo {volume} {69}},\ \bibinfo
  {eid} {124033} (\bibinfo {year} {2004})},\ \Eprint
  {http://arxiv.org/abs/gr-qc/0404014} {gr-qc/0404014} \BibitemShut {NoStop}%
\bibitem [{\citenamefont {{Mott}}\ and\ \citenamefont
  {{Massey}}(1949)}]{MottMassey}%
  \BibitemOpen
  \bibfield  {author} {\bibinfo {author} {\bibfnamefont {N.~F.}\ \bibnamefont
  {{Mott}}}\ and\ \bibinfo {author} {\bibfnamefont {H.S.W.}\ \bibnamefont
  {{Massey}}},\ }\href@noop {} {\emph {\bibinfo {title} {The Theory of Atomic
  Collisions}}}\ (\bibinfo  {publisher} {Oxford University, Clarendon Press},\
  \bibinfo {year} {1949})\BibitemShut {NoStop}%
\bibitem [{\citenamefont {{Baryshevskii}}\ \emph {et~al.}(2004)\citenamefont
  {{Baryshevskii}}, \citenamefont {{Feranchuk}},\ and\ \citenamefont
  {{Kats}}}]{BFK2004}%
  \BibitemOpen
  \bibfield  {author} {\bibinfo {author} {\bibfnamefont {V.~G.}\ \bibnamefont
  {{Baryshevskii}}}, \bibinfo {author} {\bibfnamefont {I.~D.}\ \bibnamefont
  {{Feranchuk}}}, \ and\ \bibinfo {author} {\bibfnamefont {P.~B.}\ \bibnamefont
  {{Kats}}},\ }\bibfield  {title} {\enquote {\bibinfo {title} {{Regularization
  of the Coulomb scattering problem}},}\ }\Doi {10.1103/PhysRevA.70.052701}
  {\bibfield  {journal} {\bibinfo  {journal} {\pra},\ }\textbf {\bibinfo
  {volume} {70}},\ \bibinfo {eid} {052701} (\bibinfo {year} {2004})},\ \Eprint
  {http://arxiv.org/abs/arXiv:quant-ph/0403050} {arXiv:quant-ph/0403050}
  \BibitemShut {NoStop}%
\bibitem [{\citenamefont {Karule}(1990)}]{Karule1990}%
  \BibitemOpen
  \bibfield  {author} {\bibinfo {author} {\bibfnamefont {E.}~\bibnamefont
  {Karule}},\ }\bibfield  {title} {\enquote {\bibinfo {title} {{Integrals of
  two confluent hypergeometric functions}},}\ }\Doi {10.1103/PhysRev.187.1767}
  {\bibfield  {journal} {\bibinfo  {journal} {Journal of Physics A},\ }\textbf
  {\bibinfo {volume} {23}},\ \bibinfo {pages} {1969--1971} (\bibinfo {year}
  {1990})}\BibitemShut {NoStop}%
\bibitem [{\citenamefont {{Weinberg}}(1989)}]{wein89}%
  \BibitemOpen
  \bibfield  {author} {\bibinfo {author} {\bibfnamefont {M.~D.}\ \bibnamefont
  {{Weinberg}}},\ }\bibfield  {title} {\enquote {\bibinfo {title}
  {{Self-gravitating response of a spherical galaxy to sinking satellites}},}\
  }\Doi {10.1093/mnras/239.2.549} {\bibfield  {journal} {\bibinfo  {journal}
  {\mnras},\ }\textbf {\bibinfo {volume} {239}},\ \bibinfo {pages} {549--569}
  (\bibinfo {year} {1989})}\BibitemShut {NoStop}%
\end{thebibliography}%

\end{document}